\documentclass[12pt,oneside]{article}

\usepackage{epsfig,lscape}
\usepackage{amssymb,amsfonts,amsmath}
\usepackage{rotating}

\usepackage{amsthm}

\usepackage{color}
\usepackage[dvipsnames]{xcolor}

\usepackage{float}
\usepackage{hyperref}  

\def\me{\mathrm e}

\def\dif{\mathrm d}

\def\cov{\mathrm{cov}}
\def\N{\mathrm{N}}

\def\T{ {\mathrm{\scriptscriptstyle T}} }

\def\bbR{{\mathbb{R}}}
\def\argmin{\mathrm{argmin}}

\newcommand\mytext[1]{\text{\scriptsize{#1}}}

\newcommand\dhat[1]{\hat{\hat{#1}}}

\newenvironment{prf}
{\noindent \textbf{Proof.}}{\hfill $\Box$ \vspace{.1in}}

\newtheorem{thm}{Theorem}
\newtheorem{lem}{Lemma}
\newtheorem{pro}{Proposition}
\newtheorem{cor}{Corollary}
\newtheorem{ass}{Assumption}

\theoremstyle{definition}

\theoremstyle{definition}

\allowdisplaybreaks

\usepackage[medium,compact]{titlesec}

\titlespacing*{\section} {0pt}{1.5ex}{1ex}
\titlespacing*{\subsection} {0pt}{1.5ex}{1ex}
\titlespacing*{\subsubsection} {0pt}{1ex}{1ex}

\begin{document}

\setlength{\abovedisplayskip}{5pt}
\setlength{\belowdisplayskip}{5pt}


\begin{titlepage}

\begin{center}
{\Large Model-assisted sensitivity analysis for treatment effects under unmeasured confounding via regularized calibrated estimation}

\vspace{.1in} Zhiqiang Tan\footnotemark[1]

\vspace{.1in}
\today
\end{center}

\footnotetext[1]{Department of Statistics, Rutgers University. Address: 110 Frelinghuysen Road,
Piscataway, NJ 08854. E-mail: ztan@stat.rutgers.edu.}

\paragraph{Abstract.}
Consider sensitivity analysis for estimating average treatment effects under unmeasured confounding, assumed to satisfy a marginal sensitivity model.
At the population level,
we provide new representations for the sharp population bounds and doubly robust estimating functions, recently derived by Dorn, Guo, and Kallus.
We also derive new, relaxed population bounds, depending on weighted linear outcome quantile regression.
At the sample level, we develop new methods and theory for obtaining not only doubly robust point estimators for the relaxed population bounds
with respect to misspecification of a propensity score model or an outcome mean regression model,
but also model-assisted confidence intervals which are valid if the propensity score model is correctly specified, but
the outcome quantile and mean regression models may be misspecified.
The relaxed population bounds reduce to the sharp bounds if outcome quantile regression is correctly specified.
For a linear outcome mean regression model, the confidence intervals are also doubly robust.
Our methods involve regularized calibrated estimation, with Lasso penalties but carefully chosen loss functions,
for fitting propensity score and outcome mean and quantile regression models.
We present a simulation study and
an empirical application to an observational study on the effects of right heart catheterization.

\end{titlepage}

\section{Introduction} \label{sec:intro}

Drawing inferences about average treatment effects (ATEs) from non-randomized, observational studies
has been of tremendous interest in statistics, economics, epidemiology and other fields.
When a reasonable collection of covariates are measured to control for confounding,
estimation of average treatment effects is often performed under the assumption of no unmeasured confounding (or unconfoundedness),
which treats the observational data as if they were from conditionally randomized experiments given the measured covariates.
However, this assumption can potentially be violated for various reasons such as a limited scope of the measured covariates or a limited understanding of
the confounding mechanism. It is important to conduct sensitivity analysis
to assess how the estimates might change under unmeasured confounding.

There have been various methods proposed for sensitivity analysis in causal inference.
Each sensitivity analysis method involves a parametric or semiparametric sensitivity model,
often with a sensitivity parameter encoding the magnitude of unmeasured confounding.
For example, the latent sensitivity model of Rosenbaum (2002) assumes that the odds of receiving the treatment
can differ at most by a factor of $\Gamma $ between individuals with the same covariates but different values of an unmeasured confounder.
Alternatively, the marginal sensitivity model of Tan (2006) assumes that
the odds of receiving the treatment given an unmeasured confounder and the covariates
can differ at most by a factor of $\Lambda $ from that implied by the propensity score (PS) (Rosenbaum \& Rubin 1983),
that is, the probability of receiving the treatment given the covariates only, regardless of the unmeasured confounder.
Compared with Rosenbaum's model, this model is motivated to directly constrain the density ratio of potential outcomes and
facilitate the use of inverse probability weighted (IPW) estimation and related techniques.
A linear programming method was proposed in Tan (2006) to obtain (conservative) point bounds for ATEs, but without providing quantification of sampling variation.
See Section \ref{sec:setup} for further discussion of these sensitivity models.

Recently, substantial progress has been made to develop sensitivity analysis methods using the marginal sensitivity model.
Valid bootstrap confidence intervals for ATEs were proposed by Zhao et al.~(2019) under a correctly specified PS model,
although the confidence intervals are inherently conservative, without fully accounting for the constraints under the sensitivity model.
The sharp population bounds under the marginal sensitivity model were solved by Dorn \& Guo (2022) depending on outcome quantile regression.
Furthermore, point bounds and confidence intervals were developed in Dorn et al.~(2021) with the following properties. \vspace{-.05in}
\begin{itemize}\addtolength{\itemsep}{-.05in}
\item[(i)] The point bounds are asymptotically conservative for the sharp population bounds if
either a PS model or an outcome mean regression model is correctly specified, but outcome quantile regression may be misspecified.
Moreover, these bounds are consistent for the sharp bounds if, further, outcome quantile regression is correctly specified.

\item[(ii)] For $0<c<1$,
the $(1-c)$-confidence intervals are of asymptotic levels $1-c$ (i.e., with coverage probabilities $ \ge 1-c$) for the sharp population bounds
if both a PS model and an outcome mean regression model are correctly specified, but outcome quantile regression may be misspecified.
Moreover, these confidence intervals are of asymptotic sizes $1-c$ (i.e., with coverage probabilities $1-c$)
if, further, outcome quantile regression is correctly specified.
\end{itemize} \vspace{-.05in}
The PS and outcome regression models above can be represented by machine learning algorithms.
By correct specification, these algorithms are assumed to be consistent for the true parameters (or true covariate functions) 
and, in case of (ii), at sufficiently fast rates.
The point bounds are doubly robust, but the confidence intervals
are valid only when both PS and outcome mean regression models are correctly specified.
The lack of robustness can be attributed to the fact that the nuisance parameters are estimated by maximum likelihood or related methods,
while ignoring 
the impact of sampling variations of the fitted functions on the point bounds; see Section \ref{sec:cal} for further discussion.

In this work, we build on the recent progress and make several contributions to sensitivity analysis using the marginal sensitivity model.
First, we provide new formulas for the sharp population bounds in Dorn \& Guo (2022),
which reveal a duality relationship of these bounds with the objective values from weighted outcome quantile regression.
We also derive new, relaxed population bounds, depending on weighted linear outcome quantile regression,
which correspond to the population version of the linear programming method in Tan (2006)
using a finite subset of inherent constraints under the marginal sensitivity model.
While the sharp bounds are achievable only when the true outcome quantile function is identified,
these relaxed bounds represent the optimal population bounds possibly obtained when using
outcome quantile regression with pre-specified regressors.
Furthermore, our formulas for the population bounds shed new light on the role of outcome predictive modeling in sensitivity analysis:
how wide the sensitivity intervals are depends on how accurately outcomes can be predicted through quantile regression.
See Section \ref{sec:DR-bound} for further discussion.

Second, we show that the estimating functions used in Dorn et al.~(2021) are doubly robust for relaxed population bounds,
if either a PS model or an outcome mean regression model is correctly specified, while outcome quantile regression may be misspecified.
These relaxed population bounds depend on the limits of the fitted quantile functions, and reduce to the sharp bounds if
outcome quantile regression is correctly specified. This viewpoint not only clarifies the roles of the  
PS and outcome mean and quantile regression models involved,
but also makes it possible to study point estimation and confidence intervals for
relaxed population bounds as target population quantities.

Third, we develop new methods for obtaining doubly robust point bounds and confidence intervals for the ATEs,
by extending regularized calibrated estimation for fitting the PS and outcome regression models (Tan 2020b; Ghosh \& Tan 2022)
while using the doubly robust estimating functions in high-dimensional settings.
The Lasso penalties are employed (Tibishirani 1996; Buhlmann \& van de Geer 2011),
but the loss functions are carefully chosen such that, with possible misspecification of the working models,
the resulting point bounds admit simple asymptotic expansions
of order $O_p(n^{-1/2})$ under suitable sparsity conditions.
For linear outcome quantile and mean regression models, we show that
our point bounds and Wald confidence intervals, derived from the simple asymptotic expansions, achieve both of the following properties:\vspace{-.05in}
\begin{itemize}\addtolength{\itemsep}{-.05in}
\item[(iii)] The point bounds are doubly robust, in being consistent for the relaxed population bounds if
either a PS model or an outcome mean regression model is correctly specified, but outcome quantile regression may be misspecified.
The relaxed population bounds reduce to the sharp bounds if, further, outcome quantile regression is correctly specified.

\item[(iv)] For $0<c<1$,
the $(1-c)$-confidence intervals are doubly robust, in achieving asymptotic sizes $1-c$ for the relaxed population bounds
if either a PS model or an outcome mean regression model are correctly specified, but outcome quantile regression may be misspecified.
\end{itemize} \vspace{-.05in}
Compared with properties (i)--(ii) from Dorn et al.~(2021), our results lead to doubly robust point bounds and confidence intervals,
with the relaxed population bounds as target parameters.
For binary outcomes with logistic outcome regression used, our confidence intervals are valid
if a PS model is correctly specified, but logistic outcome regression may be misspecified. In this sense, they are said to be
PS model-based and outcome model-assisted.

\vspace{.05in}
\noindent{\bf Related works.} There is a vast and growing literature on estimation of average treatment
effects. We discuss directly related works to ours, in addition to the earlier discussion.

The marginal sensitivity model (Tan 2006) and its variations have been used in various settings, in addition to ATE estimation.
Examples include estimation of conditional ATEs (Kallus et al.~2019; Lee et al.~2020),
estimation of individual treatment effects (Jin et al.~2021; Yin et al.~2022),
decision rule or policy learning (Kallus \& Zhou 2021; Sahoo et al. 2022), and experimental design (Rosenman \& Owen 2021), among others.

The latent sensitivity model (Rosenbaum 2002) is well suited to matched observational studies,
and various methods have traditionally been developed with that focus (e.g., Fogarty \& Small 2016).
Recently, sharp population bounds and semiparametric estimation for ATEs were derived by Yadlowsky et al.~(2022)
in unmatched, cross-sectional studies. Valid confidence intervals for the ATEs can be obtained if
three nuisance parameters (including the PS and two outcome regression functions) are consistently estimated at rate $o_p (n^{-1/4})$.
This would require that all the working models are correctly specified
with the true parameters satisfying certain sparsity conditions in our setting of high-dimensional regression models.
Moreover, as discussed in the Supplement Section \ref{sec:comparison-model}, the estimating functions used in Yadlowsky et al.~(2022) are not doubly robust.

The marginal sensitivity model is also related to the selection odds model of Robins et al.~(2000), as discussed in Tan (2006, Section 4.2).
Recently, tractable parametric specifications of the model have been proposed by Franks et al.~(2020). Semiparametric estimation
was derived by Scharfstein et al.~(2021) such that valid confidence intervals for ATEs can be obtained
if three nuisance parameters (including the PS and two outcome regression functions) are consistently estimated at sufficiently fast rates.
The estimating functions are constructed from the efficient influence functions, but may not be doubly robust, similarly as in Yadlowsky et al.~(2022).

The estimation approaches in Dorn et al.~(2021), Yadlowsky et al.~(2022), and Scharfstein et al.~(2021) are related to double machine learning
based on estimating functions satisfying Neyman orthogonality (Chernozhukov et al.~2018).
Double robustness and Neyman orthogonality are two distinct concepts.
For estimating functions directly converted from efficient influence functions,
the Neyman orthogonality is automatically satisfied at the true values of nuisance parameters,
but double robustness may not be achieved. See Tan (2019) for another such example in a non-causal inference setting.

Calibrated estimation for fitting PS and outcome regression models was proposed by Kim \&  Haziza (2014) and Vermeulen \& Vansteelandt (2015)
for ATE estimation under unconfoundedness in low-dimensional settings.
Recently, regularized calibrated estimation was developed by Tan (2020b) to obtain model-assisted (sometimes doubly robust) confidence intervals
for ATEs under unconfoundedness in high-dimensional settings.
The method was extended to any semiparametric two-stage estimation provided a doubly robust estimating function
is available with respect to two nuisance parameters.
Our work represents a new development to tackle ATE estimation under the marginal sensitivity model,
where three nuisance parameters are involved as described earlier.

\section{Setup and sensitivity model} \label{sec:setup}

Suppose that
$\{(Y_i, T_i, X_i): i=1,\ldots,n\}$ are independent and identically distributed (IID) observations of $(Y,T,X)$ defined on a probability space $\Omega$, where
$Y: \Omega \to \mathcal Y$ is an outcome variable, $T: \Omega \to \{0,1\}$ is a treatment variable, and $X: \Omega \to \mathcal X$ is a vector of measured
covariates. In the potential outcomes framework (Neyman 1923; Rubin 1974), let
$(Y^0, Y^1)$ be potential outcomes that would be observed under treatment 0 or 1 respectively. By
consistency, assume that $Y$ is either $Y^0$ if $T=0$ or $Y^1$ if $T=1$, that is,
$Y= (1-T) Y^0 + T Y^1$.
There are two causal parameters commonly of interest: the average treatment effect (ATE), defined as
$E(Y^1- Y^0) = \mu^1 - \mu^0$ with $\mu^t= E(Y^t)$, and the average treatment effect on the treated (ATT),
defined as $E(Y^1 - Y^0 |T=1) = \nu^1 - \nu^0$ with $\nu^t = E(Y^t | T=1)$ for $t=0,1$.
For concreteness, we mainly discuss estimation of $\mu^1$, and defer to Supplement Section \ref{sec:ATE} the discussion about estimation of ATE and ATT.

To study inference about $\mu^1$ under unmeasured confounding, we adopt a sensitivity model from Tan (2006),
which characterizes the magnitude of unmeasured confounding as follows. For a sensitivity parameter $\Lambda \ge 1$,
assume that almost surely,
\begin{align}
\Lambda^{-1} \le \lambda^*_1(Y^1,X)  \le \Lambda, \label{eq:sen-model}
\end{align}
where $\lambda^*_1( Y^1, X)$ is the
Radon--Nikodym derivative (or density ratio) between the  conditional distributions, $P(Y^1|T=0,X)$ and $P(Y^1|T=1,X)$,
or, by Bayes' rule, the odds ratio associated with the probabilities $\varpi^*_1(Y^1,X)=P(T=1|Y^1,X)$ and $\pi^*(X) =P(T=1|X)$:
\begin{align}
 \lambda^*_1(Y^1,X) =\frac{\dif P(Y^1| T=0,X)}{\dif P(Y^1 |T=1,X)} = \frac{ \pi^*(X) (1-\varpi^*_1(Y^1,X)) }{(1-\pi^*(X)) \varpi^*_1(Y^1,X)}. \label{eq:lambda1}
\end{align}
By definition, $\lambda^*_1(Y^1,X)$ provides a link between
 identifiable $P(Y^1|T=1,X)$ and $ \pi^*(X) =P(T=1|X)$
and unidentifiable $P(Y^1|T=0,X)$ and $\varpi^*_1(Y^1,X)=P(T=1|Y^1,X)$.
The extreme case of $\Lambda=1$ leads to unconfoundedness (i.e., $T$ and $Y^1$ being independent given $X$), whereas
the deviation of $\lambda^*_1 (Y^1,X)$ from 1 indicates unmeasured confounding.
For any $\Lambda>1$, the magnitude of unmeasured confounding is
restricted under (\ref{eq:sen-model}) such that the density of $P(Y^1|T=0,X)$ differs from that of $P(Y^1 |T=1,X)$ or equivalently
the odds of $T=1$ given $(Y^1,X)$ differs from that given $X$ by at most a factor of $\Lambda$.

Recently, model (\ref{eq:sen-model}) has been called a marginal sensitivity model (Zhao et al.~2019).
There is a variation of the model, which assumes that for a sensitivity parameter $ \Lambda\ge 1$,
\begin{align}
 \Lambda^{-1} \le \frac{ \pi^*(X) P(T=0 |U^1,X)}{(1-\pi^*(X)) P(T=1|U^1,X)} \le  \Lambda. \label{eq:sen-model2}
\end{align}
where $U^1$ is an unmeasured confounder such that $T \perp Y^1 | (X,U^1)$, referred to as latent unconfoundedness.
A slightly stronger version is used in Dorn \& Guo (2022) and Dorn et al.~(2021), where a similar restriction as (\ref{eq:sen-model2})
is assumed, but with $U^1$ replaced by an unmeasured confounder $U$ such that $T \perp (Y^0,Y^1) |(X,U)$.
Nevertheless, their results can be directly extended to model (\ref{eq:sen-model2}).
We show that models (\ref{eq:sen-model}) and (\ref{eq:sen-model2}) are equivalent.

\begin{lem} \label{lem:sen-model2}
(i) For any  $\Lambda\ge 1$, if model (\ref{eq:sen-model}) holds, then
there exists an unmeasured confounder $U^1$ such that $T \perp Y^1 | (X,U^1)$ and (\ref{eq:sen-model2}) holds.
(ii) For any $ \Lambda\ge 1$, if model (\ref{eq:sen-model2}) holds with an unmeasured confounder $U^1$ such that $T \perp Y^1 | (X,U^1)$,
then model (\ref{eq:sen-model}) also holds.
\end{lem}

In Supplement Section \ref{sec:comparison-model}, we compare models (\ref{eq:sen-model}) and (\ref{eq:sen-model2}) with
the latent sensitivity model of Rosenbaum (2002), including recent results from Yadlowsky et al.~(2022).
The sharp bounds of $\mu^1$ under the latent sensitivity model can be recovered from an extension of model (\ref{eq:sen-model}) with
certain covariate-dependent (instead of constant) upper and lower bounds on $\lambda^*_1 (Y^1,X)$.
The estimating functions in Yadlowsky et al.~(2022) for the sharp bounds of $\mu^1$
are not doubly robust with respect to misspecification (or inconsistent estimation) of the nuisance parameters,
even though derived from the efficient influence function and shown to satisfy the Neyman orthogonality (Chernozhukov et al.~2018).

\section{Population sensitivity analysis} \label{sec:pop-SA}

We provide new population bounds on $\mu^1$ under the sensitivity model (\ref{eq:sen-model}),
while assuming that certain parts of the joint distribution of the observed data $(TY, T, X)$ were known,
such as the propensity score $\pi^*(X)$ or the conditional distribution $P(Y|T=1,X)$.
There are two main steps in our development.
First, we derive bounds on $\mu^1$ using IPW estimation,
while allowing that $\lambda^*_1(Y^1, X)$ is unknown but satisfies the sensitivity model (\ref{eq:sen-model}).
Then we derive doubly robust bounds on $\mu^1$ using augmented IPW estimation under sensitivity model (\ref{eq:sen-model}).

\subsection{Bounds based on propensity scores} \label{sec:IPW-bound}

We first describe how $\mu^1$ can be point identified using IPW estimation (or identification) with known propensity score $\pi^*(X)$
if the density or odds ratio $\lambda^*_1( Y^1, X)$ were also known.
As discussed in Tan (2006), the distribution $P(Y^1|X)$ can be decomposed as
\begin{align*}
\dif P(Y^1|X) = \pi^*(X) \dif P(Y^1|T=1,X) + (1-\pi^*(X)) \dif P(Y^1|T=0,X).
\end{align*}
By the definition $\lambda_1(Y^1,X) = \dif P(Y^1| T=0,X) / \dif P(Y^1 |T=1,X)$
and the law of iterated expectations, 
it can be easily shown that
$ \mu^1 = E \{ \psi_{\mytext{IPW}} ( O; \pi^*, \lambda^*_1) \}$ with
\begin{align}
\psi_{\mytext{IPW}} ( O; \pi, \lambda^*_1)= TY + T \frac{1-\pi(X)}{\pi(X)} \lambda^*_1 (Y,X) Y , \label{eq:potential-IPW}
\end{align}
where $O=(TY,T,X)$ and $\pi(X)$ is a covariate function, which represents a putative probability of $T=1$ given $X$
and may differ from the true propensity score $\pi^*$.
In the unconfoundedness case where $\lambda^*_1 \equiv 1$, the function $\psi_{\mytext{IPW}}$ reduces
to the usual IPW estimating function, $TY /\pi(X)$.
Henceforth, $\psi_{\mytext{IPW}}$ can be seen as a potential IPW estimating function for $\mu^1$.

Next, we discuss how $\mu^1$ can be bounded using IPW estimation with known propensity score $\pi^*(X)$, while allowing $\lambda^*_1(Y^1,X)$ to be unknown but satisfy
the sensitivity model (\ref{eq:sen-model}) for any pre-specified $\Lambda$.
In principle, we are interested in optimizing 
\begin{align}
E \{ \psi_{\mytext{IPW}} ( O; \pi^*, \lambda^*_1) \} = E\left\{ TY+ T \frac{1-\pi^*(X)}{\pi^*(X)} \lambda^*_1 (Y,X) Y  \right\} \label{eq:IPW-obj}
\end{align}
over all possible $\lambda^*_1(Y^1,X)$ under model (\ref{eq:sen-model}).
This problem has been elegantly solved by Dorn \& Guo (2022),
and their result can be recapitulated in terms of $\lambda^*_1(Y^1,X)$ as follows.
As discussed in Tan (2006), there are inherent constraints induced by the definition of $\lambda^*_1( Y^1, X)$,
in addition to the range constraint in (\ref{eq:sen-model}).
The expression of $\lambda^*_1 (Y^1,X)$ as a density ratio in (\ref{eq:lambda1}) implies that
$\lambda^*_1 (Y^1,X) \dif P (Y^1 | T=1,X) = \dif P(Y^1 | T=0,X)$ and hence
$E [ \{ T/\pi^{*}(X) \} \lambda^*_1(Y, X) | X ] \equiv 1 $  
Conversely, any function $\lambda_1 (Y^1,X)$ satisfying
\begin{align}
& \Lambda^{-1} \le \lambda_1 (Y^1,X) \le \Lambda, \label{eq:lambda1-givenX-a} \\
& E \left\{ \frac{T}{\pi^*(X)} \lambda_1(Y, X) \Big| X \right\} \equiv  1, \label{eq:lambda1-givenX-b}
\end{align}
can be used to properly define $\dif P(Y^1 | T=0,X) = \lambda_1 (Y^1,X) \dif P (Y^1 | T=1,X)$ and hence such $\lambda_1(Y^1,X)$ can be a valid choice
of $\lambda^*_1(Y^1,X)$ under model (\ref{eq:sen-model}).
Therefore, the set of possible values of $\mu^1$ under model (\ref{eq:sen-model}) is an interval $[\mu^{1-}, \mu^{1+}]$, where
\begin{align}
& \mu^{1-} = \min_{\lambda_1} E\left\{ TY+ T \frac{1-\pi^*(X)}{\pi^*(X)} \lambda_1 (Y,X) Y  \right\}, \label{eq:mu1-}\\
& \mu^{1+} = \max_{\lambda_1} E\left\{ TY+ T \frac{1-\pi^*(X)}{\pi^*(X)} \lambda_1 (Y,X) Y  \right\}, \label{eq:mu1+}
\end{align}
subject to $\lambda_1=\lambda_1(Y^1,X)$ satisfying the constraints (\ref{eq:lambda1-givenX-a})--(\ref{eq:lambda1-givenX-b}).
The bounds $\mu^{1-}$ and $\mu^{1+}$ are said to be sharp,
because any value in the interval
$[\mu^{1-}, \mu^{1+}]$ including the two endpoints can be realized by some joint distribution of $(Y^1,T,X)$ under model (\ref{eq:sen-model}).
The following result has been obtained by Dorn \& Guo (2022), except for simplification of the expressions for $\mu^{1-}$ and $\mu^{1+}$
in terms of optimal objective values from (weighted) quantile regression (Koenker \& Basset 1978).
These expressions in (\ref{eq:mu1-Q})--(\ref{eq:mu1+Q}) are also simpler and more informative than in Dorn et al.~(2021), Section 2.1.
For any $\tau \in(0,1)$, denote the $\tau$-quantile of $P(Y | T=1,X)$ as
$ q^*_{1,\tau} (X) = \inf \{ u: P( Y \le u | T=1,X) \ge \tau \}$, which is a minimizer to $E \{ T \rho_\tau( Y,u) |X \}$ over $u\in \bbR$,
where $\rho_\tau (y,u) = \tau (y-u)_+ + (1-\tau) (u-y)_+$, known as the ``check" function, and $c_+ = c$ for $c \ge 0$ or $0$ for $c <0$.

\begin{pro} \label{pro:DG}
For any $\Lambda\ge1$, the (sharp) bounds $\mu^{1-}$ and $\mu^{1+}$ from (\ref{eq:mu1-})--(\ref{eq:mu1+}) can be determined as follows:
\begin{align}
& \mu^{1-} =\mu^{1-} (q^*_{1,1-\tau})
= E \left\{\frac{T}{\pi^*(X)} Y - (\Lambda-\Lambda^{-1})  T \frac{1-\pi^*(X)}{\pi^*(X)} \rho_{1-\tau} (Y, q^*_{1,1-\tau} (X)) \right\} , \label{eq:mu1-Q} \\
& \mu^{1+} =\mu^{1+} (q^*_{1,\tau})
= E \left\{\frac{T}{\pi^*(X)} Y + (\Lambda-\Lambda^{-1})  T \frac{1-\pi^*(X)}{\pi^*(X)} \rho_\tau (Y, q^*_{1,\tau} (X)) \right\} , \label{eq:mu1+Q}
\end{align}
where $\tau 
= \Lambda/(\Lambda+1)$,
and $\mu^{1-} (q_1)$ and $\mu^{1+} (q_1)$ denote the right-hand sides of (\ref{eq:mu1-Q}) and (\ref{eq:mu1+Q}) respectively,
with $q^*_{1,1-\tau}$ and $q^*_{1,\tau}$ replaced by any covariate function $q_1(X)$.
\end{pro}

In Dorn \& Guo (2022), the conditional version of the objective (\ref{eq:IPW-obj}) given $X=x$
is optimized subject to constraints (\ref{eq:lambda1-givenX-a})--(\ref{eq:lambda1-givenX-b}) by
using a generalized Neyman--Pearson lemma (Dantzig \& Wald 1951), separately for each $x\in\mathcal X$.
Alternatively, we provide a new result by directly optimizing the objective (\ref{eq:IPW-obj})
while relaxing the constraint (\ref{eq:lambda1-givenX-b}) to finitely many unconditional constraints, as in Tan (2006).
Let $h(x) = (h_0(x), h_1(x),  \ldots, h_m(x) )^\T$ be a vector of pre-specified covariate functions, with $h_0\equiv 1$.
Then condition (\ref{eq:lambda1-givenX-b}) implies that
\begin{align}
E \left\{ T \frac{1-\pi^*(X)}{\pi^*(X)} \lambda_1(Y, X) h (X)  \right\} =
E  \left\{ T \frac{1-\pi^*(X)}{\pi^*(X)} h (X)  \right\}  . \label{eq:lambda1-marginX}
\end{align}
By replacing (\ref{eq:lambda1-givenX-b}) with the unconditional, weaker constraints (\ref{eq:lambda1-marginX}), define
\begin{align}
& \mu^{1-}_h = \min_{\lambda_1} E\left\{ TY+ T \frac{1-\pi^*(X)}{\pi^*(X)} \lambda_1 (Y,X) Y  \right\}, \label{eq:mu1-h}\\
& \mu^{1+}_h = \max_{\lambda_1} E\left\{ TY+ T \frac{1-\pi^*(X)}{\pi^*(X)} \lambda_1 (Y,X) Y  \right\}, \label{eq:mu1+h}
\end{align}
subject to $\lambda_1=\lambda_1(Y^1,X)$ satisfying the constraints (\ref{eq:lambda1-givenX-a}) and (\ref{eq:lambda1-marginX}).
Then both $\mu^1$ and $[\mu^{1-}, \mu^{1+}]$ are contained in the interval $[\mu^{1-}_h, \mu^{1+}_h]$.
The optimization in (\ref{eq:mu1-h})--(\ref{eq:mu1+h}) can be seen as infinite-dimensional (population) linear programming,
and a sample version of (\ref{eq:mu1-h})--(\ref{eq:mu1+h}) corresponds to the (sample) linear programming method in Tan (2006).
See Section \ref{sec:relax} for further discussion.
The following result characterizes solutions to (\ref{eq:mu1-h})--(\ref{eq:mu1+h}) through finite-dimensional optimization for
weighted (W) quantile regression (Koenker \& Basset 1978).  

\begin{pro} \label{pro:mu1-h}
For any $\Lambda\ge1$, let
\begin{align}
&\bar\beta_{\mytext{W},1-} = \underset{\beta\in\bbR^{1+m}}{\argmin} \; E \left\{ T \frac{1-\pi^*(X)}{\pi^*(X)} \rho_{1-\tau} (Y, h^\T(X) \beta ) \right\}, \label{eq:beta1-}\\
&\bar\beta_{\mytext{W},1+} = \underset{\beta\in\bbR^{1+m}}{\argmin} \; E \left\{ T \frac{1-\pi^*(X)}{\pi^*(X)} \rho_\tau (Y, h^\T(X) \beta) \right\}, \label{eq:beta1+}
\end{align}
where $\tau = \Lambda/(\Lambda+1)$. Then $\mu^{1-}_h$ and $\mu^{1+}_h$ from (\ref{eq:mu1-h})--(\ref{eq:mu1+h}) are
\begin{align}
& \mu^{1-}_h =  \mu^{1-} (h^\T\bar\beta_{\mytext{W},1-} )
=E \left\{\frac{T}{\pi^*(X)} Y - (\Lambda-\Lambda^{-1})  T \frac{1-\pi^*(X)}{\pi^*(X)} \rho_{1-\tau} (Y, h^\T(X) \bar\beta_{\mytext{W},1-} ) \right\} , \label{eq:mu1-Qh} \\
& \mu^{1+}_h =  \mu^{1+} (h^\T\bar\beta_{\mytext{W},1+} )
= E \left\{\frac{T}{\pi^*(X)} Y + (\Lambda-\Lambda^{-1})  T \frac{1-\pi^*(X)}{\pi^*(X)} \rho_\tau (Y, h^\T(X) \bar\beta_{\mytext{W},1+} ) \right\} . \label{eq:mu1+Qh}
\end{align}
Moreover, $\mu^{1-}_h \le \mu^{1-} \le \mu^1 \le \mu^{1+} \le \mu^{1+}_h$.
\end{pro}

From the proof of Proposition \ref{pro:mu1-h},
the choices of $\lambda_1(Y^1,X)$ corresponding to the lower and upper bounds $\mu^{1-}_h$ and $\mu^{1+}_h$ are of the following forms.

\begin{cor} \label{cor:mu1-h}
In the context of Proposition \ref{pro:mu1-h}, a solution, $\bar\lambda_{1-}$, to the optimization problem in (\ref{eq:mu1-h}) subject to the constraints
(\ref{eq:lambda1-givenX-a}) and (\ref{eq:lambda1-marginX}) can be obtained such that
$ \bar\lambda_{1-}(Y^1, X) = \Lambda $ if $Y^1 < h^\T(X) \bar\beta_{\mytext{W},1-}$ or
$\Lambda^{-1}$ if $Y^1 > h^\T(X) \bar\beta_{\mytext{W},1-}$.
A solution, $\bar\lambda_{1+}$, to the optimization problem in (\ref{eq:mu1+h}) subject to the constraints (\ref{eq:lambda1-givenX-a}) and (\ref{eq:lambda1-marginX})
can be obtained such that
$\bar\lambda_{1+}(Y^1, X) = \Lambda$ if $ Y^1 > h^\T(X) \bar\beta_{\mytext{W},1+}$ or
$\Lambda^{-1}$ if  $Y^1 < h^\T(X) \bar\beta_{\mytext{W},1+}$.
\end{cor}

We provide some remarks about underpinnings and implications of our results.
First, our results extend previous results about duality relationships in generalized Neyman--Pearson problems.
The formulas (\ref{eq:mu1-Q})--(\ref{eq:mu1+Q}) for the sharp bounds can be, for example, stated as
\begin{align*}
& \quad \max_{\lambda_1} E\left\{ TY+ T \frac{1-\pi^*(X)}{\pi^*(X)} \lambda_1 (Y,X) Y  \right\} \\
& = \min_{q_{1,\tau}(\cdot)} E \left\{ \frac{T}{\pi^*(X)} Y + (\Lambda-\Lambda^{-1})   T \frac{1-\pi^*(X)}{\pi^*(X)} \rho_\tau (Y,q_{1,\tau}(X) ) \right\}
\end{align*}
subject to $\lambda_1=\lambda_1(Y^1,X)$ satisfying (\ref{eq:lambda1-givenX-a}) and (\ref{eq:lambda1-givenX-b}) and any covarite function $q_{1,\tau}$.
The duality relationship underlying Proposition \ref{pro:mu1-h} can be stated as
\begin{align*}
& \quad \max_{\lambda_1} E\left\{ TY+ T \frac{1-\pi^*(X)}{\pi^*(X)} \lambda_1 (Y,X) Y  \right\} \\
& = \min_{\beta\in\bbR^{1+m}} E \left\{ \frac{T}{\pi^*(X)} Y + (\Lambda-\Lambda^{-1})   T \frac{1-\pi^*(X)}{\pi^*(X)} \rho_\tau (Y, h^\T(X)\beta) \right\}
\end{align*}
subject to $\lambda_1=\lambda_1(Y^1,X)$ satisfying (\ref{eq:lambda1-givenX-a}) and (\ref{eq:lambda1-marginX}).
Given the left-hand side as a primal problem, the expression of the dual problem on the right-hand side
in terms of weighted quantile regression appears to be new.
In fact, Proposition \ref{pro:mu1-h} is closely related to Francis \& Wright (1969),
with an extension to allow convex but non-differentiable optimization, using subgradients of expectations of convex functions (Bertsekas 1973).
Although the specific setting of sensitivity analysis is studied here, our results can be extended to general settings as in Francis \& Wright (1969).

Second, the optimization problems in Proposition \ref{pro:mu1-h} differ from those in Proposition \ref{pro:DG} based on Dorn \& Guo (2022)
only in that the marginal constraint (\ref{eq:lambda1-marginX}) is imposed on $\lambda_1(Y^1,X)$
instead of the conditional constraint (\ref{eq:lambda1-givenX-b}).
Conceptually, Proposition \ref{pro:DG} can be seen either as the limit version of Proposition \ref{pro:mu1-h} as
the covariate vector $h(X)$ is expanded infinitely, or as the extreme version of Proposition  \ref{pro:mu1-h}
with the true quantile function $q^*_{1,1-\tau}(X)$ or $q^*_{1,\tau} (X)$ included as a component in $h(X)$.
While the formulas (\ref{eq:mu1-Q})--(\ref{eq:mu1+Q}) deliver the sharp bounds $\mu^{1-}$ and $\mu^{1+}$
depending on the true quantile functions $q^*_{1,1-\tau}(X)$ and $q^*_{1,\tau} (X)$,
the relaxed bounds in (\ref{eq:mu1-Qh})--(\ref{eq:mu1+Qh}) is more prescriptive in showing the optimal bounds
when a linear quantile model, $h^\T(X) \beta$, is used with pre-specified regressors $h(X)$.
In fact, from Proposition \ref{pro:mu1-h}, the following comparison can be established about different ways of obtaining bounds on $\mu^1$ with
estimated quantile functions. For simplicity,
we consider only upper bounds on $\mu^1$.

\begin{cor} \label{cor:comparison}
(i) For any covariate function $ q_{1,\tau}(X)$,
\begin{align*}
\mu^{1+} (  q_{1,\tau}) \ge \mu^{1+} ( q_{1,\tau} \bar\beta_{q_{1,\tau}} ) \ge \mu^{1+} ( (1, q_{1,\tau})^\T \bar\beta_{(1, q_{1,\tau})^\T}  )
\ge \mu^{1+} ( q^*_{1,\tau}) ,
\end{align*}
where $\mu^{1+} ( q_{1,\tau})$ denotes the right-hand side of (\ref{eq:mu1+Q}) with $q^*_{1,\tau}$ replaced by
$ q_{1,\tau}$, and $\mu^{1+}(h^\T \bar\beta_h)=\mu^{1+} (h^\T \bar\beta_{\mytext{W},1+} )$ denotes the right-hand side of (\ref{eq:mu1+Qh}),
with $\bar\beta_h= \bar\beta_{\mytext{W},1+}$ depending on $h$, for example, $h= q_{1,\tau}$.
In general, the above inequalities hold strictly if $ q_{1,\tau} \not= q^*_{1,\tau}$.

\vspace{-.08in}
(ii) For any vector of covariate functions $h(x)$,
\begin{align*}
\mu^{1+} ( h^\T \bar\beta_{\mytext{U},1+} ) \ge \mu^{1+} (h^\T \bar\beta_{\mytext{W},1+} )
\ge \mu^{1+} ( q^*_{1,\tau}) ,
\end{align*}
where $\bar\beta_{\mytext{U},1+}$ is the coefficient vector from the population version of unweighted (U) $\tau$-quantile regression of $Y$ on $h(X)$
in the treated group,
$\bar\beta_{\mytext{U},1+} = \argmin_{\beta\in\bbR^{1+m}} \; E  \{ T \rho_\tau (Y, h^\T(X) \beta) \}$.
In general, the above inequality holds strictly if $q^*_{1,\tau} \not\in \{h^\T \beta: \beta\in\bbR^{1+m} \}$.
\end{cor}
\vspace{-.07in}

Corollary \ref{cor:comparison}(i) indicates that substituting $ q_{1,\tau}$, interpreted as the limit of an estimated quantile function,
for $q^*_{1,\tau}$ in (\ref{eq:mu1+Q})
yields a higher (worse) upper bound on $\mu^1$ than using formula (\ref{eq:mu1+Qh}) with $ q_{1,\tau}$ included as component of $h$,
unless $ q_{1,\tau} = q^*_{1,\tau}$ (i.e., $q^*_{1,\tau}$ is consistently estimated).
Moreover, for any choice of regressors $h$,
Corollary \ref{cor:comparison}(ii) shows that substituting the fitted function from unweighted quantile regression in
(\ref{eq:mu1+Q}) also yields a worse upper bound on $\mu^1$ than  using formula (\ref{eq:mu1+Qh}),
unless the quantile regression model $h^\T \beta$ is correctly specified.
These results demonstrate the advantage of (\ref{eq:mu1+Qh}) over using (\ref{eq:mu1+Q}) by substitution.

Third, Propositions \ref{pro:DG} and \ref{pro:mu1-h} also shed new light on the role of outcome predictive modeling
given the covariates in sensitivity analysis.
The bounds $\mu^{1-}_h$ and $\mu^{1+}_h$ in (\ref{eq:mu1-Qh})--(\ref{eq:mu1+Qh}) are expressed as the expectation of the usual IPW estimator shifted
by the optimized objective values from weighted outcome quantile regression in the treated group (up to a factor $\Lambda-\Lambda^{-1}$).
For any fixed $\Lambda > 1$,
the smaller the optimized objective values in quantile regression
(i.e., the outcome can be predicted more accurately from the covariates),
the narrower the sensitivity interval from the lower to the upper bound becomes.
In the extreme situation where $Y$ can be perfectly predicted in the treated group as $Y = h^\T(X) \beta$ for some $\beta$,
both the bounds $\mu^{1-}_h$ and $\mu^{1+}_h$ would reduce to the expected IPW estimator and hence $\mu^1$ would be point identified
regardless of unmeasured confounding under model (\ref{eq:sen-model}).
This phenomenon can be attributed to the fact that the distribution $P(Y^1 | T=0,X)$ is absolutely
continuous with respect to $P(Y^1 | T=1,X)$ under model (\ref{eq:sen-model})  with any finite $\Lambda$.
While this absolute continuity might be violated, the preceding discussion shows that improvement in outcome predictive modeling
can be translated into narrower bounds in sensitivity analysis under model (\ref{eq:sen-model}).
A similar property can be expected for sensitivity analysis under the latent sensitivity model of Rosenbaum (2022)
as discussed in Supplement Section \ref{sec:comparison-model}.
The absolute continuity mentioned above is also satisfied (Yadlowsky et al.~2022, Lemma 2.1).

\subsection{Doubly robust bounds}  \label{sec:DR-bound}

The bounds on $\mu^1$ in Section \ref{sec:IPW-bound} depend on the true propensity score $\pi^*$ being known.
We discuss doubly robust bounds on $\mu^1$ such that
they are valid if either the propensity score $\pi^*$ or the conditional distribution $P(Y|T=1,X)$, denoted as $\eta^*_1$, is known.

First, we study doubly robust bounds derived in Dorn et al.~(2021).
To upper bound $\mu^1$, the estimating functions in Dorn et al.~(2021) can be simplified using Propositions \ref{pro:DG} as
\begin{align}
 \varphi_{+,\mytext{IPW}} (O; \pi, q_{1,\tau})
& = \frac{T}{\pi(X)} Y + (\Lambda-\Lambda^{-1})  T \frac{1-\pi(X)}{\pi(X)} \rho_\tau (Y, q_{1,\tau} (X)) \nonumber \\
& = T \left\{ Y - \tilde Y_{+} (q_{1,\tau}) \right\}  + \frac{T}{\pi(X)}  \tilde Y_{+} (q_{1,\tau}), \nonumber \\
\varphi_{+} (O; \pi,\eta_1, q_{1,\tau})
& =\varphi_{+,\mytext{IPW}} (O; \pi, q_{1,\tau})
- \left\{\frac{T}{\pi(X)}-1\right\} E_{\eta_1} \{ \tilde Y_{+} (q_{1,\tau})| T=1,X \} , \label{eq:AIPW+DG}
\end{align}
where $\tilde Y_{+} (q_{1,\tau}) = Y+ (\Lambda-\Lambda^{-1})  \rho_\tau( Y, q_{1,\tau}(X))$,
$q_{1,\tau}(X)$ is a covariate function, representing a putative $\tau$-quantile of $Y$ given $T=1$ and $X$,
$\pi(X)$ is a covariate function, representing a putative probability of $T=1$ given $X$ as in (\ref{eq:potential-IPW}),
$\eta_1$ denotes a family of probability distributions on $\mathcal Y$ indexed by $x\in\mathcal X$, representing
a putative conditional distribution of $Y$ given $T=1$ and $X$, and
$E_{\eta_1}(\cdot)$ denotes the conditional expectation taken under $\eta_1$.
In the subsequent discussion, $\eta_1$ and $\eta^*_1$ can be weakened to encode only the putative and true conditional means of
$\tilde Y_{+} (q_{1,\tau})$ given $T=1$ and $X$, so that $\eta_1 = \eta^*_1$ indicates that
$E_{\eta_1} \{ \tilde Y_{+} (q_{1,\tau})| T=1,X \} = E \{ \tilde Y_{+} (q_{1,\tau})| T=1,X \}$,
but the conditional distribution $P(Y|T=1,X)$ may not be identified.

The following bounds have been shown in Propositions 5--6 of Dorn et al.~(2021):
\begin{align}
E \left\{ \varphi_{+} (O; \pi^*,\eta_1, q_{1,\tau}) \right\} \ge \mu^{1+}, \quad
E \left\{ \varphi_{+} (O; \pi,\eta^*_1, q_{1,\tau}) \right\}  \ge \mu^{1+},  \label{eq:DG-DR}
\end{align}
where equality holds in each case if $q_{1,\tau} = q^*_{1,\tau}$.
Hence the mean of $\varphi_{+} (O; \pi,\eta_1, q_{1,\tau})$ delivers a doubly robust upper bound on $\mu^1$ if
either $\pi=\pi^*$ or $\eta_1 = \eta^*_1$, and the bound reduces to the sharp bound $\mu^{1+}$
if, further, $q_{1,\tau} = q^*_{1,\tau}$.
We point out that a stronger relationship holds: the two upper bounds in  (\ref{eq:DG-DR}) are identical to each other and to
$\mu^{1+} ( q_{1,\tau} )$ as defined in Proposition \ref{pro:DG}.  
Similar results can be obtained for the lower bound on $\mu^1$ in Dorn et al.~(2021). Let \vspace{-.1in}
\begin{align}
 \varphi_{-,\mytext{IPW}} (O; \pi, q_{1,1-\tau})
& = \frac{T}{\pi(X)} Y - (\Lambda-\Lambda^{-1})  T \frac{1-\pi(X)}{\pi(X)} \rho_{1-\tau} (Y, q_{1,1-\tau} (X)) \nonumber \\
& = T \left\{ Y - \tilde Y_{-} (q_{1,1-\tau}) \right\}  + \frac{T}{\pi(X)}  \tilde Y_{-} (q_{1,1-\tau}), \nonumber \\
\varphi_{-} (O; \pi,\eta_1, q_{1,1-\tau})
& =\varphi_{-,\mytext{IPW}} (O; \pi, q_{1,1-\tau})
- \left\{\frac{T}{\pi(X)}-1\right\} E_{\eta_1} \{ \tilde Y_{-} (q_{1,1-\tau})| T=1,X \} , \label{eq:AIPW-DG}
\end{align}
where $\tilde Y_{-} (q_{1,1-\tau}) = Y- (\Lambda-\Lambda^{-1})  \rho_{1-\tau}( Y, q_{1,1-\tau}(X))$.

\begin{pro} \label{pro:DG-DR}
For any $\Lambda \ge 1$, if either $\pi=\pi^*$ or $\eta_1=\eta^*_1$, then \vspace{-.05in}
\begin{align*}
& \mu^{1-} ( q_{1,1-\tau} ) = E \left\{ \varphi_{-} (O; \pi,\eta_1, q_{1,1-\tau}) \right\}
 \le \mu^{1-} , \\
& \mu^{1+} ( q_{1,\tau} ) = E \left\{ \varphi_{+} (O; \pi,\eta_1, q_{1,\tau}) \right\}
 \ge \mu^{1+} ,
\end{align*}
where $\tau = \Lambda/(\Lambda+1)$,
$\mu^{1-} ( q_{1,1-\tau} ) = E\{  \varphi_{-,\mytext{IPW}} (O; \pi^*, q_{1,1-\tau}) \}$, and
$\mu^{1+} ( q_{1,\tau} ) = E\{ \varphi_{+,\mytext{IPW}} $  $(O; \pi^*, q_{1,\tau}) \}$
as in Proposition \ref{pro:DG}.  
The two inequalities above reduce to equality if $q_{1,1-\tau}=q^*_{1,1-\tau}$ or $q_{1,\tau}=q^*_{1,\tau}$ respectively.
\end{pro}

From Proposition \ref{pro:DG-DR}, the estimating function $\varphi_{+} $ is doubly robust for the relaxed population bound
$\mu^{1+} ( q_{1,\tau} )$, a well-defined population quantity which is no smaller than the sharp bound $\mu^{1+}$
for any putative quantile function $q_{1,\tau}$.
We provide some remarks to help understand this result.
First, in the unconfoundedness case of $\Lambda=1$, the function $\varphi_{+} $ reduces to the usual augmented IPW estimating function
(Robins et al.~1994),
\begin{align}
\varphi_1( O; \pi, \eta_1) =  \frac{TY }{\pi(X)} - \left\{\frac{T}{\pi(X)} - 1\right\} E_{\eta_1} ( Y |T=1, X ), \label{eq:AIPW-Robins}
\end{align}
which no longer depends on 
$q_{1,\tau}$ with $\tau=1/2$.
The double robustness of $\varphi_{+}$ can be explained as follows:
if the term $ T \{ Y - \tilde Y_{+} (q_{1,\tau}) \} =
- (\Lambda-\Lambda^{-1}) T \rho_\tau (Y, q_{1,\tau} (X))$, which does not involve $\pi$ or $\eta_1$, were removed,
then $\varphi_{+} (O; \pi,\eta_1, q_{1,\tau})$ in (\ref{eq:AIPW+DG}) would be exactly obtained from (\ref{eq:AIPW-Robins})
with $Y$ replaced by $\tilde Y_{+} (q_{1,\tau})$.
Second, the fact that $\mu^{1+} ( q_{1,\tau} ) \ge \mu^{1+}$ for any $q_{1,\tau}$
can be seen directly from the expression (\ref{eq:mu1+Q}):
$\mu^{1+} = \min_{q_{1,\tau}} \mu^{1+} ( q_{1,\tau} )$ over all possible covariate functions $q_{1,\tau} (X)$,
by the characterization of the $\tau$-quantile function in terms of the loss function $\rho_\tau$.

Next, by augmented IPW estimation, we derive doubly robust estimating functions
related to the bounds of $\mu^1$ in Proposition \ref{pro:mu1-h}, depending on a vector of covariate functions $h$.
For a putative propensity score $\pi(X)$ which may differ from $\pi^*(X)$, let
\begin{align*}
&\bar\beta_{\mytext{W},1-}(\pi) = \underset{\beta\in\bbR^{1+m}}{\argmin} \; E \left\{ T \frac{1-\pi(X)}{\pi(X)} \rho_{1-\tau} (Y, h^\T(X) \beta ) \right\}, \\
&\bar\beta_{\mytext{W},1+}(\pi) = \underset{\beta\in\bbR^{1+m}}{\argmin} \; E \left\{ T \frac{1-\pi(X)}{\pi(X)} \rho_\tau (Y, h^\T(X) \beta) \right\},
\end{align*}
where the dependency of $\bar\beta_{\mytext{W},1-}(\pi)$ and $\bar\beta_{\mytext{W},1+}(\pi)$ on $\pi$ is made explicit in the notation.
Note that $\bar\beta_{\mytext{W},1-} = \bar\beta_{\mytext{W},1-}(\pi^*)$ and $\bar\beta_{\mytext{W},1+} = \bar\beta_{\mytext{W},1+}(\pi^*)$
by definition (\ref{eq:beta1-}) and (\ref{eq:beta1+}).

\begin{pro} \label{pro:mu1-h-DR}
For any $\Lambda \ge 1$, if either $\pi=\pi^*$ or $\eta_1=\eta^*_1$, then
\begin{align*}
& \mu^{1-} (h^\T \bar\beta_{\mytext{W},1-} (\pi) )
= E \left\{ \varphi_{-} (O; \pi,\eta_1, h^\T \bar\beta_{\mytext{W},1-} (\pi) ) \right\}  \le \mu^{1-} ,\\
& \mu^{1+} (h^\T \bar\beta_{\mytext{W},1+} (\pi) )
= E \left\{ \varphi_{+} (O; \pi,\eta_1, h^\T \bar\beta_{\mytext{W},1+} (\pi) ) \right\}  \ge \mu^{1+} ,
\end{align*}
where $\tau = \Lambda/(\Lambda+1)$, and $\mu^{1-}(\cdot)$ and $\mu^{1+}(\cdot)$ are defined as in Proposition \ref{pro:DG}. 
The two inequalities reduces to equality if $q^*_{1,1-\tau} \in\{h^\T\beta: \beta\in\bbR^{1+m}\}$ or $q^*_{1,\tau} \in\{h^\T\beta: \beta\in\bbR^{1+m}\}$. 
\end{pro}

From Proposition \ref{pro:mu1-h-DR}, the estimating function $\varphi_{+} (O; \pi,\eta_1, h^\T \bar\beta_{\mytext{W},1+} (\pi) )$
is doubly robust for the relaxed population bound $\mu^{1+} (h^\T \bar\beta_{\mytext{W},1+} (\pi) )$, which is no smaller than the sharp bound $\mu^{1+}$
and may differ from  $\mu^{1+} (h^\T \bar\beta_{\mytext{W},1+} )$ in Proposition \ref{pro:mu1-h} unless $\pi=\pi^*$.
Moreover, the population bound $\mu^{1+} (h^\T \bar\beta_{\mytext{W},1+} (\pi) )$ can be seen as a specific choice of
$\mu^{1+} (q_{1,\tau})$, with $q_{1,\tau} = h^\T \bar\beta_{\mytext{W},1+} (\pi)$
determined from weighted $\tau$-quantile regression of $Y$ on $h(X)$ in the treated group.
We comment that this particular choice leads to two interesting advantages, compared with the usual choice
$\mu^{1+} (h^\T \bar\beta_{\mytext{U},1+})$, where $\bar\beta_{\mytext{U},1+}$ is obtained from unweighted $\tau$-quantile regression.

The first advantage of the bound $\mu^{1+} (h^\T \bar\beta_{\mytext{W},1+} (\pi) )$ follows from  Corollary \ref{cor:comparison}(ii).
If $\pi=\pi^*$, then
$\mu^{1+} (h^\T \bar\beta_{\mytext{W},1+} (\pi) )=  \mu^{1+} (h^\T \bar\beta_{\mytext{W},1+} ) $
yields a tighter upper bound on $\mu^1$  than $\mu^{1+} (h^\T \bar\beta_{\mytext{U},1+} )$, that is,
$ \mu^{1+} \le \mu^{1+} (h^\T \bar\beta_{\mytext{W},1+} ) \le \mu^{1+} (h^\T \bar\beta_{\mytext{U},1+} )$.
In fact, this can also be seen from the expression (\ref{eq:mu1+Qh}) in Proposition \ref{pro:mu1-h}:
$\mu^{1+} (h^\T \bar\beta_{\mytext{W},1+}  ) = \min_{\beta\in\bbR^{1+m}} \mu^{1+} (h^\T \beta)$
by the definition of $\bar\beta_{\mytext{W},1+}$ from weighted $\tau$-quantile regression.
For completeness, if $\pi \not=\pi^*$ but $\eta_1=\eta^*_1$, then
$\mu^{1+} (h^\T \bar\beta_{\mytext{W},1+} (\pi) )$ and $\mu^{1+} (h^\T \bar\beta_{\mytext{U},1+} )$
remain valid upper bounds of $\mu^{1+}$, but their relative magnitudes may vary across different data settings.
The second advantage of targeting the population bound $\mu^{1+} (h^\T \bar\beta_{\mytext{W},1+} (\pi) )$ instead of
$\mu^{1+} (h^\T \bar\beta_{\mytext{U},1+} )$ is related to construction of confidence intervals based on
these population bounds in sample sensitivity analysis, and will be discussed in Section \ref{sec:sample-SA}.

\section{Sample sensitivity analysis} \label{sec:sample-SA}

The population bounds in Section \ref{sec:pop-SA} can be estimated, with associated confidence intervals, from sample data,
under sensitivity model (\ref{eq:sen-model}).
We develop new methods both in low-dimensional settings in Section \ref{sec:cal}
and in high-dimensional settings in Section \ref{sec:rcal}.

\subsection{Calibrated estimation} \label{sec:cal}

We study estimation of the population bounds in Section \ref{sec:DR-bound} from the perspective of two-stage semiparametric estimation.
For simplicity, only upper bounds on $\mu^1$ are handled.
Consider an augmented IPW estimator using the estimating function $\varphi_{+}$ in (\ref{eq:AIPW+DG}):
\begin{align}
\hat\mu^{1+} ( \gamma,\alpha,\beta) = \tilde E \left\{ \varphi_{+} (O; \pi(\cdot;\gamma),\eta_{1+} (\cdot;\alpha), q_{1,\tau}(\cdot;\beta) ) \right\}, \label{eq:mu1+class}
\end{align}
which depends on three working models: \vspace{-.05in}
\begin{itemize}\addtolength{\itemsep}{-.1in}
\item A propensity score model,  $\pi^*(X) = \pi(X;\gamma)$;
\item An outcome $\tau$-quantile regression model, $q^*_{1,\tau}(X) = q_{1,\tau}(X;\beta) $
\item An outcome mean regression model, $E \{ \tilde Y_{+} ( q_{1,\tau}(\cdot;\beta) )| T=1,X \} = \eta_{1+}(X; \alpha)$,
\end{itemize} \vspace{-.05in}
where $\eta_{1+}(X;\alpha)$ is taken to represent $E_{\eta_1} \{ \tilde Y_{+} ( q_{1,\tau}(\cdot;\beta) )| T=1,X \} $ for $\eta_1=\eta_{1+}(\cdot;\alpha)$.
Throughout $\tilde E(\cdot)$ denotes the sample average over $O_i=(T_i Y_i, T_i,X_i)$, $i=1,\ldots,n$.
We say that these models are correctly specified if
there exists a true value $\gamma^*$, $\beta^*_{1+}$, or $\alpha^*_{1+} (\beta)$ such that
$\pi^*(X) = \pi(X;\gamma^*)$, $q^*_{1,\tau}(X) = q_{1,\tau}(X;\beta^*_{1+}) $, or
$E \{ \tilde Y_{+} ( q_{1,\tau}(\cdot;\beta) )| T=1,X \} = \eta_{1+}(X; \alpha^*_{1+} (\beta) )$ respectively. Otherwise, we say that these models are misspecified.
To focus on main ideas, until Section \ref{sec:binary}, we discuss the case where these models are linear in the original or logistic scale:
\begin{align}
& \pi^*(X) = \pi(X; \gamma) = \{1 + \exp( - f^\T(X) \gamma) \}^{-1}, \label{eq:logit+pi}\\
& q^*_{1,\tau}(X) = q_{1,\tau} (X; \beta) = h^\T(X) \beta,  \label{eq:linear+Q} \\
& E \{ \tilde Y_{+} ( q_{1,\tau}(\cdot;\beta) )| T=1,X \} = \eta_{1+}(X; \alpha) = f^\T(X) \alpha , \label{eq:linear+eta}
\end{align}
where $f(x)=(f_0(x),\ldots,f_p(x))^\T$ and $h(x)=(h_0(x), \ldots, h_m(x))^\T$ are vectors of covarite functions (or regressors),
with $f_0 =h_0 \equiv 1$, and
$\gamma=(\gamma_0,\ldots,\gamma_p)^\T$ and similarly $\alpha$ and $\beta$ are vectors of unknown parameters.
The basis vectors in (\ref{eq:logit+pi}) and (\ref{eq:linear+eta}) are restricted to be the same;
otherwise, the respective basis vectors from the two models can be combined.

Application of (\ref{eq:mu1+class}) involves two stages: first constructing some estimators $(\hat\gamma,\hat\alpha_{1+}, \hat\beta_{1+})$,
and then substituting them into (\ref{eq:mu1+class}) to obtain $\hat\mu^{1+} (\hat\gamma, \hat\alpha, \hat\beta_{1+})$.
While this appears straightforward, a central issue is how to conduct first-stage estimation such that
valid inference (confidence intervals and hypothesis tests) can be derived in the second stage using
$\hat\mu^{1+} (\hat\gamma, \hat\alpha, \hat\beta_{1+})$.  
For the linear and logistic models (\ref{eq:linear+Q})--(\ref{eq:linear+eta}), the conventional estimators are defined as 
\begin{align*}
& \hat\gamma_{\mytext{ML}} = \argmin_{\gamma \in \bbR^{1+p}} \tilde E \left[ - T f^\T(X) \gamma + \log \{1+\exp(f^\T(X)\gamma) \} \right], \\
& \hat\beta_{\mytext{U},1+} = \argmin_{\beta \in\bbR^{1+m}} \tilde E \left\{ T \rho_\tau( Y, h^\T(X) \beta) \right\}, \\
& \hat\alpha_{\mytext{U},1+} =  \argmin_{\alpha \in \bbR^{1+p}} \tilde E \left[ T \{ \tilde Y_{+} ( h^\T(X) \hat\beta_{\mytext{U},1+}  ) - f^\T(X) \alpha\}^2 \right] /2,
\end{align*}
by maximum likelihood (ML), unweighted quantile regression, and unweighted least squares.
Alternatively, we propose calibrated (CAL) estimation for $(\gamma,\alpha,\beta)$ by integrating the two stages of estimation such that 
doubly robust confidence intervals can be derived in a simple manner
from the resulting estimator $\hat\mu^{1+} (\hat\gamma, \hat\alpha, \hat\beta_{1+})$.
See Supplement Section \ref{sec:add-cal} for a more elaborate discussion about conventional and calibrated estimation.
The discussion extends Tan (2020b) and Ghosh \& Tan (2022), which involve doubly robust estimating functions with only two nuisance parameters
corresponding to $(\gamma,\alpha)$ here.

The calibrated estimators, $(\hat\gamma_{\mytext{CAL}},\hat\alpha_{\mytext{WL},1+}, \hat\beta_{\mytext{WL},1+})$, are
defined jointly as solutions to the estimating equations, called (sample) calibration equations: \vspace{-.05in}
\begin{align}
& \tilde E \left( \frac{\partial \varphi_{+}}{\partial \gamma} \right) =0 , \quad
\tilde E \left(  \frac{\partial \varphi_{+}}{\partial \alpha} \right)  =0 , \quad
\tilde E \left( \frac{\partial \varphi_{+}}{\partial \beta} \right) =0 , \label{eq:cal-eq}
\end{align}
where $\varphi_{+} = \varphi_{+} (O; \pi(\cdot;\gamma),\eta_{1+} (\cdot;\alpha), q_{1,\tau}(\cdot;\beta) )$ as in (\ref{eq:mu1+class}).
These estimators can also be obtained by sequential minimization of three convex objective functions defined in Section \ref{sec:rcal}.
The estimators $\hat\alpha_{\mytext{WL},1+}$ and $\hat\beta_{\mytext{WL},1+}$ are also called weighted loss (WL) estimators,
because the objective functions correspond to weighted quantile regression and weighted least squares.

In the classical setting where $p$ and $m$ are fixed as $n$ tends to $\infty$,  
the estimator $\hat\mu^{1+}_{\mytext{CAL}} 
=\mu^{1+} (\hat\gamma_{\mytext{CAL}},\hat\alpha_{\mytext{WL},1+}, \hat\beta_{\mytext{WL},1+})$ can be shown to achieve the following properties.
First, $\hat\mu^{1+}_{\mytext{CAL}}$  is (pointwise) doubly robust
for the upper bound $ \mu^{1+} ( h^\T \bar\beta_{\mytext{WL},1+})$
if either model (\ref{eq:logit+pi}) or (\ref{eq:linear+eta}) with $\beta=\bar \beta_{\mytext{WL},1+}$ is correctly specified.
Moreover, with possible misspecification of models (\ref{eq:logit+pi})--(\ref{eq:linear+eta}),  
$\hat\mu^{1+}_{\mytext{CAL}}$ admits the simple asymptotic expansion\vspace{-.05in}
\begin{align}
& \hat\mu^{1+} (\hat\gamma_{\mytext{CAL}},\hat\alpha_{\mytext{WL},1+}, \hat\beta_{\mytext{WL},1+})
= \tilde E (\varphi_{+}) \big|_{(\gamma,\alpha,\beta)=(\bar\gamma_{\mytext{CAL}},\bar\alpha_{\mytext{WL},1+}, \bar\beta_{\mytext{WL},1+})} + o_p(n^{-1/2}) , \label{eq:cal-expan}
\end{align}
where $(\bar\gamma_{\mytext{CAL}},\bar\alpha_{\mytext{WL},1+}, \bar\beta_{\mytext{WL},1+})$
are the limit (or target) values of $(\hat\gamma_{\mytext{CAL}},\hat\alpha_{\mytext{WL},1+}, \hat\beta_{\mytext{WL},1+})$.
Doubly robust, Wald confidence intervals can be readily obtained using (\ref{eq:cal-expan}).
In contrast,  $\hat\mu^{1+}(\hat\gamma_{\mytext{ML}},$ $\hat\alpha_{\mytext{U},1+}, \hat\beta_{\mytext{U},1+})$ is doubly robust
for $ \mu^{1+} ( h^\T \bar\beta_{\mytext{U},1+})$,
but would not achieve an asymptotic expansion similar to (\ref{eq:cal-expan}) 
unless all three models (\ref{eq:logit+pi})--(\ref{eq:linear+eta}) are correctly specified.

We point out that calibrated estimation of $\beta$ is aligned with the choice of
targeting the population bounds on $\mu^1$ via weighted quantile regression in Proposition \ref{pro:mu1-h-DR}.
The upper bound $ \mu^{1+} ( h^\T \bar\beta_{\mytext{WL},1+})$ targeted by the estimator $\hat\mu^{1+} (\hat\gamma_{\mytext{CAL}},\hat\alpha_{\mytext{WL},1+}, \hat\beta_{\mytext{WL},1+})$ is the same as  $ \mu^{1+} ( h^\T \bar\beta_{\mytext{W},1+}( \pi ) )$
with $\pi$ set to $\bar\pi_{\mytext{CAL}} = \pi(\cdot;\bar\gamma_{\mytext{CAL}})$.
Hence, by Proposition \ref{pro:mu1-h-DR}, if model (\ref{eq:logit+pi}) is correctly specified with $\bar\pi_{\mytext{CAL}} = \pi^*$,
then $ \mu^{1+} ( h^\T \bar\beta_{\mytext{WL},1+})$ is a tighter upper bound on $\mu^1$
than $ \mu^{1+} ( h^\T \bar\beta_{\mytext{U},1+})$, which is targeted by the
estimator $\hat\mu^{1+} (\hat\gamma_{\mytext{ML}},\hat\alpha_{\mytext{U},1+}, \hat\beta_{\mytext{U},1+})$
using conventional estimation.
The derivation of doubly robust confidence intervals using the simple expansion (\ref{eq:cal-expan})
can be seen as the second advantage of targeting the bound $ \mu^{1+} ( h^\T \bar\beta_{\mytext{W},1+}(\pi))$ with $\pi=\bar\pi_{\mytext{CAL}}$
instead of the usual bound $ \mu^{1+} ( h^\T \bar\beta_{\mytext{U},1+})$, as mentioned in  Section \ref{sec:DR-bound}.

\subsection{Regularized calibrated estimation} \label{sec:rcal}

We propose regularized calibrated estimation for fitting
the propensity score and outcome quantile and mean regression models as in (\ref{eq:logit+pi})--(\ref{eq:linear+eta}),
such that doubly robust confidence intervals can be obtained for population bounds on $\mu^1$ and hence for $\mu^1$ itself under sensitivity model (\ref{eq:sen-model}),
in the high-dimensional setting where the numbers of regressors used are close to
or greater than the sample size.
We describe our method for handling upper bounds on $\mu^1$. Estimation for lower bounds on $\mu^1$ is discussed in Supplement Section \ref{sec:ATE}.

Given propensity score and outcome models (\ref{eq:logit+pi})--(\ref{eq:linear+eta}), our upper bound of $\mu^1$ is defined as
$\hat\mu^{1+}_{\mytext{RCAL}} = \hat\mu^{1+} (\hat\gamma_{\mytext{RCAL}},\hat\alpha_{\mytext{RWL},1+}, \hat\beta_{\mytext{RWL},1+})$ using
the augmented IPW estimator (\ref{eq:mu1+class}), i.e.,
\begin{align}
\hat\mu^{1+}_{\mytext{RCAL}} 
& = \tilde E \left\{ \varphi_{+} (O; \pi(\cdot;\hat\gamma_{\mytext{RCAL}}),\eta_{1+} (\cdot;\hat\alpha_{\mytext{RWL},1+}), q_{1,\tau}(\cdot;\hat\beta_{\mytext{RWL},1+}) ) \right\}, \label{eq:mu1+rcal}
\end{align}
where $(\hat\gamma_{\mytext{RCAL}},\hat\alpha_{\mytext{RWL},1+}, \hat\beta_{\mytext{RWL},1+})$ are regularized versions of the calibrated estimators
$(\hat\gamma_{\mytext{CAL}}, $ $ \hat\alpha_{\mytext{WL},1+}, \hat\beta_{\mytext{WL},1+})$ in Section \ref{sec:cal},
hence called regularized calibrated estimators (RCAL).
The estimator $\hat\gamma_{\mytext{RCAL}}$ is a regularized calibrated estimator of $\gamma$ (Tan 2020a),
defined as a minimizer of the penalized objective function $\ell_{\mytext{CAL}} (\gamma) + \lambda_\gamma \|\gamma_{1:p} \|_1$,
where $\lambda_\gamma \ge 0$ is a tuning parameter, $\| \cdot\|_1$ denotes the $L_1$ norm,
$\gamma_{1:p}=(\gamma_1,\ldots,\gamma_p)^\T$ excluding the intercept, and
\begin{align}
\ell_{\mytext{CAL}} (\gamma) &= \tilde E \left\{ T \me^{- f^\T(X) \gamma} + (1-T) f^\T(X) \gamma \right\} . \label{eq:gam-loss}
\end{align}
The estimator $\hat\beta_{\mytext{RWL},1+}$ is a regularized calibrated or more descriptively regularized weighted loss (RWL) estimator
of $\beta$, defined as a minimizer of
$\ell_{\mytext{WL},1+} (\beta; \hat\gamma_{\mytext{RCAL}}) + \lambda_\beta \|\beta_{1:p} \|_1$, where
$\beta_{1:p} = (\beta_1, \ldots, \beta_p)^\T$ excluding the intercept, and
\begin{align}
\ell_{\mytext{WL},1+} (\beta; \gamma) &= \tilde  E \left\{ T \frac{1-\pi(X;\gamma)}{\pi(X;\gamma)} \rho_\tau (Y, h^\T(X) \beta) \right\} . \label{eq:beta-loss}
\end{align}
This is the loss function in weighted $\tau$-quantile regression of $Y$ against $h(X)$ in the treated group, with weight 
$w(X;\gamma)= (1-\pi(X;\gamma))/\pi(X;\gamma) = \me^{-f^\T(X) \gamma}$.
The estimator $\hat\alpha_{\mytext{RWL},1+}$ is a regularized calibrated or weighted loss estimator
of $\alpha$, defined as a minimizer of
$\ell_{\mytext{WL},1+} (\alpha; \hat\gamma_{\mytext{RCAL}}, $ $ \hat\beta_{\mytext{RWL},1+} ) + \lambda_\alpha \|\alpha_{1:p} \|_1$, where
$\alpha_{1:p} = (\alpha_1, \ldots, \alpha_p)^\T$ excluding the intercept, and
\begin{align}
\ell_{\mytext{WL},1+} (\alpha; \gamma,\beta) &= \tilde  E \left[ T \frac{1-\pi(X;\gamma)}{\pi(X;\gamma)}
\left\{ \tilde Y_{+} (h^\T \beta ) - f^\T(X) \alpha \right\}^2 \right] /2 . \label{eq:alpha-loss}
\end{align}
This is the loss function in weighted least square regression of $\tilde Y_{+} (h^\T \beta )$ against $f(X)$ in the treated group, with
the same weight $w(X;\gamma)$ as in $\ell_{\mytext{WL},1+} (\beta;  \gamma ) $.
See Supplement Section \ref{sec:KKT} for implications of the Karush–Kuhn–Tucker conditions for the proposed estimators.

The loss functions $\ell_{\mytext{CAL}}(\gamma)$, $\ell_{\mytext{WL},1+}(\beta; \gamma)$, or $\ell_{\mytext{WL},1+}(\alpha; \gamma,\beta)$
is easily shown to be convex in $\gamma$, in $\beta$ for fixed $\gamma$, or in $\alpha$ for fixed $(\gamma,\alpha)$ respectively.
These loss functions are derived from calibration equation (\ref{eq:cal-eq}) in that
\begin{align}
& \tilde E \left( \frac{\partial \varphi_{+}}{\partial \alpha} \right) = \frac{\partial \ell_{\mytext{CAL}} }{\partial\gamma}(\gamma), \quad
 \tilde E \left( \frac{\partial \varphi_{+}}{\partial \gamma} \right) = \frac{\partial \ell_{\mytext{WL},1+} }{\partial\alpha}(\alpha; \gamma,\beta) , \label{eq:ell-phi-a} \\
&  \tilde E \left( \frac{\partial \varphi_{+}}{\partial \beta} \right) = (\Lambda-\Lambda^{-1}) \frac{\partial \ell_{\mytext{WL},1+}}{\partial\beta} (\beta; \gamma) ,
\label{eq:ell-phi-b}
\end{align}
where the partial derivatives of $\varphi_{+}$ are calculated in Supplement (\ref{eq:deriv-gam})--(\ref{eq:deriv-beta}).
Hence sequentially minimizing the three loss functions without penalties recovers the calibrated estimators in Section \ref{sec:cal}.
Interestingly, the roles of $\gamma$ and $\alpha$ are interchanged when taking derivatives in
the two equations in (\ref{eq:ell-phi-a}), whereas derivatives are taken with respect to only $\beta$ in (\ref{eq:ell-phi-b}).

We provide theoretical analysis of the proposed method in high-dimensional settings in Supplement Section \ref{sec:theory}.
With possible misspecification of models (\ref{eq:logit+pi})--(\ref{eq:linear+eta}), our main result, Theorem \ref{thm:mu1+expan}, shows that
under the sparsity condition $( s_{\bar\gamma}+ s_{\bar\alpha}+ s_{\bar\beta} ) \sqrt{\log(p^\dag)} = o(\sqrt{n})$,
the proposed estimator $\hat\mu^{1+}_{\mytext{RCAL}}$ admits the asymptotic expansion similar to (\ref{eq:cal-expan}):
\begin{align}
\hat\mu^{1+}_{\mytext{RCAL}}
& = \tilde E \left\{ \varphi_{+} (O; \pi(\cdot;\bar\gamma_{\mytext{CAL}}),\eta_{1+} (\cdot;\bar\alpha_{\mytext{WL},1+}), q_{1,\tau}(\cdot;\bar\beta_{\mytext{WL},1+}) ) \right\}  + o_p(n^{-1/2})  , \label{eq:rcal-expan}
\end{align}
where  $p^\dag = 1+\max(p,m)$, $(s_{\bar\gamma},s_{\bar\alpha},s_{\bar\beta})$ are the sizes of the subsets of
$(\bar\gamma_{\mytext{CAL}},\bar\alpha_{\mytext{WL},1+}, \bar\beta_{\mytext{WL},1+})$ including the intercept and other nonzero components, and
$(\bar\gamma_{\mytext{CAL}},\bar\alpha_{\mytext{WL},1+}, \bar\beta_{\mytext{WL},1+})$
are the target values defined as the minimizers of the population versions of
$\ell_{\mytext{CAL}}(\gamma)$, $\ell_{\mytext{WL},1+}(\beta; \bar\gamma_{\mytext{CAL}})$, and $\ell_{\mytext{WL},1+}(\alpha; \bar\gamma_{\mytext{CAL}} \bar\beta_{\mytext{WL},1+})$, as in Section \ref{sec:cal}.
If either model (\ref{eq:logit+pi}) or (\ref{eq:linear+eta}) with $\beta=\bar \beta_{\mytext{WL},1+}$ is correctly specified,
then a doubly robust confidence interval of asymptotic size $1-c$ can be obtained for
$ \mu^{1+} ( h^\T \bar\beta_{\mytext{WL},1+})$ as \vspace{-.1in}
\begin{align}
\Big(-\infty,\; \hat\mu^{1+}_{\mytext{RCAL}} + z_c \sqrt{ \hat V^{1+}_{\mytext{RCAL}} / n} \,\Big], \label{eq:upper-ci}
\end{align}
where $z_c$ is the $(1-c)$ quantile of $\N(0,1)$, and\vspace{-.05in}
\begin{align}
\hat V^{1+}_{\mytext{RCAL}} = \tilde E \left[ \left\{  \varphi_{+} (O; \pi(\cdot;\hat\gamma_{\mytext{RCAL}}),\eta_{1+} (\cdot;\hat\alpha_{\mytext{RWL},1+}), q_{1,\tau}(\cdot;\hat\beta_{\mytext{RWL},1+}) ) - \hat\mu^{1+}_{\mytext{RCAL}} \right\}^2 \right]. \label{eq:est-variance}
\end{align}
By Proposition \ref{pro:mu1-h-DR}, (\ref{eq:upper-ci}) is also a doubly robust confidence interval at asymptotic level (not size) $1-c$ for
the sharp bound $\mu^{1+}$ and the true value $\mu^1$ under model (\ref{eq:sen-model}).
If, further, model (\ref{eq:linear+Q}) is correctly specified, then
$ \mu^{1+} ( h^\T \bar\beta_{\mytext{WL},1+})$ coincides with the sharp bound $\mu^{1+}$,
and hence (\ref{eq:upper-ci}) is a doubly robust, asymptotic size $1-c$ confidence interval for $\mu^{1+}$.

It is instructive to compare our method with regularized maximum likelihood-related estimation (RML) in Dorn et al.~(2002).
The upper bound for $\mu^1$ based on RML is
\begin{align}
\hat\mu^{1+}_{\mytext{RML}} &= \hat\mu^{1+} (\hat\gamma_{\mytext{RML}},\hat\alpha_{\mytext{RU},1+}, \hat\beta_{\mytext{RU},1+}), \label{eq:mu1+rml}
\end{align}
where $(\hat\gamma_{\mytext{RML}},\hat\alpha_{\mytext{RU},1+}, \hat\beta_{\mytext{RU},1+})$ are the Lasso penalized versions of
$(\hat\gamma_{\mytext{ML}},\hat\alpha_{\mytext{U},1+}, \hat\beta_{\mytext{U},1+})$ defined in Section \ref{sec:cal}.
By extending the properties of $(\hat\gamma_{\mytext{ML}},\hat\alpha_{\mytext{U},1+}, \hat\beta_{\mytext{U},1+})$ discussed in 
Supplement Section \ref{sec:add-cal} to high-dimensional settings, 
if all three models (\ref{eq:logit+pi})--(\ref{eq:linear+eta}) are correctly specified,
then under suitable sparsity conditions, $\hat\mu^{1+}_{\mytext{RML}}$ can be shown to admit an asymptotic expansion in the form of (\ref{eq:rcal-expan})
and valid Wald confidence intervals can be obtained similarly as (\ref{eq:upper-ci}).
But such properties would in general fail if any of the models (\ref{eq:logit+pi})--(\ref{eq:linear+eta}) is misspecified.
See Tan (2020b, Section 3.2) for related discussion.

\section{Further development}

\subsection{Dual formulation and relaxed estimation} \label{sec:relax}

The Lasso penalties in Section \ref{sec:rcal} are mainly introduced to facilitate the control of estimation errors
when applying the augmented IPW estimator $\hat\mu^{1+}(\gamma,\alpha,\beta)$ in (\ref{eq:mu1+class})
in high-dimensional, sparse settings. While such use of Lasso penalties can be commonly found in high-dimensional statistics (Buhlmann \& van de Geer 2011),
we return to the population bound $\mu^{1+}_h$ in (\ref{eq:mu1+h})
and show that the Lasso penalized weighted estimator $\hat\beta_{\mytext{RWL},1+}$ for outcome quantile regression (\ref{eq:linear+Q}) can
also be derived in a dual formulation, through a relaxation of the sample version of (\ref{eq:mu1+h}).
Subsequently, we propose a simple modification to $\hat\mu^{1+}_{\mytext{RCAL}}$,
to improve finite-sample performance while achieving similar convergence properties.

First, a sample version of $\mu^{1+}_h$ can be obtained as follows. For an estimator $\hat\gamma$ in model (\ref{eq:logit+pi}) and
a vector of covariate functions $h(x) = (h_0(x), \ldots, h_m(x))^\T$ with $h_0\equiv 1$, let
\begin{align}
\hat \mu_h^{1+} (\hat\gamma) = \max_{  \lambda_{1i}: i=1,\ldots,n }
\frac{1}{n} \sum_{i=1}^n \left\{ T_i Y_i + T_i \frac{1-\pi(X_i;\hat\gamma)}{\pi(X_i;\hat\gamma)} \lambda_{1i} Y_i \right\}, \label{eq:hat-mu1+h}
\end{align}
where $\{ \lambda_{1i}: i=1,\ldots,n \}$ are real numbers satisfying the constraints
\begin{align}
& \Lambda^{-1} \le \lambda_{1i} \le \Lambda, \quad i=1,\ldots,n,  \label{eq:hat-mu1+h-contr1} \\
& \sum_{i=1}^n  T_i \frac{1-\pi(X_i;\hat\gamma)}{\pi(X_i;\hat\gamma)} \lambda_{1i} h(X_i) =  \sum_{i=1}^n T_i \frac{1-\pi(X_i;\hat\gamma)}{\pi(X_i;\hat\gamma)} h(X_i).  \label{eq:hat-mu1+h-contr2}
\end{align}
Then applying Proposition \ref{pro:mu1-h} with the expectation taken under the empirical distribution (which is allowed) shows that
$\hat \mu_h^{1+} (\hat\gamma)$ can be identified  as
\begin{align}
 \hat \mu_h^{1+} (\hat\gamma) & = \tilde E \left\{ \frac{T}{\pi(X;\hat\gamma)} Y + (\Lambda-\Lambda^{-1})
    T \frac{1-\pi(X;\hat\gamma)}{\pi(X;\hat\gamma)} \rho_\tau (Y, h^\T \hat\beta_{\mytext{WL},1+}(\hat\gamma) ) \right\},  \label{eq:dual-hat-mu1+h}
\end{align}
where  $\hat\beta_{\mytext{WL},1+}(\hat\gamma) = \argmin_{\beta}\; \ell_{\mytext{WL},1+} (\beta; \hat\gamma)$, with $\ell_{\mytext{WL},1+} (\beta; \hat\gamma)$
defined in Section \ref{sec:rcal}, i.e., $\hat\beta_{\mytext{WL},1+}(\hat\gamma)$ is defined by (non-penalized)
weighted $\tau$-quantile regression in the treated group with weight
$w(X;\hat\gamma) = (1-\pi(X;\hat\gamma))/\pi(X;\hat\gamma)$.
For $\hat\gamma=\hat\gamma_{\mytext{CAL}}$, the estimator $\hat\beta_{\mytext{WL},1+}(\hat\gamma_{\mytext{CAL}})$
recovers the (non-penalized) estimator $\hat\beta_{\mytext{WL},1+}$ discussed in Section \ref{sec:cal}.

The estimator $\hat \mu_h^{1+} (\hat\gamma) $ from the primal definition (\ref{eq:hat-mu1+h}) can be seen
as a variant of the linear programming method in Tan (2006) for sensitivity analysis under model (\ref{eq:sen-model}).
In fact, let $\hat H_1$ be a discrete distribution supported on $\{ (Y_i,X_i): T_i=1, i=1,\ldots,n\}$
with $\hat H_1( \{Y_i,X_i\}) = n^{-1} \pi^{-1}(X_i;\hat\gamma)$,
where $\hat\gamma$ is chosen, for example $\hat\gamma=\hat\gamma_{\mytext{CAL}}$, such that $n^{-1} \sum_{i=1}^n T_i/\pi(X_i;\hat\gamma) = 1$.
Then (\ref{eq:hat-mu1+h}) can be rewritten in a similar form as in Tan (2006):
\begin{align*}
\hat \mu_h^{1+} (\hat\gamma) =\max_{\lambda_1} \int \{ \pi(x;\hat\gamma) + (1-\pi(x;\hat\gamma)) \lambda_1 \} y \,\dif \hat H_1,
\end{align*}
over $\lambda_1 = \lambda_1(y,x)$ subject to the constraints
\begin{align*}
& \Lambda^{-1} \le \lambda_1(y,x) \le \Lambda \quad \text{for any } (y,x), \\
& \int (1-\pi(x;\hat\gamma)) \lambda_1 h(x) \,\dif \hat H_1 = \int (1-\pi(x;\hat\gamma)) h(x) \,\dif \hat H_1.
\end{align*}
Hence (\ref{eq:dual-hat-mu1+h}) provides a dual representation of the sample bound $\hat \mu_h^{1+} (\hat\gamma)$ through weighted $\tau$-quantile regression
with parameter $\beta \in \bbR^{1+m}$.
For fixed $p$ and $m$ and growing $n$ as in Supplement Section \ref{sec:add-cal}, this representation can be used
to study sampling properties of $\hat \mu_h^{1+} (\hat\gamma)$,
which was not addressed in Tan (2006).
The estimator $\hat \mu_h^{1+} (\hat\gamma)$ can be shown to be consistent for the population upper bound $\mu^{1+} ( h^\T \bar\beta_{\mytext{WL},1+}(\gamma^*) )$
if model (\ref{eq:logit+pi}) is correctly specified,
where $ \bar\beta_{\mytext{WL},1+}(\gamma^*)$ is a minimizer of the population version of $\ell_{\mytext{WL},1+} (\beta; \gamma^*)$.
The bound  $\mu^{1+} ( h^\T \bar\beta_{\mytext{WL},1+}(\gamma^*) )$ reduces to the sharp bound $\mu^{1+}$ if, further, model (\ref{eq:linear+Q})
is correctly specified.
An asymptotic expansion similar to (\ref{eq:general-expan}) can also be obtained for $\hat \mu_h^{1+} (\hat\gamma)$.

In high-dimensional settings with large $m$ (as well as $p$) relative to $n$,
a similar approach as in Section \ref{sec:rcal} is to use the sample bound $\hat \mu_h^{1+} (\hat\gamma)$ in (\ref{eq:dual-hat-mu1+h}), but replace
$\hat\beta_{\mytext{WL},1+}(\hat\gamma)$ (as well as $\hat\gamma$) with a Lasso penalized version.
Alternatively, we consider a relaxation of $\hat \mu_h^{1+} (\hat\gamma) $
in the primal definition (\ref{eq:hat-mu1+h}) as follows. For a tuning parameter $\lambda_\beta\ge 0$, let
\begin{align}
\dhat \mu_h^{1+} (\hat\gamma) = \max_{ \lambda_{1i}: i=1,\ldots,n }
\frac{1}{n} \sum_{i=1}^n \left\{ T_i Y_i + T_i \frac{1-\pi(X_i;\hat\gamma)}{\pi(X_i;\hat\gamma)} \lambda_{1i} Y_i \right\}, \label{eq:dhat-mu1+h}
\end{align}
where $\{ \lambda_{1i}: i=1,\ldots,n \}$ are real numbers satisfying (\ref{eq:hat-mu1+h-contr1}) and,
as a relaxation of (\ref{eq:hat-mu1+h-contr2}),
\begin{align}
&  \sum_{i=1}^n  T_i \frac{1-\pi(X_i;\hat\gamma)}{\pi(X_i;\hat\gamma)} \lambda_{1i} = \sum_{i=1}^n T_i \frac{1-\pi(X_i;\hat\gamma)}{\pi(X_i;\hat\gamma)} , \nonumber\\
& \frac{1}{n} \left| \sum_{i=1}^n  T_i \frac{1-\pi(X_i;\hat\gamma)}{\pi(X_i;\hat\gamma)} \lambda_{1i} h_j(X_i)- \sum_{i=1}^n T_i \frac{1-\pi(X_i;\hat\gamma)}{\pi(X_i;\hat\gamma)} h_j(X_i) \right| \le (\Lambda-\Lambda^{-1}) \lambda_\beta,\;  j=1,\ldots, m. \label{eq:hat-mu1+h-contr2b}
\end{align}
The relaxation of (\ref{eq:hat-mu1+h-contr2}) to (\ref{eq:hat-mu1+h-contr2b}) is of different nature from
the relaxation of (\ref{eq:lambda1-givenX-b}) to (\ref{eq:lambda1-marginX}) when deriving population bounds, although
both are directed to the inherent constraints about $\lambda^*_1(\cdot)$.
The following result gives a dual representation of $\dhat \mu_h^{1+} (\hat\gamma)$,
related to Lasso penalized weighted quantile regression. 
The proof involves optimizing over the dual variables associated with the constraint (\ref{eq:hat-mu1+h-contr1})
in the duality theory for linear programming.
A population (and more general) result can also be derived, corresponding to a relaxation of $\mu^{1+}_h$ in (\ref{eq:mu1+h}).

\begin{pro} \label{pro:relaxed}
Let $\hat\beta_{\mytext{RWL},1+}(\hat\gamma) = \argmin_{\beta}\, \{\ell_{\mytext{WL},1+} (\beta,\hat\gamma) +\lambda_\beta \|\beta_{1:p}\|_1 \}$,
with $\ell_{\mytext{WL},1+} (\beta,\hat\gamma)$ defined in Section \ref{sec:rcal}. Then $\dhat \mu_h^{1+} (\hat\gamma)$ in (\ref{eq:dhat-mu1+h}) can be
determined as
\begin{align}
 \dhat \mu_h^{1+} (\hat\gamma) & =
\tilde E \left\{ \frac{T}{\pi(X;\hat\gamma)} Y + (\Lambda-\Lambda^{-1})
    T \frac{1-\pi(X;\hat\gamma)}{\pi(X;\hat\gamma)} \rho_\tau (Y, h^\T \hat\beta_{\mytext{RWL},1+}(\hat\gamma) ) \right\} \nonumber \\
&\quad    + (\Lambda-\Lambda^{-1}) \lambda_\beta \| \hat\beta_{\mytext{RWL},1+}(\hat\gamma)_{1:p}\|_1 .  \label{eq:dual-dhat-mu1+h}
\end{align}
\end{pro}

There is a subtle difference between the sample bound $\dhat \mu_h^{1+} (\hat\gamma)$  and
those obtained using Lasso penalized estimation in Section \ref{sec:rcal}. For IPW estimation, let
\begin{align*}
\hat\mu^{1+}_{\mytext{IPW}} (\gamma,\beta)
&= \tilde E \left\{ \varphi_{+,\mytext{IPW}} (O; \pi(\cdot; \gamma), h^\T \beta ) ) \right\} \\
&= \tilde E \left\{ \frac{T}{\pi(X;\gamma)} Y + (\Lambda-\Lambda^{-1})
    T \frac{1-\pi(X;\gamma)}{\pi(X;\gamma)} \rho_\tau (Y, h^\T \beta ) \right\} .
\end{align*}
Then $\dhat \mu_h^{1+} (\hat\gamma)$ in (\ref{eq:dual-dhat-mu1+h}) can be expressed as
\begin{align}
\dhat \mu_h^{1+} (\hat\gamma) = \hat\mu^{1+}_{\mytext{IPW}} (\hat\gamma, \hat\beta_{\mytext{RWL},1+}(\hat\gamma)) +
(\Lambda-\Lambda^{-1}) \lambda_\beta \| \hat\beta_{\mytext{RWL},1+}(\hat\gamma)_{1:p}\|_1 , \label{eq:dhat-IPW}
\end{align}
where $\hat\mu^{1+}_{\mytext{IPW}} (\hat\gamma, \hat\beta_{\mytext{RWL},1+}(\hat\gamma))$ is
a direct extension of $\hat \mu_h^{1+} (\hat\gamma)$ in (\ref{eq:dual-hat-mu1+h}) using the Lasso estimator $\hat\beta_{\mytext{RWL},1+}(\hat\gamma)$.
Hence $\dhat \mu_h^{1+} (\hat\gamma)$ differs from  $\hat\mu^{1+}_{\mytext{IPW}} (\hat\gamma, \hat\beta_{\mytext{RWL},1+}(\hat\gamma))$
in incorporating  the Lasso penalty of $\hat\beta_{\mytext{RWL},1+}(\hat\gamma)$.
An interesting property from including the particular penalty term is that, by the primal definition in (\ref{eq:dhat-mu1+h}),
$\dhat \mu_h^{1+} (\hat\gamma)$ is nondecreasing in the tuning parameter $\lambda_\beta$ for any fixed sample.
Such a property is in general not satisfied by $\hat\mu^{1+}_{\mytext{IPW}} (\hat\gamma, \hat\beta_{\mytext{RWL},1+}(\hat\gamma))$.
By this difference, the relaxation-based bound $\dhat \mu_h^{1+} (\hat\gamma)$ can be seen to reflect the impact of using the Lasso penalty more properly
than the Lasso-substituted bound $\hat\mu^{1+}_{\mytext{IPW}} (\hat\gamma, \hat\beta_{\mytext{RWL},1+}(\hat\gamma))$.

While further study is needed to compare the two approaches discussed above,
we extend the idea of incorporating a Lasso penalty term in the sample bound (not just in estimation of nuisance parameters)
to augmented IPW estimation. By mimicking (\ref{eq:dhat-IPW}), define
\begin{align}
\dhat \mu^{1+}_{\mytext{RCAL}} = \hat \mu^{1+}_{\mytext{RCAL}} +
(\Lambda-\Lambda^{-1}) \lambda_\beta \| (\hat\beta_{\mytext{RWL},1+})_{1:p} \|_1 , \label{eq:dhat-AIPW}
\end{align}
where $\hat \mu^{1+}_{\mytext{RCAL}} $ is defined in (\ref{eq:mu1+rcal}) and
$\hat\beta_{\mytext{RWL},1+}= \hat\beta_{\mytext{RWL},1+}(\hat\gamma_{\mytext{RCAL}})$, the same as in Section \ref{sec:rcal}.
Interestingly, Proposition \ref{pro:dhat-expan} shows that $\dhat \mu^{1+}_{\mytext{RCAL}}$
admits an asymptotic expansion similar to (\ref{eq:rcal-expan}) for $\hat \mu^{1+}_{\mytext{RCAL}}$,
but with an additional term as the limit of the Lasso penalty term.
Similar modification can be made to the lower bound $\hat \mu^{1-}_{\mytext{RCAL}}$.
Such sample bounds and associated confidence intervals will be said to be based on regularized calibrated estimation with relaxation.
We stress that the Lasso penalty term in (\ref{eq:dhat-AIPW}) is tied to the weighted Lasso estimator $\hat\beta_{\mytext{RWL},1+}$
as motivated by the dual representation (\ref{eq:dual-dhat-mu1+h}). If a Lasso version of the unweighted estimator $\hat\beta_{\mytext{U},1+}$
were used in place of $\hat\beta_{\mytext{RWL},1+}$ in outcome quantile regression, the connection with (\ref{eq:dual-dhat-mu1+h}) would be lost.

\begin{pro}  \label{pro:dhat-expan}
Suppose that the conditions of Theorem \ref{thm:mu1+expan} are satisfied and
$( s_{\bar\gamma}+ s_{\bar\alpha}+ s_{\bar\beta} ) \sqrt{\log(p^\dag)} = o(\sqrt{n})$.
Then the following asymptotic expansion holds:
\begin{align*}
\dhat\mu^{1+}_{\mytext{RCAL}}
& = \hat\mu^{1+} (\bar\gamma_{\mytext{CAL}},\bar\alpha_{\mytext{WL},1+}, \bar\beta_{\mytext{WL},1+}) +
(\Lambda-\Lambda^{-1}) \lambda_\beta \| (\bar\beta_{\mytext{WL},1+})_{1:p} \|_1
+ o_p(n^{-1/2}),  
\end{align*}
where $(\bar\gamma_{\mytext{CAL}},\bar\alpha_{\mytext{WL},1+}, \bar\beta_{\mytext{WL},1+})$ are the target values as before.
\end{pro}

From this result, similarly as in the discussion of (\ref{eq:upper-ci}) based on expansion (\ref{eq:rcal-expan}),
if either model (\ref{eq:logit+pi}) or (\ref{eq:linear+eta}) with $\beta=\bar \beta_{\mytext{WL},1+}$ is correctly specified, then
\begin{align}
\Big(-\infty,\; \dhat\mu^{1+}_{\mytext{RCAL}} + z_c \sqrt{ \hat V^{1+}_{\mytext{RCAL}} / n} \,\Big], \label{eq:dhat-upper-ci}
\end{align}
is a doubly robust confidence interval at asymptotic size $1-c$ for
 $\mu^{1+} (h^\T \bar\beta_{\mytext{WL},1+}) + (\Lambda-\Lambda^{-1}) \lambda_\beta \| (\bar\beta_{\mytext{WL},1+})_{1:p} \|_1$,
and hence at asymptotic level (not size) $1-c$ for the population upper bound $\mu^{1+} (h^\T \bar\beta_{\mytext{WL},1+})$
and the sharp bound $\mu^{1+}$.
The conservativeness in covering $\mu^{1+} (h^\T \bar\beta_{\mytext{WL},1+})$ can also be directly deduced because (\ref{eq:dhat-upper-ci}) contains
(\ref{eq:upper-ci}), which is a doubly robust confidence interval at asymptotic size $1-c$ for $\mu^{1+} (h^\T \bar\beta_{\mytext{WL},1+})$.
The additional term, $(\Lambda-\Lambda^{-1}) \lambda_\beta \| (\bar\beta_{\mytext{WL},1+})_{1:p} \|_1$, seems to be
nonnegligible given $\lambda_\beta$ of order $\sqrt{\log(p^\dag)/n}$.
Nevertheless, our numerical study in Section \ref{sec:simulation}
shows that confidence intervals from RCAL with relaxation are often more satisfactory than vanilla RCAL,
which may suffer under-coverage of even the sharp bounds $\mu^{1-}$ and $\mu^{1+}$. In such applications,
all Lasso tuning parameters $(\lambda_\gamma,\lambda_\beta,\lambda_\alpha)$ are selected by cross validation,
which is not taken into account in our theory in Supplement Section \ref{sec:theory}.
Further research is desired to investigate these issues.

\subsection{Binary and range-restricted outcomes} \label{sec:binary}

The use of the linear quantile regression (\ref{eq:linear+Q}) and linear mean regression (\ref{eq:linear-eta}) can be questionable
for binary and range-restricted (e.g., nonnegative) outcomes. For concreteness,
we consider the prototypical case of a binary outcome $Y$ (taking values $0$ or $1$) and discuss two options for deriving upper bounds of $\mu^{1+}$ based on
regularized calibrated estimation.

For both options, we still use the linear quantile model (\ref{eq:linear+Q}) by the following considerations.
For a binary outcome $Y$, the conditional quantile $q^*_{1,\tau}(X)$ can be determined from the conditional probability
$m^*_1(X) = E(Y | T=1,X)$ as \vspace{-.05in}
\begin{align}
& q^*_{1,\tau}(X) = \left\{ \begin{array}{cl}
1, & \text{if }  m^*_1(X) > 1-\tau,\\
0, & \text{if }  m^*_1(X) \le 1-\tau.
\end{array} \right. \label{eq:q-from-m}
\end{align}
The discontinuity of the preceding expression makes it difficult to well specify either a linear model or a nonlinear (smooth in $\beta$)
model $q_{1,\tau} (X;\beta)$ for $q^*_{1,\tau}(X)$. For the linear quantile model (\ref{eq:linear+Q}), the convexity of
the loss function, $\ell_{\mytext{WL},1+}(\beta;\gamma)$ in (\ref{eq:beta-loss}),
facilitates both numerical optimization and theoretical analysis as discussed in Supplement Section \ref{sec:theory}.
By comparison, for a nonlinear model $q_{1,\tau} (X;\beta)$, the corresponding loss function  for calibrated estimation,
$\ell_{\mytext{WL},1+}(\beta;\gamma)$ in (\ref{eq:beta-loss}) with $h^\T\beta$ replaced by $q_{1,\tau} (X;\beta)$,
would be nonconvex in $\beta$ and hence both numerical optimization and theoretical analysis would be challenging.
Importantly, although model (\ref{eq:linear+Q}) may be misspecified,
our method is developed to be robust to such misspecification:
the proposed point bounds and confidence intervals are valid (or conservative) in relation to the sharp bound $\mu^{1+}$,
provided either model (\ref{eq:logit+pi}) or (\ref{eq:linear+eta}) is correctly specified.

The two options differ in whether the linear mean model (\ref{eq:linear+eta}) is retained.
For the first option, model (\ref{eq:linear+eta}) is used and allowed to be misspecified.
In the presence of such misspecification,
the proposed point bounds and confidence intervals remain valid (or conservative) provided model (\ref{eq:logit+pi}) is correctly specified.
Alternatively, we point out another option which provides a more direct extension of
model-assisted inference with binary outcomes under unconfoundedness in Tan (2020b).
Consider a logistic regression model with binary $Y$:
\begin{align}
P (Y=1 | T=1, X) = m_1 (X;\alpha) = \{1+\exp(-f^\T(X)\alpha)\}^{-1}, \label{eq:logit+m}
\end{align}
where $f(X)$ is the same as in model (\ref{eq:logit+pi}).
By the relationship $E \{ \tilde Y_{+} ( q_{1,\tau}(\cdot;\beta) )| T=1,X \}
=m^*_1(X) + (\Lambda-\Lambda^{-1}) \{ m_1^*(X) \rho_\tau(1, q_{1,\tau}(X;\beta)) + (1-m_1^*(X))  \rho_\tau(0, q_{X,\tau}(X;\beta)) \}$,
model (\ref{eq:logit+m}) induces a nonlinear model for $E \{ \tilde Y_{+} ( q_{1,\tau}(\cdot;\beta) )| T=1,X \}$:
\begin{align}
& E \{ \tilde Y_{+} ( q_{1,\tau}(\cdot;\beta) )| T=1,X \} = \eta_1(X;\alpha) = m_1(X;\alpha) \nonumber \\
& \quad  + (\Lambda-\Lambda^{-1}) \left\{ m_1(X;\alpha) \rho_\tau(1, q_{1,\tau}(X;\beta)) + (1-m_1(X;\alpha))  \rho_\tau(0, q_{1,\tau}(X;\beta)) \right\}.
\label{eq:nonlinear+eta}
\end{align}
The calibration equation for estimating $\alpha$ can be calculated as
\begin{align}
0 &= \tilde E \left( \frac{\partial \varphi_{+}}{\partial \gamma} \right) = \tilde E \left[ - T \frac{1-\pi(X;\gamma)}{\pi(X;\gamma)}
 \left\{ \tilde Y_{+} (h^\T \beta ) - \eta_1(X;\alpha)  \right\} f(X) \right] \nonumber \\
 & =  \tilde E \left[ - T \frac{1-\pi(X;\gamma)}{\pi(X;\gamma)} v(X;\beta)
 \left\{ Y - m_1 (X;\alpha) \right\} f(X) \right],  \label{eq:alpha-eq2}
\end{align}
where $v(X; \beta) = \Lambda - (\Lambda - \Lambda^{-1}) \{ (h^\T(X)\beta )_+ - (h^\T(X)\beta -1)_+ \} \in [\Lambda^{-1} ,\Lambda]$.
This is the score equation for fitting weighted logistic regression of $Y$ given $T=1$ and $X$, with the weight
$\{(1-\pi(X;\gamma)) / \pi(X;\gamma)\} v(X;\beta)$. For regularized calibrated estimation, $\hat\alpha_{\mytext{RWL},1+}$ can be defined as in Section \ref{sec:rcal}
except with the loss function
\begin{align*}
 \ell_{\mytext{RWL},1+} (\alpha; \gamma,\beta) =  \tilde E \left[ T \frac{1-\pi(X;\gamma)}{\pi(X;\gamma)} v(X;\beta)
 \left\{ - Y f^\T (X) \alpha - m_1 (X;\alpha) \right\} \right] .
\end{align*}
For the redefined $\hat\alpha_{\mytext{RWL},1+}$ but the same $\hat\gamma_{\mytext{RCAL}}$ and $\hat\beta_{\mytext{RWL},1+}$ as in Section \ref{sec:rcal},
if model (\ref{eq:logit+pi}) is correctly specified, then
a similar result as Tan (2020b, Theorem 6) can be established such that
$\hat\mu^{1+}_{\mytext{RCAL}} = \hat\mu^{1+}(\hat\gamma_{\mytext{RCAL}}, \hat\alpha_{\mytext{RWL},1+}, \hat\beta_{\mytext{RWL},1+})$
admits the asymptotic expansion (\ref{eq:rcal-expan})
provided 
$( s_{\bar\gamma}+ s_{\bar\alpha}+ s_{\bar\beta} ) \sqrt{\log(p^\dag)} = o(\sqrt{n})$,
where $(s_{\bar\gamma}, s_{\bar\alpha}, s_{\bar\beta})$ are the non-sparsity sizes of the target values.
Under this sparsity condition, if model (\ref{eq:logit+pi}) is correctly specified but model (\ref{eq:logit+m}) may be misspecified, then
the confidence interval (\ref{eq:upper-ci}) achieves asymptotic size $1-c$ for the population upper bound $\mu^{1+} (h^\T \bar\beta_{\mytext{WL},1+})$
and the sharp bound $\mu^{1+}$.
Such a method is said to be based on model (\ref{eq:logit+pi}) and assisted by models (\ref{eq:linear+Q}) and (\ref{eq:logit+m}).
This method also has the conceptual advantage that propensity score estimation is conducted without access to any outcome data, which helps to
avoid bias from outcome modeling (Rubin 2001).

In contrast with Section \ref{sec:rcal}, the confidence intervals here are not doubly robust, i.e., may be invalid
if model (\ref{eq:logit+m}) is correctly specified but model (\ref{eq:logit+pi}) is misspecified.
The reason is that the estimator $\hat\gamma_{\mytext{RCAL}}$ is defined according to the calibration equation for estimating $\gamma$ with linear model (\ref{eq:linear+eta})
but not the nonlinear model (\ref{eq:nonlinear+eta}), which can be calculated as
\begin{align}
0 &= \tilde E \left( \frac{\partial \varphi_{+}}{\partial \alpha} \right) = \tilde E \left[ - \left\{ \frac{T}{\pi(X;\gamma)} - 1\right\}
\frac{\partial}{\partial\alpha}  \eta_1(X;\alpha)  \right] \nonumber \\
 & =  \tilde E \left[  - \left\{ \frac{T}{\pi(X;\gamma)} - 1\right\} v(X;\beta) \dot m_1(X;\alpha) f(X) \right],  \label{eq:gamma-eq2}
\end{align}
where $\dot m_1(X;\alpha) = m_1(X;\alpha)(1-m_1(X;\alpha))$.
The calibration equations (\ref{eq:alpha-eq2})--(\ref{eq:gamma-eq2}) are coupled, in each depending on $(\gamma,\alpha)$,
and hence need to be simultaneously solved, together with the calibration equation for $\beta$, to define the calibrated estimators.
This leads to both computational and statistical challenges in low- and high-dimensional settings, as discussed in Tan (2020b, Section 3.5).
In principle, the iterative approach of Ghosh \& Tan (2022) can be exploited to develop doubly robust confidence intervals,
which is not pursued here.

We briefly comment on how binary outcomes can be handled by Dorn et al.~(2021).
A quantile model can be defined by substituting $m_1(X;\alpha)$ for $m^*_1(X)$ in (\ref{eq:q-from-m}),
and then a model for $E \{ \tilde Y_{+} ( q_{1,\tau}(\cdot;\beta) )| T=1,X \}$ is obtained by
substituting the quantile model together with $m_1(X;\alpha)$ into (\ref{eq:nonlinear+eta}).
These models are in general discontinuous in $\alpha$, and hence would render calibrated estimation inapplicable.
On the other hand, in the approach of Dorn et al.~(2021),
the parameter $\alpha$ is estimated by regularized maximum likelihood for logistic regression (\ref{eq:logit+m}),
in addition to similar estimation of $\gamma$ for logistic regression (\ref{eq:logit+pi}).
However, the resulting Wald confidence intervals defined in the form of (\ref{eq:upper-ci}) would not be valid
unless both models (\ref{eq:logit+pi}) and (\ref{eq:logit+m}) are correctly specified, similarly as discussed in Section \ref{sec:rcal}.

\section{Simulation study} \label{sec:simulation}

We conduct a simulation study to evaluate the performance of the proposed method,
compared with that based on conventional estimation of nuisance functions as in Dorn et al.~(2021),
without or with regularization in low- or high-dimensional settings.

The regularized calibrated (RCAL) and likelihood-related (RML) estimation for $\gamma$ and $\alpha$
are implemented using the R package \texttt{RCAL} (Tan \& Sun 2020).
The Lasso penalized estimation for $\beta$ with or without weights in quantile regression
is implemented using the R package \texttt{quantreg} (Koenker et al.~2022).
For both RCAL and RML, the Lasso tuning parameters $(\lambda_\gamma, \lambda_\beta, \lambda_\alpha)$ are selected using
5-fold cross-validation based on the corresponding loss functions,
where the set of possible values searched is $\{ \lambda^* / 2^j: j=0,1,\ldots,10\}$
and $\lambda^*$ is a data-dependent limit such that the Lasso estimate is zero.
For RCAL, the estimators of $(\gamma,\beta,\alpha)$ are computed with Lasso tuning parameters selected sequentially
similarly as described in Tan (2020b, Section 4.1).

Our simulation settings for the observed data $(TY, T, X)$ are similar as in Tan (2020b), but unmeasured confounding is allowed.
Let $X=(X_1,\ldots,X_p)$ be multivariate normal with means 0 and covariances $\cov(X_j,X_k)= 2^{-|j-k|}$ for $1\le j,k\le p$.
In addition, let $X_j^\dag = X_j + \{(X_j+1)_+\}^2$ for $j=1,\ldots,4$.
Consider the following data-generating configurations. \vspace{-.05in}
\begin{itemize}  \addtolength{\itemsep}{-.1in}
\item[(C1)] Generate $T$ given $X$ as a Bernoulli variable with
$P(T=1 | X) = \{ 1+ \exp( -1-X_1 -0.5 X_2 - 0.25 X_3 - 0.125 X_4 ) \}^{-1}$,
and, independently, generate $Y$ given $T=1$ and $X$ as a normal variable with variance 1 and mean
$ X_1 + 0.5 X_2 + 0.25 X_3 + 0.125 X_4$.

\item[(C2)] Generate $T$ give $X$ as in (C1), but, independently, generate $Y$ given $T=1$ and $X$ as a normal variable with variance 1 and mean
$ X_1^\dag + 0.5 X_2^\dag + 0.25 X_3^\dag + 0.125 X_4^\dag$.

\item[(C3)] Generate  $Y$ given $T=1$ and $X$ as in (C1), but, independently, generate $T$ given $X$ as a Bernoulli variable with
$P(T=1 | X) = \{ 1+ \exp( -1- X_1^\dag - 0.5 X_2^\dag - 0.25 X_3^\dag -0.125 X_4^\dag  ) \}^{-1}$.
\end{itemize}\vspace{-.05in}
The three working models (\ref{eq:logit+pi})--(\ref{eq:linear+eta}) are used, with $f_j(X) =h_j(X) = X_j$ for $j=1,\ldots,p \,(=m)$.
These models can be classified as follows, depending on the data configuration above: \vspace{-.05in}
\begin{itemize} \addtolength{\itemsep}{-.1in}
\item[(C1)] models (\ref{eq:logit+pi})--(\ref{eq:linear+eta}) are correctly specified;

\item[(C2)] model (\ref{eq:logit+pi}) is correctly specified, and models (\ref{eq:linear+Q})--(\ref{eq:linear+eta}) are misspecified;

\item[(C3)] model (\ref{eq:logit+pi}) is misspecified, and models (\ref{eq:linear+Q})--(\ref{eq:linear+eta}) are correctly specified.
\end{itemize}\vspace{-.05in}
By design, the two models (\ref{eq:linear+Q})--(\ref{eq:linear+eta}) are either correctly specified or misspecified in the same way.
For simplicity, we refer to (\ref{eq:logit+pi}) as a propensity score (PS) model and models (\ref{eq:linear+Q})--(\ref{eq:linear+eta})
as outcome quantile and mean regression (OR) models.

\begin{figure} [t!]
\begin{tabular}{c}
\includegraphics[width=6in, height=2in]{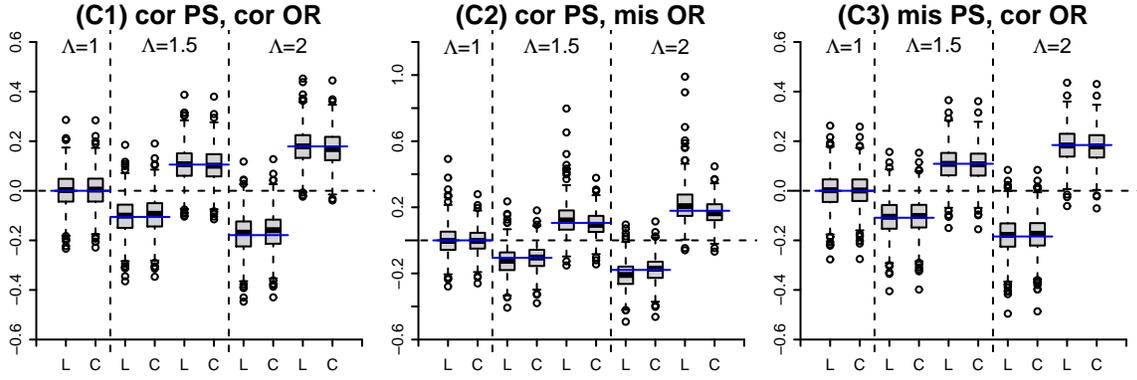} \vspace{-.1in}
\end{tabular}
\caption{\scriptsize Boxplots of point bounds on $\mu^1$ with $(n,p)=(800,10)$ from likelihood-related estimation (L)
and calibrated estimation (C), such as
$\hat\mu^{1+}(\hat\gamma_{\mytext{ML}},\hat\alpha_{\mytext{U},1+}, \hat\beta_{\mytext{U},1+})$  and
$\hat\mu^{1+}(\hat\gamma_{\mytext{CAL}},\hat\alpha_{\mytext{WL},1+}, \hat\beta_{\mytext{WL},1+})$
for three choices of $\Lambda$, separated by vertical lines.
For $\Lambda=1.5$ or 2, both lower (left) and upper (right) bounds are presented.
The sharp lower and upper bounds $\mu^{1 \mp}$ are located at the horizontal line segments (blue).
} \label{fig:simu-p10} 
\end{figure}

\begin{figure} [t!]
\begin{tabular}{c}
\includegraphics[width=6in, height=2in]{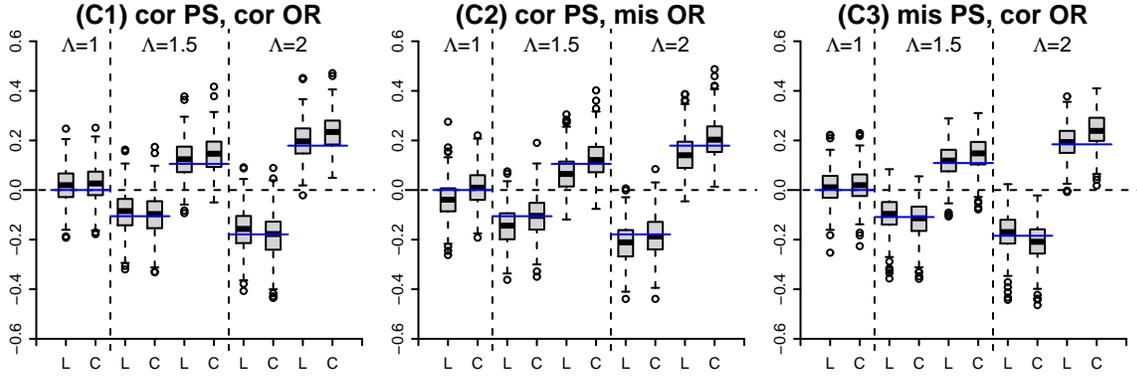} \vspace{-.1in}
\end{tabular}
\caption{\scriptsize Boxplots of point bounds on $\mu^1$ from regularized likelihood-related estimation (L)
and regularized calibrated estimation with relaxation (C) with $(n,p)=(800,200)$, similarly as in Figure \ref{fig:simu-p10}.
} \label{fig:simu-p200} 
\end{figure}

For $(n,p)=(800,10)$, Figure~\ref{fig:simu-p10}
presents the boxplots of points bounds on $\mu^1$ from non-regularized estimation for three choices of $\Lambda$ (including $\Lambda=1$),
obtained from 500 repeated simulations.
While the boxplots from calibrated (CAL) and likelihood-related (ML) estimation are similar in the configurations (C1) and (C3),
there are considerable differences in (C2):
the point bounds from calibrated estimation not only show smaller variances but also are, on average, closer to the sharp bounds
(i.e., upper/lower bounds are closer from above/below), compared with those from likelihood-related estimation.
This difference agrees with our theory in Section \ref{sec:DR-bound}:
if PS model is correct, then our population bounds based on weighted quantile regression are tighter than based on unweighted quantile regression.

For $(n,p)=(800,200)$, Figure~\ref{fig:simu-p200}
presents the boxplots of point bounds on $\mu^1$ from regularized estimation,
RCAL with relaxation and RML as in (\ref{eq:dhat-AIPW}) and (\ref{eq:mu1+rml}).
The results from vanilla RCAL (without relaxation) are presented in Supplement Section \ref{sec:add-simulation}.
Compared with the case of $p=10$, the point bounds from RCAL and RML show more noticeable biases,
both possibly due to regularization.
The biases from RCAL, if nonnegligible, are consistently conservative (i.e., positive/negative for upper/lower bounds) against the sharp bounds,
whereas those from RML may not be conservative; see for example the upper RML bounds in the case (C2).
This difference can be attributed to the relaxation used in RCAL, because vanilla RCAL bounds may fail to be conservative, as shown in
Figure \ref{fig:simu-p200-vanilla} in the Supplement.

\begin{table}[t!]
\caption{\small Coverage proportions of confidence intervals (CIs) from RML
and RCAL with relaxation for $(n,p)=(800,200)$} \label{tab:simu-p200}  \vspace{-.05in}
\scriptsize
\begin{center}
\begin{tabular*}{1\textwidth}{@{\extracolsep\fill} c lll lll lll} \hline
      & \multicolumn{3}{c}{(C1) cor PS, cor OR} & \multicolumn{3}{c}{(C2) cor PS, mis OR} & \multicolumn{3}{c}{(C3) mis PS, cor OR} \\
      & $\Lambda=1$ & $\Lambda=1.5$ & $\Lambda=2$ & $\Lambda=1$ & $\Lambda=1.5$ & $\Lambda=2$ & $\Lambda=1$ & $\Lambda=1.5$ & $\Lambda=2$ \\ \hline
      & \multicolumn{9}{c}{Lower 95\% CIs for the sharp lower bound of $\mu^1$} \\
RML  & 0.886 & 0.896 & 0.896 & 0.992 & 0.986 & 0.984 & 0.930 & 0.934 & 0.930 \\
RCAL & 0.870 & 0.906 & 0.922 & 0.916 & 0.924 & 0.926 & 0.906 & 0.964 & 0.974 \\

      & \multicolumn{9}{c}{Upper 95\% CIs for the sharp upper bound of $\mu^1$} \\
RML  & 0.976 & 0.966 & 0.960  & 0.866 & 0.834 & 0.834 & 0.972 & 0.954 & 0.952 \\
RCAL & 0.974 & 0.986 & 0.990  & 0.960 & 0.974 & 0.974 & 0.982 & 0.978 & 0.986 \\

      & \multicolumn{9}{c}{90\% CIs for the sharp lower to upper bound of $\mu^1$} \\
RML  & 0.862 & 0.862 & 0.856 & 0.858 & 0.820 & 0.818 & 0.902 & 0.888 & 0.882 \\
RCAL & 0.844 & 0.892 & 0.912 & 0.876 & 0.898 & 0.900 & 0.888 & 0.942 & 0.960 \\ \hline
\end{tabular*}\\[.05in]
\end{center}  \vspace{-.1in}
\end{table}

For $(n,p)=(800,200)$, Table~\ref{tab:simu-p200}
presents the coverage proportions of confidence intervals on the sharp bounds of $\mu^1$ from
RCAL with relaxation and RML.\
The results from vanilla RCAL (without relaxation) are presented in the Supplement,
as well as those for $(n,p)=(800,10)$.
While the coverage proportions show various deviations from the nominal probabilities, the differences between
RCAL and RML are the most pronounced in the case (C2), where RCAL two-sided confidence intervals achieve
coverage proportions close to the nominal probability 90\%, but RML confidence intervals suffer noticeable under-coverage.
Such improvement from RCAL is facilitated by the use of relaxation in Section \ref{sec:relax}. Without relaxation, RCAL confidence intervals also suffer
under-coverage in the case (C2), as seen from Table \ref{tab:simu-p200-vanilla} in the Supplement.

\section{Empirical application} \label{sec:application}

We provide an application to a medical study in Connors et al.~(1996) on the effects of right heart catheterization (RHC).
The study  included $n=5735$ critically ill patients  admitted  to  the  intensive  care  units  of  5  medical  centers.
For each patient, the data consist of treatment status $T$ ($=1$ if RHC was used
within  24  hours  of  admission  and  0  otherwise),  health  outcome $Y$
(survival  time  up  to  30  days),  and  a list of 75 covariates $X$
specified by medical specialists in critical care.
Previously, the ATE for 30-day survival probabilities was estimated under confoundedness using propensity score and outcome regression models
with main effects only (Hirano \& Imbens 2002; Vermeulen \& Vansteelandt 2015),
with interaction terms manually added (Tan 2006), or with all main effects and interactions allowed in high-dimensional settings (Tan 2020b).
Sensitivity analysis based on model (\ref{eq:sen-model}) was conducted in Tan (2006) employing IPW-type estimation mentioned in Section \ref{sec:relax}
but providing only point bounds,
and in Dorn et al. (2022) employing augmented IPW estimation with conventional estimation of nuisance functions.

We apply the proposed methods and those based on Dorn et al. (2022), as described in Section \ref{sec:simulation},
both without regularization (CAL or ML) using only main-effect working models and with regularization (RCAL or RML) using high-dimensional working models
allowing main effects and interactions. In the latter case, similarly as in Tan (2020b),
we take $f(X)=h(X)$ which consists of all main effects and two-way interactions of $X$
except those with the fractions of nonzero values less than 46 (i.e., 0.8\% of the sample size 5735).
The dimension is $p=m=1855$, excluding the constant. All regressors are standardized with sample means 0 and variances 1.
Similarly as in the simulation study,
the Lasso tuning parameters $(\lambda_\gamma,\lambda_\beta,\lambda_\gamma)$ are
selected by 5-fold cross validation over a discrete set $\{\lambda^* / 2^{j/4}: j=0,1,\ldots,24\}$,
where $\lambda^*$ is a data-dependent upper limit.

\begin{figure} [t!]
\begin{tabular}{c}
\includegraphics[width=6in, height=2in]{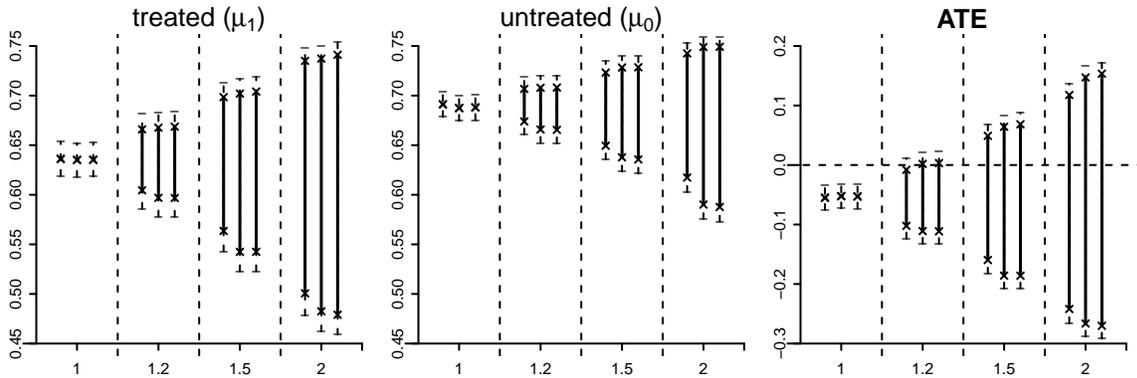} \vspace{-.1in}
\end{tabular}
\caption{\scriptsize Point bounds ($\times$) and 90\% confidence intervals on 30-day survival probabilities and ATE,
using working models with main effects and interactions, for four choices of $\Lambda$ separated by vertical lines.
For each $\Lambda$, three methods are used (from left to right): RML (Dorn et al. 2022), RCAL with relaxation using
linear outcome mean regression (Section \ref{sec:rcal}),
and RCAL with relaxation using logistic outcome mean regression (Section \ref{sec:binary}).
} \label{fig:rhc-intx} 
\end{figure}

Figure \ref{fig:rhc-intx} presents the point bounds and 90\% confidence intervals using working models with both main effects and interactions.
The numerical estimates and standard errors and all the results from using main-effect models are presented in Supplement Section \ref{sec:add-application}.
Overall, the point-bound enclosed intervals and confidence intervals from RCAL are somewhat wider than from RML.
This is consistent with the conservatives of RCAL when relaxation is used,
although, as shown in Supplement Figure \ref{fig:rhc-intx-vanilla}, the intervals from vanilla RCAL without relaxation remain to be slightly wider than those from RML.
In theory, the validity of RML confidence intervals depends on correct specification of both the propensity score and outcome models, whereas
the RCAL confidence intervals remain valid even when the outcome models are misspecified.

At $\Lambda=1.2$, the upper point bound of ATE from RML stays negative (similarly as in Tan 2006), whereas those from CAL are slightly above 0
(about $0.002$). Nevertheless, the confidence intervals from both RML and RCAL contain 0, by accounting for sampling variation.
This indicates that the negative ATE found from unconfoundedness estimation could barely hold
while allowing for unmeasure confounding at $\Lambda = 1.2$: the odds of being treated or untreated given
potential outcomes and covariates differ from those given only the covariates by at most 1.2.

Compared with those from using main-effect models, the upper point bounds for ATE slightly increase,
but the standard errors of the point bounds are consistently reduced,
from about $0.015$ to $0.011$, by regularized estimation using more flexible working models; see Supplement Tables \ref{tab:rhc-intx}--\ref{tab:rhc-main}.
The efficiency gain agrees with a similar finding in Tan (2020b).

The point-bound enclosed intervals and confidence intervals for the 30-day survival probability if treated are
noticeably wider than those for the survival probability if untreated.
This can be mainly explained by the fact that the untreated group size 3551 is about 1.6 times the treatd group size 2184,
rather than different accuracies in the prediction of the outcomes from the covariates within the two groups.
See the Supplement for further discussion.



\vspace{.3in}
\centerline{\bf\Large References}

\begin{description}\addtolength{\itemsep}{-.15in}


\item Bertsekas, D.P. (1973)  Stochastic optimization problems with nondifferentiable cost functionals, {\em Journal of Optimization Theory and Applications},
12, 218-231.


\item Buhlmann, P. and van de Geer, S. (2011) {\em Statistics for High-Dimensional Data: Methods, Theory and Applications}, New York: Springer.

\item Chernozhukov, V., Chetverikov, D., Demirer, M., Duflo, E., Hansen, C., Newey, W.K., and Robins, J.M. (2018)
Double/debiased machine learning for treatment and structural parameters, {\em Econometrics Journal}, 21, C1-C68.

\item Connors, A.F., Speroff, T., Dawson, N.V., et al. (1996) The effectiveness
of right heart catheterization in the initial care of critically ill patients,
{\em Journal of the American Medical Association}, 276, 889-897.

\item Dorn, J. and Guo, K. (2022) Sharp sensitivity analysis for inverse propensity weighting via quantile balancing, {\em Journal of the American Statistical Association}, to appear. 

\item Dorn, J., Guo, K., and Kallus, N. (2021) Doubly-valid/doubly-sharp sensitivity analysis for causal inference with unmeasured confounding, arXiv:2112.11449.

\item Fogarty, C.B. and Small, D.S. (2016) Sensitivity analysis for multiple comparisons in matched
observational studies through quadratically constrained linear programming. {\em Journal of the American Statistical Association}, 111, 1820-1830.

\item Francis, R.L. and Wright, G.P. (1969) Some duality relationships for the generalized Neyman-Pearson Problem, {\em Journal of Optimization Theory and Applications}, 4, 394-412.

\item Franks, A., D'Amour, A., and Feller, A. (2020) Flexible sensitivity analysis for observational
studies without observable implications, {\em Journal of the American Statistical Association}, 115, 1730-1746.

\item Ghosh, S. and Tan, Z. (2021) Doubly robust semiparametric inference using regularized calibrated estimation with high-dimensional data, {\em  Bernoulli}, 28, 1675-1703.

\item Hirano, K., and Imbens,  G.W. (2002) Estimation of causal effects using
propensity score weighting: An application to data on right heart catheterization,
{\em Health Services and Outcomes Research Methodology}, 2, 259-278.

\item Jin, Y., Ren, Z., and Candes, E.J. (2021) Sensitivity analysis of individual treatment effects: A robust
conformal inference approach, arXiv:2111.12161.

\item Kallus, N., and Zhou, A. (2021) Minimax-optimal policy learning under unobserved
confounding, {\em Management Science}, 67, 2870-2890.

\item Kallus, N., Mao, X., and Zhou, A. (2019) Interval estimation of individual-level causal effects
under unobserved confounding, {\em Proceedings of Machine Learning Research}, 2281-2290.

\item Kim, J.K. and Haziza, D. (2014) Doubly robust inference with missing data in survey sampling, {\em Statistics Sinica}, 24, 375-394.

\item Koenker, R. and Bassett, G.Jr. (1978) Regression quantiles, {\em Econometrica}, 46, 33-50.

\item Koenker, R., Portnoy, S., and others (2022) quantreg: Quantile regression, R package version 5.93.

\item Lee, K., Bargagli-Stoffi, F.J., and Dominici, F. (2020) Causal rule ensemble: Interpretable
inference of heterogeneous treatment effects, arXiv: 2009.09036.


\item Neyman, J. (1923) On the application of probability theory to agricultural
experiments: Essay on principles, Section 9, in {\em Statistical Science}, 1990, 5, 465-480.


\item Robins, J.M., Rotnitzky, A., and Zhao, L.P. (1994) Estimation of regression
coefficients when some regressors are not always observed, {\em Journal of the American Statistical Association}, 89, 846-866.

\item Robins, J.M., Rotnitzky, A., and Scharfstein, D.O. (2000) Sensitivity analysis for selection bias and unmeasured confounding in missing
data and causal inference models, in {\em Statistical Models in Epidemiology, the Environment, and Clinical Trials}, Berlin: Springer, 1-94.

\item Rosenbaum, P.R. (2002) {\em Observational Studies} (2nd edition). New York: Springer.

\item Rosenbaum, P.R. and Rubin, D.B. (1983) The central role of the propensity score in observational studies for causal
effects, {\em Biometrika}, 70, 41-55.

\item Rosenman, E.T.R. and Owen, A.B. (2021) Designing experiments informed by observational studies, {\em Journal of Causal Inference}, 9, 147-171.

\item Rubin, D.B. (1974) Estimating causal effects of treatments in randomized and nonrandomized Studies, {\em Journal of Educational Psychology}, 66, 688-701.

\item Rubin, D.B. (1976) Inference and missing data, {\em Biometrika}, 63, 581-590.

\item Rubin, D.B. (2001) Using propensity scores to help design observational studies: Application to the tobacco litigation, {\em Health Services \& Outcomes Research Methodology}, 2, 169-188.

\item Sahoo, R., Lei, L., and Wager, S. (2022) Learning from a biased sample, arXiv:2209.01754.

\item Scharfstein, D.,  Nabi, R., Kennedy, E.H., Huang, M.-Y., Bonvini, M., and Smid, M. (2021) Semiparametric sensitivity analysis:
Unmeasured confounding in observational studies, \\ arXiv:2104.08300.

\item Tan, Z. (2006) A distributional approach for causal inference using propensity scores,
{\em Journal of the American Statistical Association}, 101, 1619-1637.

\item Tan, Z. (2019) On doubly robust estimation for logistic partially linear models, {\em Statistics and Probability Letters}, 155, 108577.

\item Tan, Z. (2020a) Regularized calibrated estimation of propensity scores with model misspecification and high-dimensional data, {\em Biometrika}, 107, 137-158.

\item Tan, Z. (2020b) Model-assisted inference for treatment effects using regularized calibrated estimation with high-dimensional data, {\em Annals of Statistics}, 48, 811-837.

\item Tan, Z. and Sun, B. (2020) RCAL: Regularized calibrated estimation, R package version 2.0.

\item Tibshirani, R. (1996) Regression shrinkage and selection via the Lasso, {\em Journal of the Royal Statistical Society}, Ser. B, 58, 267-288.

\item Vermeulen, K. and Vansteelandt, S. (2015) Bias-reduced doubly
robust estimation, {\em Journal of the American Statistical Association}, 110, 1024-1036.


\item Yadlowsky, S., Namkoong, H., Basu, S., Duchi, J., and Tian, L. (2022) Bounds on the conditional and average treatment effect
with unobserved confounding factors, {\em Annals of Statistics}, to appear.

\item Yin, M., C. Shi, Y. Wang, and Blei, D.M. (2022) Conformal sensitivity analysis for individual treatment
effects, {\em Journal of the American Statistical Association}, to appear.

\item Zhao, Q., Small, D.S., and Bhattacharya, B.B. (2019) Sensitivity analysis
for inverse probability weighting estimators via the percentile bootstrap,
{\em Journal of the Royal Statistical Society}, Ser.~B, 81, 735-761.

\end{description}

\clearpage

\setcounter{page}{1}

\setcounter{section}{0}
\setcounter{equation}{0}

\setcounter{table}{0}
\setcounter{figure}{0}

\setcounter{pro}{0}
\renewcommand{\thepro}{S\arabic{pro}}

\setcounter{lem}{0}
\renewcommand{\thelem}{S\arabic{lem}}

\setcounter{thm}{0}
\renewcommand{\thethm}{S\arabic{thm}}

\setcounter{ass}{0}
\renewcommand{\theass}{S\arabic{ass}}

\renewcommand{\thesection}{\Roman{section}}
\renewcommand{\theequation}{S\arabic{equation}}

\renewcommand\thetable{S\arabic{table}}
\renewcommand\thefigure{S\arabic{figure}}

\begin{center}
{\Large Supplementary Material for}\\
{\Large ``Model-assisted sensitivity analysis for treatment effects under unmeasured confounding via regularized calibrated estimation"}

\vspace{.1in} Zhiqiang Tan
\end{center}

\section{Comparison of sensitivity models} \label{sec:comparison-model}

We compare marginal sensitivity models (\ref{eq:sen-model}) and (\ref{eq:sen-model2}) with
the latent sensitivity model of Rosenbaum (2002), which assumes that
for any $u$, $u^\prime$ and $x$,
\begin{align}
 \Gamma^{-1} \le \frac{ P(T=1|U^1= u^\prime, X=x) P(T=0 |U^1=u,X=x)}{P(T=0 | U^1=u^\prime,X=x) P(T=1|U^1=u,X=x)} \le  \Gamma, \label{eq:sen-model3}
\end{align}
where $\Gamma \ge 1$ is a sensitivity parameter, and $U^1$ is an unmeasured confounder such that  $T \perp Y^1 | (X,U^1)$ as in (\ref{eq:sen-model2}).
A slightly stronger version is used in Yadlowsky et al.~(2022), with $U^1$ in (\ref{eq:sen-model3}) replaced by $U$ such that $T \perp (Y^0,Y^1) | (X,U)$.
The results of Yadlowsky et al.~(2022) can be easily extended to model (\ref{eq:sen-model3}), and sharp bounds on the ATE can be obtained without
invoking any symmetry assumption.
By Yadlowsky et al.~(2022, Lemma 2.1), we show that model (\ref{eq:sen-model3}) can be reformulated in terms of
the density or odds ratio $\lambda^*_1 (Y^1,X)$.

\begin{lem} \label{lem:sen-model3}
(i) For any $ \Gamma \ge 1$, if model (\ref{eq:sen-model3}) holds with an unmeasured confounder $U^1$ such that $T \perp Y^1 | (X,U^1)$,
then there exists a covariate function $r(X)$ such that
\begin{align}
 r (X) \le \lambda^*_1 (Y^1, X) \le r(X) \Gamma, \quad \Gamma^{-1} \le r(X) \le 1 . \label{eq:sen-model3b}
\end{align}
(ii) For any $\Gamma \ge 1$, if (\ref{eq:sen-model3b}) holds with some covariate function $r(X)$,
then there exists an unmeasured confounder $U^1$ such that $T \perp Y^1 | (X,U^1)$ and (\ref{eq:sen-model3}) holds.
\end{lem}

From Lemma \ref{lem:sen-model3},
it can be shown that model (\ref{eq:sen-model}) or (\ref{eq:sen-model2})
implies model (\ref{eq:sen-model3}) or (\ref{eq:sen-model3b}) with $\Gamma=\Lambda^2$,
and model (\ref{eq:sen-model}) or (\ref{eq:sen-model2}) is implied by (hence weaker than) model (\ref{eq:sen-model3}) or (\ref{eq:sen-model3b}) with $\Gamma=\Lambda$.
See also Zhao et al.~(2019) for the result without involving (\ref{eq:sen-model3b}).
The relationship can be summarized as
\begin{align}
& \mathcal P_L (\Lambda) \subset \mathcal P_M (\Lambda) \subset  \mathcal P_L (\Lambda^2) \subset \mathcal P_M (\Lambda^2) , \label{eq:nesting}
\end{align}
where $\mathcal P_M (\Lambda)$ denotes the set of distributions allowed by model (\ref{eq:sen-model}) or (\ref{eq:sen-model2}) with parameter $\Lambda$,
and $\mathcal P_L (\Gamma)$ denotes the set of distributions allowed by model (\ref{eq:sen-model3}) or (\ref{eq:sen-model3b}) with parameter $\Gamma$.

By the reformulation (\ref{eq:sen-model3b}), the latent sensitivity model (\ref{eq:sen-model3}) imposes a lower bound $\Gamma^{-1}$ on
$\lambda^*_1 (y,x)$ and an upper bound $\Gamma$ on the ratio of the upper over the lower limits of $\lambda^*_1 (y,x)$ for each $x$.
In contrast, the marginal sensitivity model (\ref{eq:sen-model}) imposes both a lower bound $\Lambda^{-1}$ and an upper bound $\Lambda$ on $\lambda^*_1 (y,x)$
for each $x$. In the following, we provide further comparison of the marginal and latent sensitivity models. For simplicity,
only the upper bound of $\mu^1$ is discussed. The lower bound of $\mu^1$ can be handled by symmetry.

First, consider an extension of the marginal sensitivity model (\ref{eq:sen-model}),
\begin{align}
\Lambda_1(X) \le \lambda^*_1(Y^1,X)  \le \Lambda_2(X), \label{eq:sen-model-ext}
\end{align}
where $\Lambda_2(X) \ge 1 \ge \Lambda_1(X)$ are some {\it pre-specified} covariate functions.
Model (\ref{eq:sen-model}) corresponds to the special case where
$\Lambda_1(X) \equiv \Lambda^{-1}$ and $\Lambda_2(X) \equiv \Lambda$ for some constant $\Lambda \ge 1$.
Accordingly, Proposition \ref{pro:DG} can be extended such that
the sharp bound $\mu^{1+}$ under model (\ref{eq:sen-model-ext}) is
\begin{align}
& \mu^{1+}_{(\Lambda_1,\Lambda_2)}
= E \left\{\frac{T}{\pi^*(X)} Y + (\Lambda_2(X )-\Lambda_1(X) )  T \frac{1-\pi^*(X)}{\pi^*(X)} \rho_{\tau(X)} (Y, q^*_{1,\tau} (X)) \right\} \nonumber \\
& =  E \left[\frac{T}{\pi^*(X)} Y +  T \frac{1-\pi^*(X)}{\pi^*(X)} \left\{ (\Lambda_2(X) -1) (Y-q^*_{1,\tau} (X))_+ + (1-\Lambda_1(X) )(q^*_{1,\tau} (X)-Y)_+ \right\} \right], \label{eq:mu1+Q-ext}
\end{align}
where the dependency on $\Lambda_1=\Lambda_1(\cdot)$ and $\Lambda_2=\Lambda_2(\cdot)$ is made explicit in the notation,
 $\tau(X) = (\Lambda_2(X) -1)/(\Lambda_2(X) - \Lambda_1(X))$ and $q^*_{1,\tau}(X) = q^*_{1,\tau(X)} (X)$, the $\tau(X)$-quantile of $P(Y|T=1,X)$.
The bound $\mu^{1+}_{(\Lambda_1,\Lambda_2)} $  can be equivalently expressed as
\begin{align}
\mu^{1+}_{(\Lambda_1,\Lambda_2)}  =
 E \left\{T Y + (1-T) E \left( \tilde Y_{+}(q^*_{1,\tau})  | T=1, X \right) \right\}, \label{eq:mu1+Q-ext2}
\end{align}
where $\tilde Y_{+} (q_1) = Y+ (\Lambda_2(X) -1) (Y-q_{1} (X))_+ + (1-\Lambda_1(X) )(q_{1} (X)-Y)_+ $ for a covariate function $q_1(X)$.
The estimating function $\varphi_{+}(\cdot)$ can be extended as
\begin{align*}
\varphi_{+, (\Lambda_1,\Lambda_2)} (O; \pi,\eta_1,q_1)
= T Y+  T \frac{1-\pi(X)}{\pi(X)} \tilde Y_{+} (q_1) - \left\{\frac{T}{\pi(X)} -1\right\} E_{\eta_1} \left\{\tilde Y_{+} (q_1) | T=1,X\right\},
\end{align*}
where $\pi$ and $q_1$ represent a putative propensity score and a $\tau(X)$-quantile function,
and $\eta_1$ represents a conditional distribution of $Y$ given $T=1$ and $X$.
Similarly as in Proposition \ref{pro:DG-DR}, $\varphi_{+, (\Lambda_1,\Lambda_2)} (\cdot)$ is doubly robust for an upper bound of $\mu^1$:
if either $\pi=\pi^*$ or $\eta_1 = \eta^*_1$, then
\begin{align*}
& \mu^{1+}_{(\Lambda_1,\Lambda_2)} ( q_1 ) = E \left\{ \varphi_{+, (\Lambda_1,\Lambda_2)} (O; \pi,\eta_1, q_1 ) \right\}
 \ge \mu^{1+}_{(\Lambda_1,\Lambda_2)}  ,
\end{align*}
where $\mu^{1+}_{(\Lambda_1,\Lambda_2)} ( q_1 )$ denotes the right-hand side of (\ref{eq:mu1+Q-ext}) or (\ref{eq:mu1+Q-ext2}) with
$q^*_{1,\tau}$ replaced by $q_1$. The above inequality reduces to equality, hence $\varphi_{+, (\Lambda_1,\Lambda_2)} (\cdot)$ is doubly robust
is doubly robust for the sharp bound $\mu^{1+}_{(\Lambda_1,\Lambda_2)} $ if, further, $q_1 = q^*_{1,\tau}$.

The doubly robust estimating function $\varphi_{+, (\Lambda_1,\Lambda_2)} (\cdot)$ can be understood through augmented IPW estimation as in Section \ref{sec:DR-bound}.
Alternatively,  $\varphi_{+, (\Lambda_1,\Lambda_2)} (\cdot)$ can be derived
from the efficient influence function for the sharp bound $\mu^{1+}_{(\Lambda_1,\Lambda_2)} $ under model (\ref{eq:sen-model-ext}), which can be shown to be
$\varphi_{+, (\Lambda_1,\Lambda_2)} (O; \pi^*,\eta^*_1,q^*_{1,\tau})- \mu^{1+}_{(\Lambda_1,\Lambda_2)} $, which are evaluated at
the true values $\pi^*$, $\eta^*_1$ and $q^*_1$.
This is similar to the result that
the efficient influence function for the sharp bound $\mu^{1+}$ under model (\ref{eq:sen-model}) is
$\varphi_{+} (O; \pi^*,\eta^*_1,q^*_{1,\tau}) -\mu^{1+}$ as shown in Dorn et al.~(2021).
We comment that doubly robust estimating functions and efficient influence functions are two distinct concepts,
and one does not automatically lead to the other (even though they are often directly related as in the preceding setting).
See Tan (2019) for a nontrivial example of doubly robust estimating functions.

Next, we turn to the latent sensitivity model (\ref{eq:sen-model3}) or (\ref{eq:sen-model3b}).
The sharp upper bound for $\mu^1$ is shown in Yadlowsky et al.~(2022) to be
\begin{align}
\mu^{1+}_\Gamma = E \left\{ TY + (1-T) q^*_1 (X) \right\}, \label{eq:mu1+Q-latent}
\end{align}
where $q^*_1(X)$ is determined such that $ E \{ \Gamma (Y- q^*_1(X))_+ - (q^*_1(X) - Y)_+ | T=1, X \})= 0$ under some moment condition (which is assumed henceforth).
Moreover, the efficient influence function for $\mu^{1+}_\Gamma$ is shown to be
\begin{align*}
\psi_{+} (O; \pi^*, q^*_1, r^*) =
TY + (1-T) q^*_1 (X) +  T \frac{1-\pi^*(X)}{\pi^*(X)} r^*(X) \left\{ \Gamma (Y- q^*_1(X))_+ - (q^*_1(X) - Y)_+  \right\},
\end{align*}
where $r^* (X) = \{ P( Y \le q^*_1(X) | T=1,X) + \Gamma P( Y > q^*_1(X) | T=1,X)\}^{-1} $
satisfying $ \Gamma^{-1} \le r^*(X) \le 1$.
We show that the sharp bound $\mu^{1+}_\Gamma$ and the efficient influence function above
can be matched with those under model (\ref{eq:sen-model-ext}) for a particular choice of $(\Lambda_1(X), \Lambda_2(X))$.

\begin{pro} \label{pro:matching}
Let
$\Lambda^*_1 (X) = r^*(X)$, $\Lambda^*_2 (X) = r^*(X) \Gamma$, and $\tau^*(X) =
(\Lambda^*_2(X) -1)/(\Lambda^*_2(X) - \Lambda^*_1(X))$. Then the following results hold.
\begin{itemize}
\item[(i)] $\tau^*(X) = P( Y \le q^*_1(X) | T=1, X)$,

\item[(ii)] $ \mu^{1+}_\Gamma =  \mu^{1+}_{(\Lambda^*_1,\Lambda^*_2)}$,

\item[(iii)] $ \psi_{+} (O; \pi^*, q^*_1, r^* ) = \varphi_{+, (\Lambda^*_1,\Lambda^*_2)} (O; \pi^*,\eta^*_1,q^*_1 ) $.
\end{itemize}
In other words, $q^*_1(X)$ is a $\tau^*(X)$-quantile of $Y$ given $T=1$ and $X$, i.e., $q^*_1(X) = q^*_{1,\tau^*(X)} (X)$,
and $ \mu^{1+}_\Gamma$ and $ \psi_{+} (O; \pi^*, q^*_1)$ coincide with the sharp upper bound of $\mu^1$ and the efficient influence function
under model (\ref{eq:sen-model-ext}) with $(\Lambda_1(X), \Lambda_2(X))$ set to $(\Lambda^*_1(X), \Lambda^*_2(X))$.
\end{pro}

Proposition \ref{pro:matching} establishes an interesting relationship between models (\ref{eq:sen-model}) and (\ref{eq:sen-model3}).
But a careful interpretation is needed.
The particular choice $(\Lambda^*_1(X), \Lambda^*_2(X))$ is unknown, depending on the covariate functions $q^*_1(X)$ and $r^*(X)$
associated with the sharp bound $\mu^{1+}_\Gamma$ and the efficient influence function.
Hence model (\ref{eq:sen-model3}) cannot be treated as model (\ref{eq:sen-model}) with any pre-specified choice $(\Lambda_1(X), \Lambda_2(X))$.
Nevertheless, once $q^*_1(X)$ and $r^*(X)$ are determined (or consistently estimated),
the sharp bound of $\mu^1$ and efficient influence function under model (\ref{eq:sen-model3}) can be equivalently obtained from model (\ref{eq:sen-model})
with $(\Lambda_1(X), \Lambda_2(X))$ set to $(\Lambda^*_1(X), \Lambda^*_2(X))$.
In particular, the sharp bound $\mu^{1+}_\Gamma$ in (\ref{eq:mu1+Q-latent}) can also be expressed in the form of (\ref{eq:mu1+Q-ext2}) as
\begin{align}
\mu^{1+}_\Gamma  =
 E \left\{T Y + (1-T) E \left( \tilde Y_{+}(q^*_1)  | T=1, X \right) \right\},  \label{eq:mu1+Q-latent2}
\end{align}
where $\tilde Y_{+} (q^*_1) = Y+ (\Lambda^*_2(X) -1) (Y-q^*_{1} (X))_+ + (1-\Lambda^*_1(X) )(q^*_{1} (X)-Y)_+ $.
In contrast with (\ref{eq:mu1+Q-latent}),
the expression (\ref{eq:mu1+Q-latent2}) reveals how $\mu^{1+}_\Gamma $ depends on the optimal objective values from outcome quantile regression,
albeit at covariate-dependent quantile levels.

The estimating function for the sharp bound $\mu^{1+}_\Gamma$ used in Yadlowsky et al.~(2022) is defined as
$ \psi_{+} (O; \pi, q_1, r )$, that is, $\psi_{+} (O; \pi^*, q^*_1, r^*)$ with
$(\pi^*, q^*_1, r^*)$ replaced by some putative (or estimated) functions $(\pi, q_1, r )$.
However, $ \psi_{+} (O; \pi, q_1, r )$ seems to be only singly robust.
It can be easily shown that
$ E \{ \psi_{+} (O; \pi, q^*_1, r )\} = \mu^{1+}_\Gamma$,
with $(\pi, r)$ possibly different from from $(\pi^*, r^*)$.
But $\psi_{+} (O; \pi, q_1, r )$ may fail to be unbiased for $\mu^{1+}_\Gamma$
if $q_1 \not = q^*_1$ and $\pi=\pi^*$ or $r = r^*$.
We point out that the lack of double robustness for the estimating function $ \psi_{+} (O; \pi, q_1, r )$
is separate from and compatible with the fact that
$ \psi_{+} (O; \pi, q_1, r)$ is shown in Yadlowsky et al.~(2022) to satisfy the Neyman orthogonality (Chernozhukov et al.~2018)
at the true values $(\pi^*, q^*_1, r^*)$.
In fact, the Neyman orthogonality is automatically satisfied at the true values of nuisance parameters
by estimating functions directly converted from efficient influence functions.
Fundamentally, double robustness and the Neyman orthogonality are two distinct concepts.
See Tan (2019) for related discussion in the context of logistic partially linear models.

\section{Additional discussion on calibrated estimation} \label{sec:add-cal}

To supplement Section \ref{sec:cal},
we describe some general, high-level asymptotic results and compare conventional and calibrated estimation in more details,
in the classical setting where the parameter dimensions $p$ and $m$ are fixed
as the sample size $n$ tends to $\infty$.
For completeness, parts of the discussion in Section \ref{sec:cal} are reproduced here.

Suppose that $(\hat\gamma,\hat\alpha_{1+}, \hat\beta_{1+})$ are
M-estimators defined as minimizers of sample objective functions satisfying the consistency condition:
\begin{itemize}
\item[(B)] The population objective functions for $(\hat\gamma,\hat\alpha_{1+}, \hat\beta_{1+})$ are minimized by the true parameter values under correctly specified models respectively.
\end{itemize}
Then the following asymptotic properties can be shown under suitable regularity conditions (White 1982; Manski 1988).
The estimator $\hat\gamma$ converges at rate $O_p(n^{-1/2})$ to a target value $\bar\gamma$,
which minimizes the population objective function for $\hat\gamma$.
The target value $\bar\gamma$ coincides with the true value $\gamma^*$ (i.e., $\hat\gamma$ is consistent) if model (\ref{eq:logit+pi}) is correctly specified, but
remains well-defined  if model (\ref{eq:logit+pi}) is misspecified.
Similarly, $\hat\beta_{1+}$ converges at rate $O_p(n^{-1/2})$ to a target value $\bar\beta_{1+}$,
which coincides with the true value $\beta^*_{1+}$
if model (\ref{eq:linear+Q}) is correctly specified, but
remains well-defined  if model (\ref{eq:linear+Q}) is misspecified.
Moreover, $\hat\alpha_{1+}$  converges at rate $O_p(n^{-1/2})$ to a target value
$\bar\alpha_{1+} = \bar\alpha_{1+} (\bar\beta_{1+})$,
which coincides with the true value $\alpha^*_{1+}  = \alpha^*_{1+} (\bar\beta_{1+})$
if model (\ref{eq:linear+eta}) with $\beta=\bar \beta_{1+}$ is correctly specified, but
remains well-defined otherwise.

We distinguish two types of asymptotic properties for the resulting estimator $\hat\mu^{1+} (\hat\gamma, \hat\alpha, \hat\beta_{1+})$.
First, $\hat\mu^{1+} (\hat\gamma, \hat\alpha, \hat\beta_{1+})$ can be shown to converge in probability to
$ E \{ \varphi_{+} ( O; \bar\pi, \bar\eta_{1+}, \bar q_{1,\tau}) \}$,
where $\bar \pi(X) = \pi(X;\bar\gamma)$, $\bar q_{1,\tau}(X) = h^\T (X) \bar\beta_{1+}$, and $\bar \eta_{1+}(X) = \eta_{1+}(X; \bar\alpha_{1+})$.
Then by the double robustness of estimating function $\varphi_{+}$ in Proposition \ref{pro:DG-DR},
the estimator $\hat\mu^{1+} (\hat\gamma, \hat\alpha, \hat\beta_{1+})$ is (pointwise) doubly robust for the population upper bound
$ \mu^{1+} ( \bar q_{1,\tau} )$,
i.e., remains consistent for $ \mu^{1+} ( \bar q_{1,\tau} )$ if either model (\ref{eq:logit+pi}) or (\ref{eq:linear+eta}) with $\beta=\bar \beta_{1+}$
is correctly specified.
The bound $ \mu^{1+} ( \bar q_{1,\tau} )$ reduces to the sharp bound $\mu^{1+}$ if, further, model (\ref{eq:linear+Q}) is correctly specified.
Hence the three models (\ref{eq:logit+pi})--(\ref{eq:linear+eta}) play different roles.
Correct specification of either model (\ref{eq:logit+pi}) or model (\ref{eq:linear+eta})
ensures that $\hat\mu^{1+} (\hat\gamma, \hat\alpha, \hat\beta_{1+})$ converges to a valid upper bound $ \mu^{1+} ( \bar q_{1,\tau} )$,
whereas correct specification of  model (\ref{eq:linear+Q}) implies that the upper bound $ \mu^{1+} ( \bar q_{1,\tau} )$ is sharp.

Second, with possible misspecification of models (\ref{eq:logit+pi})--(\ref{eq:linear+eta}),
the estimator $\hat\mu^{1+} (\hat\gamma, \hat\alpha_{1+}, \hat\beta_{1+})$ can be shown to admit the following asymptotic expansion:
\begin{align}
& \hat\mu^{1+} (\hat\gamma, \hat\alpha_{1+}, \hat\beta_{1+})
= \tilde E (\varphi_{+} )
+ E^\T \left( \frac{\partial \varphi_{+}}{\partial \gamma}\right) (\hat\gamma-\bar\gamma) \nonumber \\
& \quad + E^\T \left( \frac{\partial \varphi_{+}}{\partial \alpha}\right) (\hat\alpha_{1+} - \bar\alpha_{1+} )
+ E^\T \left( \frac{\partial \varphi_{+}}{\partial \beta}\right) (\hat\beta_{1+}-\bar\beta_{1+}) + o_p(n^{-1/2}) , \label{eq:general-expan}
\end{align}
where $\varphi_{+} = \varphi_{+} (O; \pi(\cdot;\gamma),\eta_{1+} (\cdot;\alpha), q_{1,\tau}(\cdot;\beta) )$ as in (\ref{eq:mu1+class}),
and $\varphi_{+}$ and its partial derivatives are evaluated at $(\bar\gamma, \bar \alpha_{1+}, \bar\beta_{1+})$.
The preceding expansion (\ref{eq:general-expan}) shows how the asymptotic behavior of $\hat\mu^{1+} (\hat\gamma, \hat\alpha_{1+}, \hat\beta_{1+})$ is
affected by the estimators $(\hat\gamma, \hat\alpha_{1+}, \hat\beta_{1+})$ through the second to fourth terms on the right-hand side.
In fact, removing these three terms in (\ref{eq:general-expan}) yields the asymptotic expansion of the infeasible estimator
$\hat\mu^{1+} (\bar\gamma, \bar\alpha_{1+}, \bar\beta_{1+})$ using the target values.

From the expression in (\ref{eq:AIPW+DG}), the partial derivatives in (\ref{eq:general-expan}) can be calculated as
\begin{align}
& \frac{\partial \varphi_{+}}{\partial \gamma} = - T \frac{1-\pi(X;\gamma)}{\pi(X;\gamma)}
 \left\{ \tilde Y_{+} (h^\T \beta ) - f^\T(X) \alpha \right\} f(X), \label{eq:deriv-gam}\\
& \frac{\partial \varphi_{+}}{\partial \alpha} = - \left\{ \frac{T}{\pi(X;\gamma)} -1 \right\} f(X) ,\label{eq:deriv-alpha} \\
& \frac{\partial \varphi_{+}}{\partial \beta} = (\Lambda-\Lambda^{-1})  T\frac{1-\pi(X;\gamma)}{\pi(X;\gamma)}
\frac{\partial}{\partial\beta} \rho_\tau (Y, h^\T(X)\beta),\label{eq:deriv-beta}
\end{align}
where $\varphi_{+} = \varphi_{+} (O; \pi(\cdot;\gamma),\eta_{1+} (\cdot;\alpha), q_{1,\tau}(\cdot;\beta) )$ as in (\ref{eq:mu1+class}), and
$\partial \rho_\tau (Y, h^\T(X)\beta) /\partial \beta$ denotes a subgradient of $\rho_\tau (Y, h^\T(X)\beta)$, which is convex in $\beta$.
Remarkably, if model (\ref{eq:linear+eta}), (\ref{eq:logit+pi}), or (\ref{eq:linear+Q}) is correctly specified, then,
{\it respectively}, the following identities are satisfied:
\begin{align}
& E \left( \frac{\partial \varphi_{+}}{\partial \gamma} \right) \Big|_{\alpha=\alpha^*_{1+}(\beta)} =0 , \quad
E \left(  \frac{\partial \varphi_{+}}{\partial \alpha} \right) \Big|_{\gamma=\gamma^*} =0 , \quad
E \left( \frac{\partial \varphi_{+}}{\partial \beta} \right) \Big|_{\beta=\beta^*_{1+}} =0 , \label{eq:cal-iden}
\end{align}
which will be called calibration identities.
The first two identities in (\ref{eq:cal-iden}) are due to double robustness of $\varphi_{+}$ with respect to $(\gamma, \alpha)$ for any fixed $\beta$,
similarly as discussed in Tan (2020b) and Ghosh \& Tan (2022).
The third identity in (\ref{eq:cal-iden}) is new and can be attributed to the fact that
if model (\ref{eq:linear+Q}) is correctly specified, then $E(\varphi_{+})$ at $\beta= \beta^*_{1+}$
achieves a minimum over all possible $\beta$ for any fixed $(\gamma,\alpha)$, by the standard characterization of quantile regression.


A direct implication of the calibration identities in (\ref{eq:cal-iden})
is that if all three models (\ref{eq:logit+pi})--(\ref{eq:linear+eta}) are correctly specified, then
the asymptotic expansion (\ref{eq:general-expan}) reduces to
\begin{align}
& \hat\mu^{1+} (\hat\gamma, \hat\alpha_{1+}, \hat\beta_{1+})
= \tilde E (\varphi_{+}) \big|_{(\gamma,\alpha,\beta)=(\bar\gamma, \bar \alpha_{1+}, \bar\beta_{1+})} + o_p(n^{-1/2}) , \label{eq:desired-expan}
\end{align}
because the second to fourth terms on the right-hand side of (\ref{eq:general-expan}) vanishes.
Given the simple expansion (\ref{eq:desired-expan}), valid variance estimation
and Wald confidence intervals can be derived
by ignoring the variation in $(\hat\gamma,\hat\alpha_{1+}, \hat\beta_{1+})$.
This simplification is computationally appealing, but may in general fail unless all three models (\ref{eq:logit+pi})--(\ref{eq:linear+eta}) are correctly specified.

All the preceding discussion can be applied to the conventional estimators $(\hat\gamma_{\mytext{ML}},\hat\alpha_{\mytext{U},1+},$ $ \hat\beta_{\mytext{U},1+})$.
Then $\hat\mu^{1+}(\hat\gamma_{\mytext{ML}},\hat\alpha_{\mytext{U},1+}, \hat\beta_{\mytext{U},1+})$ is pointwise doubly robust,
but valid variance estimation in general needs to be derived using expansion (\ref{eq:general-expan}) and
influence functions for $(\hat\gamma_{\mytext{ML}},\hat\alpha_{\mytext{U},1+}, \hat\beta_{\mytext{U},1+})$, allowing for model misspecification.
The simple expansion (\ref{eq:desired-expan}) would be invalid unless  all three models (\ref{eq:logit+pi})--(\ref{eq:linear+eta}) are correctly specified.
Alternatively, we propose calibrated (CAL) estimation for $(\gamma,\alpha,\beta)$
such that the simple expansion (\ref{eq:desired-expan}) can be achieved with possible misspecification of models (\ref{eq:logit+pi})--(\ref{eq:linear+eta}).
The calibrated estimators, $(\hat\gamma_{\mytext{CAL}},\hat\alpha_{\mytext{WL},1+}, \hat\beta_{\mytext{WL},1+})$, are
defined jointly as solutions to the estimating equations:
\begin{align}
& \tilde E \left( \frac{\partial \varphi_{+}}{\partial \gamma} \right) =0 , \quad
\tilde E \left(  \frac{\partial \varphi_{+}}{\partial \alpha} \right)  =0 , \quad
\tilde E \left( \frac{\partial \varphi_{+}}{\partial \beta} \right) =0 , \label{Seq:cal-eq}
\end{align}
which are converted from the calibration identities (\ref{eq:cal-iden}) and will be called (sample) calibration equations.
The pair (\ref{eq:cal-iden}) and (\ref{Seq:cal-eq}) are reminiscent of the score identity and score equation in likelihood theory.
The system of equations (\ref{Seq:cal-eq}) can be solved sequentially:
$\hat\gamma_{\mytext{CAL}}$ is solved from the second equation,
$\hat\beta_{\mytext{WL},1+}$ from the third equation, and
then $\hat\alpha_{\mytext{WL},1+}$ from the first equation.
These estimators can also be obtained by sequential minimization of three convex objective functions
$\ell_{\mytext{CAL}}(\gamma)$, $\ell_{\mytext{WL},1+}(\beta; \gamma)$, and $\ell_{\mytext{WL},1+}(\alpha; \gamma,\beta)$,
as defined in Section \ref{sec:rcal}.
The estimators $\hat\alpha_{\mytext{WL},1+}$ and $\hat\beta_{\mytext{WL},1+}$ are also called weighted loss (WL) estimators,
because the objective functions correspond to weighted quantile regression and weighted least squares.

The calibrated estimators $(\hat\gamma_{\mytext{CAL}},\hat\alpha_{\mytext{WL},1+}, \hat\beta_{\mytext{WL},1+})$ can be easily seen to satisfy consistency condition B,
by the calibration identities (\ref{eq:cal-iden}) and the convexity of the objective functions.
Let $(\bar\gamma_{\mytext{CAL}},\bar\alpha_{\mytext{WL},1+}, \bar\beta_{\mytext{WL},1+})$ be solutions to the population versions
of the calibration equations (\ref{Seq:cal-eq})  or equivalently minimizers of the population versions of
$\ell_{\mytext{CAL}}(\gamma)$, $\ell_{\mytext{WL},1+}(\beta; \bar\gamma_{\mytext{CAL}})$, and $\ell_{\mytext{WL},1+}(\alpha; \bar\gamma_{\mytext{CAL}}, \bar\beta_{\mytext{WL},1+})$.
Note that $\bar\beta_{\mytext{WL},1+}$ may differ from $\bar\beta_{\mytext{W},1+}$ based on $\pi^*$ in Proposition \ref{pro:mu1-h}.
Then by the second identity in (\ref{eq:cal-iden}),
$\gamma^*$ satisfies the stationary condition for minimizing the population version of $\ell_{\mytext{CAL}}(\gamma)$
if model (\ref{eq:logit+pi}) is correctly specified.
By the third identity in (\ref{eq:cal-iden}), $\beta^*_{1+}$ minimizes the population version of
$\ell_{\mytext{WL},1+}(\beta; \bar\gamma_{\mytext{CAL}})$ if model (\ref{eq:linear+Q}) is correctly specified.
By the first identity in (\ref{eq:cal-iden}), $\alpha^*_{1+}(\bar\beta_{\mytext{WL},1+})$ minimizes the population version of
$\ell_{\mytext{WL},1+}(\alpha; \bar\gamma_{\mytext{CAL}}, \bar\beta_{\mytext{WL},1+})$ if model (\ref{eq:linear+eta})
with $\beta=\bar\beta_{\mytext{WL},1+}$ is correctly specified.

From the previous discussion based on condition B, it can be shown under suitable regularity conditions that
$\hat\mu^{1+} (\hat\gamma_{\mytext{CAL}},\hat\alpha_{\mytext{WL},1+}, \hat\beta_{\mytext{WL},1+})$ is not only doubly robust, but also admits
the asymptotic expansion (\ref{eq:general-expan}), with
$(\bar\gamma, \bar \alpha_{1+}, \bar\beta_{1+})$ replaced by the limit values $(\bar\gamma_{\mytext{CAL}},\bar\alpha_{\mytext{WL},1+},$ $ \bar\beta_{\mytext{WL},1+})$, which by definition satisfy the population versions of the calibration equations (\ref{Seq:cal-eq}).
Hence the second to fourth terms on the right-hand side of (\ref{eq:general-expan}) vanishes,
and $\hat\mu^{1+} (\hat\gamma_{\mytext{CAL}},\hat\alpha_{\mytext{WL},1+}, \hat\beta_{\mytext{WL},1+})$ satisfies
the simple asymptotic expansion (\ref{eq:desired-expan}).

\begin{pro} \label{pro:cal-expan}
Suppose that $(\hat\gamma_{\mytext{CAL}},\hat\alpha_{\mytext{WL},1+}, \hat\beta_{\mytext{WL},1+})$
converge to $(\bar\gamma_{\mytext{CAL}},\bar\alpha_{\mytext{WL},1+}, \bar\beta_{\mytext{WL},1+})$ at rate $O_p(n^{-1/2})$, and
$\hat\mu^{1+} (\hat\gamma_{\mytext{CAL}},\hat\alpha_{\mytext{WL},1+}, \hat\beta_{\mytext{WL},1+})$ satisfies asymptotic
expansion (\ref{eq:general-expan}) as $p$ and $m$ are fixed and $n\to\infty$.
Then $\hat\mu^{1+} (\hat\gamma_{\mytext{CAL}},\hat\alpha_{\mytext{WL},1+}, \hat\beta_{\mytext{WL},1+})$ is
doubly robust for the upper bound
$ \mu^{1+} $ $( h^\T \bar\beta_{\mytext{WL},1+})$,
i.e., remains consistent if either model (\ref{eq:logit+pi}) or (\ref{eq:linear+eta}) with $\beta=\bar \beta_{\mytext{WL},1+}$
is correctly specified.
Moreover, $\hat\mu^{1+} (\hat\gamma_{\mytext{CAL}},\hat\alpha_{\mytext{WL},1+}, \hat\beta_{\mytext{WL},1+})$ satisfies
the asymptotic expansion (\ref{eq:desired-expan}), with $(\bar\gamma, \bar \alpha_{1+}, \bar\beta_{1+})$ replaced by
$(\bar\gamma_{\mytext{CAL}},\bar\alpha_{\mytext{WL},1+}, \bar\beta_{\mytext{WL},1+})$.
\end{pro}

Compared with $\hat\mu^{1+}(\hat\gamma_{\mytext{ML}},\hat\alpha_{\mytext{U},1+}, \hat\beta_{\mytext{U},1+})$,
the estimator $\hat\mu^{1+} (\hat\gamma_{\mytext{CAL}},\hat\alpha_{\mytext{WL},1+}, \hat\beta_{\mytext{WL},1+})$ using calibrated estimation
achieves the simple expansion (\ref{eq:desired-expan}), while allowing for misspecification of models (\ref{eq:logit+pi})--(\ref{eq:linear+eta}).
Hence valid Wald confidence intervals can be derived in the usual manner, by ignoring the sampling variation in
$(\hat\gamma_{\mytext{CAL}},\hat\alpha_{\mytext{WL},1+}, \hat\beta_{\mytext{WL},1+})$.
The resulting confidence intervals are doubly robust for the upper bound
$ \mu^{1+} ( h^\T \bar\beta_{\mytext{WL},1+})$
if either model (\ref{eq:logit+pi}) or (\ref{eq:linear+eta}) with $\beta=\bar \beta_{\mytext{WL},1+}$ is correctly specified.
In the current fixed-dimensional setting, this benefit of calibrated estimation can be mainly computational, because
doubly robust confidence intervals can also be obtained
using expansion (\ref{eq:general-expan}) for
$\hat\mu^{1+}(\hat\gamma_{\mytext{ML}},\hat\alpha_{\mytext{U},1+}, \hat\beta_{\mytext{U},1+})$ and
influence functions for $(\hat\gamma_{\mytext{ML}},\hat\alpha_{\mytext{U},1+}, \hat\beta_{\mytext{U},1+})$.
In Section \ref{sec:rcal}, we study regularized calibrated estimation to obtain doubly robust confidence intervals
in high-dimensional settings with  $p$ and $m$ close to or greater than $n$ as $n\to\infty$,
where the influence function based approach is not applicable.

\section{Implications of KKT conditions} \label{sec:KKT}

There are simple and interesting implications from the construction of our RCAL estimators, in addition to
desirable statistical properties under sparsity conditions discussed in Section \ref{sec:rcal}.
First, as in Tan (2020a), the KKT (or stationary) condition from the definition of $\hat\gamma_{\mytext{RCAL}}$ shows that
the fitted propensity score $\hat\pi_{\mytext{RCAL}} (X) = \pi(X; \hat\gamma_{\mytext{RCAL}} )$ satisfies
\begin{align}
& \frac{1}{n} \sum_{i=1}^n \frac{T_i}{\hat \pi_{\mytext{RCAL}} (X_i) } = 1, \label{eq:KKT-gam-1} \\
& \frac{1}{n} \left| \sum_{i=1}^n \frac{T_i f_j(X_i)}{\hat \pi_{\mytext{RCAL}} (X_i) } - \sum_{i=1}^n f_j(X_i) \right| \le \lambda_\gamma, \quad j=1,\ldots,p , \label{eq:KKT-gam-2}
\end{align}
where equality holds in (\ref{eq:KKT-gam-2}) for any $j$ such that the $j$th estimate $(\hat\gamma_{\mytext{RCAL}})_j$ is nonzero.
While (\ref{eq:KKT-gam-1}) says that the inverse weights, $\hat \pi^{-1}_{\mytext{RCAL}} (X_i) $, add to $n$ in the treated group,
(\ref{eq:KKT-gam-2}) represents box constraints on the differences between the weighted averages in the treated group
and the overall sample average of $f_j$'s.
Second, by the KKT (or stationary) condition for the intercept in $\hat\alpha_{\mytext{RWL},1+}$, the fitted mean regression function
$ \hat\eta_{\mytext{RWL},1+} (X) = \eta_{1+}(X; \hat\alpha_{\mytext{RWL},1+})$ satisfies
\begin{align*}
\frac{1}{n} \sum_{i=1}^n T_i \frac{1-\hat \pi_{\mytext{RCAL}}(X_i) }{\hat \pi_{\mytext{RCAL}}(X_i) }
\left\{\tilde Y_{+,i} (h^\T \hat\beta_{\mytext{RWL},1+})  - \hat \eta_{\mytext{RWL},1+}(X_i) \right\}= 0,
\end{align*}
where $\tilde Y_{+,i} (h^\T \hat\beta_{\mytext{RWL},1+}) = Y_i + (\Lambda-\Lambda^{-1}) \rho_\tau (Y_i, h^\T(X_i) \hat\beta_{\mytext{RWL},1+})$.
Then, by direct calculation, our estimator $\hat\mu^{1+}_{\mytext{RCAL}} $ in (\ref{eq:mu1+rcal}) can be simplified to
\begin{align}
\hat\mu^{1+}_{\mytext{RCAL}} = \frac{1}{n} \sum_{i=1}^n \left\{ T_i Y_i + (1-T_i) \hat\eta_{\mytext{RWL},1+} (X_i) \right\}.  \label{eq:mu1+imput}
\end{align}
This shows that the potential outcome $Y^1_i$ is imputed as $\hat\eta_{\mytext{RWL},1+} (X_i)$ for each subject $i$ in the untreated group
in our method, according to the expression $\mu^1 = E(TY+(1-T)Y^1)$.
Third, the Lasso penalty for $\hat\beta_{\mytext{RWL},1+}$ can also be introduced in a dual formulation in terms of $\lambda_1(Y^1,X)$,
where equality constraints (\ref{eq:lambda1-marginX})
are relaxed to box constraints in the sample version of the optimization in (\ref{eq:mu1+h}).
See Section \ref{sec:relax} for further discussion.

\section{High-dimensional analysis} \label{sec:theory}

We provide theoretical analysis to justify the proposed method in Section \ref{sec:rcal} in the high-dimensional, sparse setting
where the numbers of regressors in working models (\ref{eq:logit+pi})--(\ref{eq:linear+eta})
are large, but only small, unknown subsets of the regressors are associated with nonzero {\it target} coefficients
(defined as minimizers of the population objective functions) with possible model misspecification.
Our analysis shows that under suitable sparsity conditions, the desired asymptotic expansion (\ref{eq:rcal-expan}) is satisfied and hence
valid doubly robust confidence intervals can be obtained for upper (and similarly lower) bounds of $\mu^1$ from sensitivity model (\ref{eq:sen-model}).

There are two main steps in our theoretical analysis. We first study the convergences of the first-stage estimators
$(\hat\gamma_{\mytext{RCAL}},\hat\alpha_{\mytext{RWL},1+}, \hat\beta_{\mytext{RWL},1+})$, which are then used to study
convergence of the second-stage estimator $\hat\mu^{1+}_{\mytext{RCAL}}$.
While convergence of $\hat\gamma_{\mytext{RCAL}}$ can be directly taken from Tan (2020a), high-dimensional analysis of
$(\hat\alpha_{\mytext{RWL},1+}, \hat\beta_{\mytext{RWL},1+})$ and $\hat\mu^{1+}_{\mytext{RCAL}}$ involves nontrivial complications,
which are absent in related analysis of regularized calibrated estimation (Tan 2020b) as well as
previous analyses concerned with single-stage estimation (e.g., Bickel et al.~2009; Buhlmann \& van de Geer 2011).
In fact, the loss functions for $\hat\beta_{\mytext{RWL},1+}$ and $\hat\alpha_{\mytext{RWL},1+}$
correspond to weighted quantile regression and weighted least squares,
with the weight depending on $\hat\gamma_{\mytext{RCAL}}$ from fitting the propensity score model.
Moreover, the loss function $\ell_{\mytext{WL},1+} (\beta; \hat\gamma_{\mytext{RCAL}} )$
from (\ref{eq:beta-loss}) is not twice differentiable, so that analysis in Tan (2020b) is not applicable.
The loss function $\ell_{\mytext{WL},1+} (\alpha; \hat\gamma_{\mytext{RCAL}}, \hat\beta_{\mytext{RWL},1+} )$
from (\ref{eq:alpha-loss}) involves a data-dependent response variable
$\tilde Y_{+} ( h^\T \hat\beta_{\mytext{RWL},1+})$, depending on $\hat\beta_{\mytext{RWL},1+}$ from fitting weighted quantile regression,
which introduces additional variation compared with the setting in Tan (2020b).
Convergence of $\hat\mu^{1+}_{\mytext{RCAL}}$ is affected by random variation from all the first-stage estimators
$(\hat\gamma_{\mytext{RCAL}},\hat\alpha_{\mytext{RWL},1+}, \hat\beta_{\mytext{RWL},1+})$.
Nevertheless, we carefully tackle the foregoing issues by building on and extending previous techniques.
See further discussion after Theorems \ref{thm:beta-conv}, \ref{thm:alpha-conv}, and \ref{thm:mu1+expan}.

First, we describe relevant results from Tan (2020a) about convergence of $\hat\gamma_{\mytext{RCAL}}$ to the target value
$\bar\gamma_{\mytext{CAL}}$, which is defined as a minimizer of the expected loss function $E \{ \ell_{\mytext{CAL}} (\gamma)\}$ with
$\ell_{\mytext{CAL}} (\gamma)$ in (\ref{eq:gam-loss}).
Suppose that the Lasso tuning parameter is specified for $\hat\gamma_{\mytext{RCAL}}$ as $\lambda_\gamma = A_0 \lambda_0$,
where $A_0>0$ is a scaling constant, $\epsilon \in(0,1)$ is a tail probability for error bounds,
and $\lambda_0 = \sqrt{\log(p^\dag /\epsilon)/n}$ is a base rate, depending on $p^\dag = 1+\max(p,m)$
such that, for convenience, the same quantity $\lambda_0$ can be used in later results about $(\hat\alpha_{\mytext{RWL},1+}, \hat\beta_{\mytext{RWL},1+})$.

For a matrix $\Sigma$ with row indices $\{0,1,\ldots,k\}$,
a compatibility condition (Buhlmann \& van de Geer 2011) is said to hold with a subset $S \in \{0,1,\ldots,k\}$ and constants $\nu >0$ and $\xi>1$ if
$\nu^2  (\sum_{j\in S} |b_j|)^2 \le |S| ( b^\T \Sigma b )$
for any vector $b=(b,b_1,\ldots,b_k)^\T \in \mathbb R^{1+k} $ satisfying $\sum_{j\not\in S} |b_j| \le \xi \sum_{j\in S} |b_j| $,
where $|S|$ denotes the size of $S$.

\begin{ass} \label{ass:gam}
Suppose that the following conditions are satisfied: \vspace{-.05in}
\begin{itemize} \addtolength{\itemsep}{-.05in}
\item[(i)] $\max_{j=0,1,\ldots,p} |f_j(X)| \le C_0$ almost surely for a constant $C_0 \ge 1$;

\item[(ii)] $f^\T(X) \bar\gamma_{\mytext{CAL}}\ge B_0$ almost surely for a constant $B_0 \in \mathbb R$, that is,
$\pi(X; \bar\gamma_{\mytext{CAL}})$ is bounded from below by $( 1+\me^{-B_0})^{-1}$;

\item[(iii)] the compatibility condition holds for $\Sigma_f $ with
the subset $S_{\bar\gamma} = \{0\} \cup \{j: (\bar\gamma_{\mytext{CAL}})_j \not=0, j=1,\ldots,p\}$ and some constants $\nu_0>0$ and $\xi_0>1$,
where $\Sigma_f= E \{ T w(X; \bar \gamma_{\mytext{CAL}}) f(X) f^\T (X) \}$ and
$w(X;\gamma)= \{1-\pi(X;\gamma)\} /\pi(X;\gamma)=\me^{-f^\T(X)\gamma}$.

\item[(iv)] $|S_{\bar\gamma}| \lambda_0 \le \varrho_0$ for a sufficiently small constant $\varrho_0 >0$, depending only on $(A_0,B_0,C_0,\xi_0,\nu_0)$.
\end{itemize} \vspace{-.05in}
\end{ass}

\begin{thm}[Tan 2020a] \label{thm:gam-conv}
Suppose that Assumption~\ref{ass:gam} holds. Then for $A_0 > C_{01} (\xi_0+1)/(\xi_0-1)$, we have
with probability at least $1-4\epsilon$,
\begin{align*}
& \tilde E \left\{ T w(X; \bar\gamma_{\mytext{CAL}}) |f^\T(X) (\hat\gamma_{\mytext{RCAL}} - \bar\gamma_{\mytext{CAL}})|^2 \right\}
\le M_0 |S_{\bar\gamma}| \lambda_0^2,\\
&  \| \hat\gamma_{\mytext{RCAL}} - \bar\gamma_{\mytext{CAL}} \|_1  \le M_0 |S_{\bar\gamma}| \lambda_0,
\end{align*}
where $C_{01} = \sqrt{2} (\me^{-B_0}+1) C_0$ and $M_0>0$ is a constant depending only on $(A_0, B_0, C_0, \nu_0, \xi_0, \varrho_0)$.
\end{thm}

Next, we study the convergence of $\hat\beta_{\mytext{RWL},1+}$ to the target value $\bar\beta_{\mytext{WL},1+}$,
which is defined as a minimizer of the expected loss function $E \{\ell_{\mytext{WL},1+} (\beta; \bar\gamma_{\mytext{CAL}} )\}$ with
$\ell_{\mytext{WL},1+} (\beta; \gamma )$ in (\ref{eq:beta-loss}).
Suppose that the Lasso tuning parameter is specified for $\hat\beta_{\mytext{RWL},1+}$ as $\lambda_\beta = A_1 \lambda_0$,
where $A_1>0 $ is a scaling constant and $\lambda_0$ is defined as before.

\begin{ass} \label{ass:beta}
Suppose that the following conditions are satisfied:\vspace{-.05in}
\begin{itemize}\addtolength{\itemsep}{-.05in}
\item[(i)] $\max_{j=0,1,\ldots,m} |h_j(X)| \le C_1$ almost surely for a constant $C_1 \ge 1$;

\item[(ii)] the margin condition holds for some constants $\kappa_1>0$ and $\zeta_1>0$:
$E \{ \ell (\beta; \bar\gamma_{\mytext{CAL}}) - \ell (\bar\beta_{\mytext{WL},1+}; \bar\gamma_{\mytext{CAL}}) \}
\ge \kappa_1 (\beta-\bar\beta_{\mytext{WL},1+})^\T \Sigma_{\bar\beta} (\beta-\bar\beta_{\mytext{WL},1+})$
for any $\beta$ satisfying $\| \beta-\bar\beta_{\mytext{WL},1+}\|_1 \le \zeta_1$,
where $\Sigma_h= E \{ T w(X; \bar \gamma_{\mytext{CAL}}) h(X) h^\T (X) \}$;

\item[(iii)] the compatibility condition holds for $\Sigma_h$ with
the subset $S_{\bar\beta} = \{0\} \cup \{j: (\bar\beta_{\mytext{WL},1+})_j \not=0, j=1,\ldots,p\}$ and some constants $\nu_1>0$ and $\xi_1>1$;

\item[(iv)] $|S_{\bar\beta}| \lambda_0 \le \varrho_1$ for a sufficiently small constant $\varrho_1>0$ such that
$A_{11}^{-1} ( \kappa_1^{-1} \xi_{11}^{-2} M_{01} \varrho_0 +  \kappa_1^{-1} \nu_1^{-2} \xi_{12}^2 \varrho_1 ) \le \zeta_1$,
depending on $(\varrho_0,\varrho_1)$ and other constants, where $(A_{11}, M_{01}, \xi_{11}, \xi_{12})$ are defined in Lemma \ref{lem:beta-global} of the Supplement.
\end{itemize}\vspace{-.05in}
\end{ass}

The four conditions in Assumption \ref{ass:beta} are stated in terms of target values to allow possible model misspecification, although they are
comparable to those in analysis of Lasso penalized estimation with Lipschitz loss functions,
for example, Lemma 6.8 about penalized logistic regression in Buhlmann \& van de Geer (2011).
The margin condition can be justified from more primitive conditions about conditional densities of $Y$ given $T=1$ and $X$. See the discussion of
Condition D1 in Belloni \& Chernozhukov (2011).
The growth condition, Assumption \ref{ass:beta}(iv), can be roughly interpreted as
$ (|S_{\bar\gamma}|+|S_{\bar\beta}|) \sqrt{\log (p^\dag) } = o (\sqrt{n})$. This condition may be relaxed to only require that
$( |S_{\bar\gamma}|+|S_{\bar\beta}|) \log (p^\dag) = o(n)$, possibly with additional side conditions,
by conducting more refined analysis, for example, similarly as in Belloni \& Chernozhukov (2011).
On the other hand, this condition is already weaker than the sparsity condition,
$ (|S_{\bar\gamma}|+|S_{\bar\beta}|)  \log (p^\dag)  = o (\sqrt{n})$, required in our later result to obtain valid confidence intervals based on
$\hat\mu^{1+}_{\mytext{RCAL}}$.

\begin{thm} \label{thm:beta-conv}
Suppose that Assumptions~\ref{ass:gam}--\ref{ass:beta} hold and $\lambda_0\le 1$. Then for $A_1 > (M_{02}+C_{11}) (\xi_1+1)/(\xi_1-1)$, we have
with probability at least $1-8 \epsilon$,
\begin{align}
& \tilde E \left\{ T w(X; \bar\gamma_{\mytext{CAL}}) |h^\T(X) (\hat\beta_{\mytext{RWL},1+} - \bar\beta_{\mytext{WL},1+})|^2 \right\}
\le M_1 (|S_{\bar\gamma}| + |S_{\bar\beta}|) \lambda_0^2,  \label{eq:beta-error1} \\
& \| \hat\beta_{\mytext{RWL},1+} - \bar\beta_{\mytext{WL},1+} \|_1 \le M_1 (|S_{\bar\gamma}| + |S_{\bar\beta}|) \lambda_0, \label{eq:beta-error2}
\end{align}
where $C_{11} = 8 \me^{-B_0} C_1$, $M_{02}$ is defined in Lemma \ref{lem:depend-gam} of the Supplement,
and $M_1>0$ is a constant depending only on $(A_1, C_1, \kappa_1, \zeta_1, \nu_1, \xi_1)$ as well as $(M_0, B_0, C_0, \varrho_0)$.
\end{thm}

Theorem \ref{thm:beta-conv} demonstrates the convergence of $\hat\beta_{\mytext{RWL},1+}$ to $\bar\beta_{\mytext{WL},1+}$
at the rate $s \log(p^\dag )/n$ in the (in-sample) weighted $L_2$ prediction error for the linear predictor $ h^\T(X) \hat\beta_{\mytext{RWL},1+} $
and at the rate $s \sqrt{ \log(p^\dag )/n}$ in the $L_1$ coefficient error, where $s=|S_{\bar\gamma}| + |S_{\bar\beta}|$
depends on the sparsity sizes from both target values $\bar\gamma_{\mytext{CAL}}$ and $\bar\beta_{\mytext{WL},1+}$.
For simplicity, the same constant $M_1$ is stated in the error bounds (\ref{eq:beta-error1})--(\ref{eq:beta-error2}),
even though different multiplicative constants are involved in our technical results. See Lemma \ref{lem:beta-global} and
Corollary \ref{cor:beta-global} of the Supplement, where error bounds are also provided for the in-sample excess loss
$\ell_{\mytext{WL},1+} (\hat\beta_{\mytext{RWL},1+}; \bar\gamma_{\mytext{CAL}}) - \ell_{\mytext{WL},1+} (\bar\beta_{\mytext{WL},1+}; \bar\gamma_{\mytext{CAL}} )$
and the out-of-sample weighted $L_2$ prediction error and excess loss.

While the convergence rates in Theorem \ref{thm:beta-conv} are standard, up to the sparsity size, in high-dimensional statistics
(Buhlmann \& van de Geer 2011), our results are obtained after tackling several technical issues mentioned earlier.
We extend a strategy of deriving and inverting an expanded quadratic inequality in Tan (2020b, Remark 7) to handle the data-dependent weight
$w(\cdot;\hat\gamma_{\mytext{RCAL}})$, while exploiting concentration inequalities and localized analysis for Lipschitz, convex loss functions
as in Buhlmann \& van de Geer (2011, Section 6.7) to handle the non-smooth loss function $\ell_{\mytext{WL},1+} (\beta; \hat\gamma_{\mytext{RCAL}} )$.
See Lemmas \ref{lem:emp-concen}, \ref{lem:depend-gam}, \ref{lem:beta-local}
and \ref{lem:beta-global} in the Supplement.
Hence our analysis complements previous analysis of Lasso-penalized weighted outcome regression in Tan (2020b),
where the loss function is assumed to be twice differentiable.
The difference between the two technical approaches is similar to that in
classical asymptotic theory of M-estimation based on convexity or smoothness respectively (e.g., Pollard 1991).

We also study the convergence of $\hat\alpha_{\mytext{RWL},1+}$ to the target value $\bar\alpha_{\mytext{WL},1+}$,
which is defined as a minimizer of the expected loss function $E \{\ell_{\mytext{WL},1+} (\alpha; \bar\gamma_{\mytext{CAL}},\bar\beta_{\mytext{WL},1+}  )\}$ with
$\ell_{\mytext{WL},1+} (\alpha; \gamma,\beta )$ in (\ref{eq:alpha-loss}).
Suppose that the Lasso tuning parameter is specified for $\hat\alpha_{\mytext{RWL},1+}$ as $\lambda_\alpha = A_2 \lambda_0$,
where $A_2>0$ is a scaling constant and $\lambda_0$ is defined as before.

\begin{ass} \label{ass:alpha}
Suppose that the following conditions are satisfied:\vspace{-.05in}
\begin{itemize} \addtolength{\itemsep}{-.05in}
\item[(i)] $\varepsilon_{1+} = \tilde Y_{+} (h^\T \bar\beta_{\mytext{WL},1+}) - f^\T(X) \bar\alpha_{\mytext{WL},1+}$
is uniformly sub-Gaussian given $T=1$ and $X$:
$ D_0^2 E\{ \exp(\varepsilon_{1+}^2  /D_0^2) -1 |T=1, X\} \le D_1^2$
for some positive constants $(D_0,D_1)$, where
$\tilde Y_{+} (h^\T \bar\beta_{\mytext{WL},1+}) = Y + (\Lambda-\Lambda^{-1}) \rho_\tau(Y, h^\T \bar\beta_{\mytext{WL},1+})$.

\item[(ii)] the compatibility condition holds for $\Sigma_f$ with
the subset $S_{\bar\alpha}= \{0\} \cup \{j: (\bar\alpha_{\mytext{WL},1+})_j \not=0, j=1,\ldots,p\}$ and some constants $\nu_2>0$ and $\xi_2>1$;

\item[(iii)] $(1+\xi_2)^2 \nu_2^{-2} C_{02} |S_{\bar\alpha}| \lambda_0 \le \varrho_2$ for a constant $0< \varrho_2 <1$, where $C_{02}= 4 \me^{-B_0} C_0^2$.
\end{itemize} \vspace{-.05in}
\end{ass}

\begin{thm} \label{thm:alpha-conv}
Suppose that Assumptions~\ref{ass:gam}--\ref{ass:alpha} hold and $\lambda_0\le 1$. Then for $A_2 > C_{21} (\xi_2+1)/(\xi_2-1)$, we have
with probability at least $1-12\epsilon$,
\begin{align*}
& \tilde E \left\{ T w(X; \bar\gamma_{\mytext{CAL}}) |f^\T(X) (\hat\alpha_{\mytext{RWL},1+} - \bar\alpha_{\mytext{WL},1+})|^2 \right\}
\le M_2 (|S_{\bar\gamma}| + |S_{\bar\beta}| + |S_{\bar\alpha}|) \lambda_0^2,\\
& \| \hat\alpha_{\mytext{RWL},1+} - \bar\alpha_{\mytext{WL},1+} \|_1 \le M_2 (|S_{\bar\gamma}| + |S_{\bar\beta}| + |S_{\bar\alpha}|) \lambda_0,
\end{align*}
where $C_{21} = \me^{-B_0} C_0 \sqrt{ 8 (D_0^2 + D_1^2) } $ and $M_2>0$ is a constant depending only on
$(A_2, D_0, D_1, \nu_2, $ $\xi_2, \varrho_2)$ as well as $(M_1, M_0, B_0, C_0, \varrho_0)$.
\end{thm}

The three conditions in Assumption \ref{ass:alpha} are similar to those in Tan (2020b, Theorem 2) for
Lasso-penalized weighted linear outcome regression, except that the
response variable is constructed here using weighted quantile regression.
Our analysis extends the strategy of deriving and inverting an expanded quadratic inequality in Tan (2020b)
to handle both the data-dependent weight $w(\cdot;\hat\gamma_{\mytext{RCAL}})$ and
the data-dependent response variable $\tilde Y_{+} (h^\T \hat\beta_{\mytext{RWL},1+})$
and establish suitable error bounds for $\hat\alpha_{\mytext{RWL},1+}$,
which depend on sparsity sizes from the three target values $(\bar\gamma_{\mytext{CAL}}, \bar\alpha_{\mytext{WL},1+}, \bar\beta_{\mytext{WL},1+})$.
Note that the loss function
$\ell_{\mytext{WL},1+} (\alpha; \hat\gamma_{\mytext{RCAL}}, \hat\beta_{\mytext{RWL},1+} )$ for  $\hat\alpha_{\mytext{RWL},1+}$
is twice differentiable but non-Lipschitz in $\alpha$, so that our analysis is more aligned with Tan (2020b) than
Lipschitz-based analysis for Theorem \ref{thm:beta-conv}.
Nevertheless, the Lipschitz property of the quantile loss function $\rho_\tau(y,u)$ is instrumental
for handling the data-dependent response variable $\tilde Y_{+} (h^\T \hat\beta_{\mytext{RWL},1+})$. See Lemma \ref{lem:depend-gam2} of the Supplement.

Finally, we study the augmented IPW estimator
$\hat \mu^{1+}_{\mytext{RCAL}}$, depending on the first-stage regularized estimators
$(\hat\gamma_{\mytext{RCAL}},\hat\alpha_{\mytext{RWL},1+}, \hat\beta_{\mytext{RWL},1+})$.
The following result gives an error bound for $\hat \mu^{1+}_{\mytext{RCAL}}$, allowing that
all three models (\ref{eq:logit+pi})--(\ref{eq:linear+eta}) may be misspecified.

\begin{thm} \label{thm:mu1+expan}
Suppose that Assumptions~\ref{ass:gam}--\ref{ass:alpha} hold and $\lambda_0\le 1$.
Then for suitably large choices $(A_0,A_1,A_2)$ as in Theorems \ref{thm:gam-conv}--\ref{thm:alpha-conv}, we have with probability at least $1-14\epsilon$,
\begin{align}
& \left| \hat \mu^{1+}_{\mytext{RCAL}}-\hat\mu^{1+} (\bar\gamma_{\mytext{CAL}},\bar\alpha_{\mytext{WL},1+}, \bar\beta_{\mytext{WL},1+}) \right|
\le M_{31} |S_{\bar\gamma}| \lambda_0^2 + M_{32} |S_{\bar\beta}| \lambda_0^2  + M_{33} |S_{\bar\alpha}| \lambda_0^2 , \label{eq:mu1+expan}
\end{align}
where $M_{31}>0$ is a constant depending only on $(M_0,M_1,M_2)$,
$M_{32} >0$ is a constant depending only on $(M_1,M_2)$, and
$M_{33} >0$ is a constant depending only on $M_2$, each in addition to $(B_0,C_0,D_0,D_1,\varrho_0)$.
\end{thm}

Two types of asymptotic properties can be obtained from Theorem \ref{thm:mu1+expan}.
First, combining (\ref{eq:mu1+expan}) with Proposition \ref{pro:mu1-h-DR} shows that
$\hat \mu^{1+}_{\mytext{RCAL}}$ is pointwise doubly robust for the upper bound $\mu^{1+} (h^\T \bar\beta_{\mytext{WL},1+})$
if either model (\ref{eq:logit+pi}) or (\ref{eq:linear+eta}) with $\beta=\bar \beta_{\mytext{WL},1+}$ is correctly specified,
provided $ s \lambda_0^2 = o(1)$, that is,
$s \log(p^\dag) = o(n)$, where $s=|S_{\bar\gamma}|+ |S_{\bar\beta}| + |S_{\bar\alpha}|$.
Second, inequality (\ref{eq:mu1+expan}) also shows that $\hat \mu^{1+}_{\mytext{RCAL}}$ admits the desired
$n^{-1/2}$ asymptotic expansion (\ref{eq:rcal-expan}), provided $s \lambda_0^2 = o(n^{-1/2})$,
that is, $s \log(p^\dag) = o(\sqrt{n})$.
Even if both models (\ref{eq:logit+pi}) and (\ref{eq:linear+eta}) with $\beta=\bar \beta_{\mytext{WL},1+}$ are misspecified,
the point limit of $\hat \mu^{1+}_{\mytext{RCAL}}$ may differ from $\mu^{1+} (h^\T \bar\beta_{\mytext{WL},1+})$,
but the asymptotic expansion (\ref{eq:rcal-expan}) remains valid under $s \log(p^\dag) = o(\sqrt{n})$.
These results extend Tan (2020b, Theorem 3) under unconfoundedness to sensitivity analysis under model (\ref{eq:sen-model}).

To obtain the error bound (\ref{eq:mu1+expan}), our analysis carefully controls how $\hat \mu^{1+}_{\mytext{RCAL}}$ is affected by
the variations from all three estimators $(\hat\gamma_{\mytext{RCAL}},\hat\alpha_{\mytext{RWL},1+}, \hat\beta_{\mytext{RWL},1+})$, which
are characterized in Theorems \ref{thm:gam-conv}--\ref{thm:alpha-conv}.
In particular, the control of the error due to $\hat\beta_{\mytext{RWL},1+}$ is facilitated by exploiting
the Lipschitz property of the quantile loss function $\rho_\tau(y,u)$.
It is helpful to stress that the error rate $s \lambda_0^2$ achieved in (\ref{eq:mu1+expan}) depends crucially on
the construction of the estimators $(\hat\gamma_{\mytext{RCAL}},\hat\alpha_{\mytext{RWL},1+}, \hat\beta_{\mytext{RWL},1+})$
such that their target values satisfy the population versions of the calibration equations (\ref{eq:cal-eq}).
Otherwise, the difference
$\hat \mu^{1+}_{\mytext{RCAL}}-\hat\mu^{1+} (\bar\gamma_{\mytext{CAL}},\bar\alpha_{\mytext{WL},1+}, \bar\beta_{\mytext{WL},1+})$
would be of order $s \lambda_0$, as expected from the asymptotic expansion (\ref{eq:general-expan}) in the fixed-dimensional setting.
A similar reasoning is also behind the simple expansion (\ref{eq:desired-expan}) satisfied by the
$\hat \mu^{1+} (\hat\gamma_{\mytext{CAL}},\hat\alpha_{\mytext{WL},1+}, \hat\beta_{\mytext{WL},1+})$ in Section \ref{sec:cal}.

For variance estimation, the estimated variance $\hat V^{1+}_{\mytext{RCAL}}$ in (\ref{eq:est-variance})
can be shown similarly as in Tan (2020b, Theorem 4)
to converge in probability to its population version
\begin{align*}
\bar V^{1+}_{\mytext{CAL}} = \tilde E \left[ \left\{  \varphi_{+} (O; \pi(\cdot;\bar\gamma_{\mytext{CAL}}),\eta_{1+} (\cdot;\bar\alpha_{\mytext{WL},1+}), q_{1,\tau}(\cdot;\bar\beta_{\mytext{WL},1+}) ) - \bar\mu^{1+}_{\mytext{CAL}} \right\}^2 \right],
\end{align*}
provided $( |S_{\bar\gamma}|+ |S_{\bar\beta}| + |S_{\bar\alpha}|) \lambda_0 = o(1)$, that is,
$( |S_{\bar\gamma}|+ |S_{\bar\beta}| + |S_{\bar\alpha}|) \sqrt{\log(p^\dag)} = o(\sqrt{n})$,
which is weaker than the sparsity condition above for achieving the asymptotic expansion (\ref{eq:rcal-expan}).
Then by the Slutsky theorem, the Wald confidence interval (\ref{eq:upper-ci}) is doubly robust in
containing $\mu^{1+} (h^\T \bar\beta_{\mytext{WL},1+})$ with asymptotic probability $1-c$, if
either model (\ref{eq:logit+pi}) or (\ref{eq:linear+eta}) with $\beta=\bar \beta_{\mytext{WL},1+}$ is correctly specified
while model (\ref{eq:linear+Q}) may be misspecified, under the sparsity condition
$( |S_{\bar\gamma}|+ |S_{\bar\beta}| + |S_{\bar\alpha}|) \log(p^\dag) = o(\sqrt{n})$.
Correct specification of model (\ref{eq:linear+Q}) implies that $\mu^{1+} (h^\T \bar\beta_{\mytext{WL},1+})=\mu^{1+}$ and
(\ref{eq:upper-ci}) becomes doubly robust in covering $\mu^{1+}$ with asymptotic probability $1-c$.

\section{ATE and ATT estimation} \label{sec:ATE}

Our methods and theory are mainly presented on upper point bounds and confidence intervals for $\mu^1$, but
they can be readily extended to lower bounds for $\mu^1$ and to point bounds and confidence intervals for $\mu^0$, ATE, and ATT.

First, we describe our method for handling lower bounds on $\mu^1$.
In parallel to models (\ref{eq:linear+Q}) and (\ref{eq:linear+eta}), consider, in addition to model (\ref{eq:logit+pi}), the following models
\begin{align}
& q^*_{1,1-\tau}(X) = q_{1,1-\tau} (X; \beta) = h^\T(X) \beta,  \label{eq:linear-Q} \\
& E \{ \tilde Y_{-} ( q_{1,\tau}(\cdot;\beta) )| T=1,X \} = \eta_{1-}(X; \alpha) = f^\T(X) \alpha , \label{eq:linear-eta}
\end{align}
where $\eta_{1-}(X;\alpha)$ is taken to represent $E_{\eta_1} \{ \tilde Y_{-} ( q_{1,1-\tau}(\cdot;\beta) )| T=1,X \} $ for $\eta_1=\eta_{1-}(\cdot;\alpha)$.
The point estimator for our lower bound of $\mu^1$ is
\begin{align*}
\hat\mu^{1-}_{\mytext{RCAL}} &= \hat\mu^{1-} (\hat\gamma_{\mytext{RCAL}},\hat\alpha_{\mytext{RWL},1-}, \hat\beta_{\mytext{RWL},1-}) \nonumber \\
& = \tilde E \left\{ \varphi_{-} (O; \pi(\cdot;\hat\gamma_{\mytext{RCAL}}),\eta_{1-} (\cdot;\hat\alpha_{\mytext{RWL},1-}), q_{1,1-\tau}(\cdot;\hat\beta_{\mytext{RWL},1-}) ) \right\}, 
\end{align*}
where $\varphi_{-}$ is defined in (\ref{eq:AIPW-DG}), $\hat\gamma_{\mytext{RCAL}}$ is the same as before, and $(\hat\alpha_{\mytext{RWL},1-},\hat\beta_{\mytext{RWL},1-})$ are defined as follows.
The estimator $\hat\beta_{\mytext{RWL},1-}$ is  a minimizer of
$\ell_{\mytext{WL},1-} (\beta; \hat\gamma_{\mytext{RCAL}}) + \lambda_\beta \|\beta_{1:p} \|_1$, where
\begin{align*}
\ell_{\mytext{WL},1-} (\beta; \gamma) &= \tilde  E \left\{ T \frac{1-\pi(X;\gamma)}{\pi(X;\gamma)} \rho_{1-\tau} (Y, h^\T(X) \beta) \right\} .
\end{align*}
The estimator $\hat\alpha_{\mytext{RWL},1-}$ is a minimizer of
$\ell_{\mytext{WL},1-} (\alpha; \hat\gamma_{\mytext{RCAL}}, \hat\beta_{\mytext{RWL},1-} ) + \lambda_\alpha \|\alpha_{1:p} \|_1$, where
\begin{align*}
\ell_{\mytext{WL},1-} (\alpha; \gamma,\beta) &= \tilde  E \left[ T \frac{1-\pi(X;\gamma)}{\pi(X;\gamma)}
\left\{ \tilde Y_{-} (h^\T \beta ) - f^\T(X) \alpha \right\}^2 \right] /2 .
\end{align*}
Similarly as $\hat\mu^{1+}_{\mytext{RCAL}}$, the estimator $\hat\mu^{1-}_{\mytext{RCAL}}$ can be shown under suitable sparsity conditions
to admit a simple expansion in the form (\ref{eq:desired-expan}), with $\varphi_{+}$ replaced by $\varphi_{-}$ evaluated at the target values $(\bar\gamma_{\mytext{CAL}},\bar\alpha_{\mytext{WL},1-}, \bar\beta_{\mytext{WL},1-})$.
If either model (\ref{eq:logit+pi}) or (\ref{eq:linear-eta}) with $\beta=\bar \beta_{\mytext{WL},1-}$ is correctly specified,
then a doubly robust confidence interval of asymptotic size $1-c$ can be obtained for
$ \mu^{1-} ( h^\T \bar\beta_{\mytext{WL},1-})$ as
$[\, \hat\mu^{1-}_{\mytext{RCAL}} - z_c \sqrt{ \hat V^{1-}_{\mytext{RCAL}} / n},\;\infty)$,
where $\hat V^{1-}_{\mytext{RCAL}}$ is defined similarly as $\hat V^{1+}_{\mytext{RCAL}}$.
By combining the one-sided confidence intervals for the lower and upper bounds of $\mu^1$,
if either model (\ref{eq:logit+pi}) or models (\ref{eq:linear+eta}) with $\beta=\bar \beta_{\mytext{WL},1+}$
and (\ref{eq:linear-eta}) with $\beta=\bar \beta_{\mytext{WL},1-}$ are correctly specified, then
\begin{align}
\Big[\, \hat\mu^{1-}_{\mytext{RCAL}} - z_{c/2} \sqrt{ \hat V^{1-}_{\mytext{RCAL}} / n},\;
\hat\mu^{1+}_{\mytext{RCAL}} + z_{c/2} \sqrt{ \hat V^{1+}_{\mytext{RCAL}} / n} \,\Big] \label{eq:mu1-ci}
\end{align}
is a doubly robust confidence interval which covers $[\mu^{1-} ( h^\T \bar\beta_{\mytext{WL},1-}),\; \mu^{1+} ( h^\T \bar\beta_{\mytext{WL},1+})]$
and hence the true value $\mu^1$ with asymptotic probability at least $1-c$.

Next, point bounds and confidence intervals for $\mu^0$ can be obtained by applying those for $\mu^1$, but with
the observed data $\{O_i=(T_i Y_i, T_i, X_i): i=1,\ldots,n\}$ replaced by $\{O^{(0)}_i = ( (1-T_i)Y_i, 1-T_i, X_i): i=1,\ldots,n\} $.
The sample bound for $\mu^0$ in parallel to $\hat\mu^{1+} $ is
\begin{align*}
\hat\mu^{0+} ( \gamma,\alpha,\beta) = \tilde E \left\{ \varphi_{+} (O^{(0)}; 1-\pi(\cdot;\gamma),\eta_{0+} (\cdot;\alpha), q_{0,\tau}(\cdot;\beta) ) \right\}, 
\end{align*}
where $\varphi_{+}(\cdot)$ is defined in (\ref{eq:AIPW+DG}), but evaluated at $O^{(0)} = ( (1-T)Y, 1-T, X)$ and the associated working models such that
\begin{align*}
& \varphi_{+} (O^{(0)}; 1-\pi(\cdot;\gamma),\eta_{0+} (\cdot;\alpha), q_{0,\tau}(\cdot;\beta) ) \\
& = \frac{(1-T) Y}{1-\pi(X;\gamma)} + (\Lambda-\Lambda^{-1}) \frac{ (1-T) \pi(X;\gamma)}{1-\pi(X;\gamma)} \rho_\tau (Y, q_{0,\tau} (X;\beta))
 - \left\{\frac{1-T}{1-\pi(X;\gamma)}-1\right\} \eta_{0+} (X; \alpha) .
\end{align*}
Consider a logistic propensity score model $\pi(X;\gamma)$ as in (\ref{eq:logit+pi}), and
a linear quantile regression model $q_{0,\tau}(X;\beta)$ and a linear mean regression model $\eta_{0+}(X;\alpha)$ as follows:
\begin{align}
& q^*_{0,\tau}(X) = q_{0,\tau}(X;\beta) = h^\T(X) \beta,  \label{eq:linear+Q0}\\
& E\{ \tilde Y_{+}(q_{0,\tau}(\cdot;\beta)) |T=0,X\} = \eta_{0+}(X;\alpha) = f^\T(X) \alpha,  \label{eq:linear+eta0}
\end{align}
where $q^*_{0,\tau}(X)$ is the $\tau$-quantile of $P(Y|T=0,X)$, and
$\tilde Y_{+}(q_{0,\tau}) = Y + (\Lambda-\Lambda^{-1})  \rho_\tau (Y, q_{0,\tau}) $.
The resulting upper bound of $\mu^0$ from RCAL is
\begin{align*}
\hat\mu^{0+}_{\mytext{RCAL}} &= \hat\mu^{0+} (\hat\gamma^{(0)}_{\mytext{RCAL}},\hat\alpha_{\mytext{RWL},0+}, \hat\beta_{\mytext{RWL},0+}) ,
\end{align*}
where $(\hat\gamma^{(0)}_{\mytext{RCAL}},\hat\alpha_{\mytext{RWL},0+}, \hat\beta_{\mytext{RWL},0+})$ are defined as
$(\hat\gamma_{\mytext{RCAL}},\hat\alpha_{\mytext{RWL},1+}, \hat\beta_{\mytext{RWL},1+})$ in Section \ref{sec:rcal}, except with
$T$ and $f^\T(X) \gamma$ replaced by $1-T$ and $-f^\T(X) \gamma$.
Under suitable sparsity conditions, if model (\ref{eq:logit+pi}) or (\ref{eq:linear+eta0}) with $\beta= \bar\beta_{\mytext{WL},0+}$ is correctly specified,
then $(-\infty, \hat\mu^{0+}_{\mytext{RCAL}} + z_c \sqrt{ \hat V^{0+}_{\mytext{RCAL}} }]$
is a doubly robust confidence interval of asymptotic size $1-c$ for $\mu^{0+} (h^\T \bar\beta_{\mytext{WL},0+})$
and hence of asymptotic level $1-c$ for the sharp bound $\mu^{0+}$, where
$\hat V^{0+}_{\mytext{RCAL}}$ is defined similarly as $\hat V^{1+}_{\mytext{RCAL}}$ in (\ref{eq:est-variance}),
and $\mu^{0+}$ and $\mu^{0+} (h^\T \bar\beta_{\mytext{WL},0+})$ are defined similarly as  $\mu^{1+}$ and $\mu^{1+} (h^\T \bar\beta_{\mytext{WL},1+})$
in (\ref{eq:mu1+}) and (\ref{eq:mu1+Q}).
If model (\ref{eq:linear+Q0}) is correctly specified, then $\mu^{0+} (h^\T \bar\beta_{\mytext{WL},0+})$ reduces to $\mu^{0+}$.

From the lower bound $\hat\mu^{1-}_{\mytext{RCAL}}$ and upper bound $\hat\mu^{0+}_{\mytext{RCAL}}$, the resulting lower bound for $\mu^1-\mu^0$ is
$\hat\mu^{1-}_{\mytext{RCAL}} - \hat\mu^{0+}_{\mytext{RCAL}}$.
If model (\ref{eq:logit+pi}) or models (\ref{eq:linear+eta}) with $\beta= \bar\beta_{\mytext{WL},1+}$ and
(\ref{eq:linear+eta0}) with $\beta= \bar\beta_{\mytext{WL},0+}$ are correctly specified,
then a doubly robust confidence interval of asymptotic size $1-c$ for $\mu^{1-} (h^\T \bar\beta_{\mytext{WL},1-}) - \mu^{0+} (h^\T \bar\beta_{\mytext{WL},0+})$
and hence of asymptotic level $1-c$ for $\mu^1-\mu^0$ is
\begin{align*}
\Big[\, \hat\mu^{1-}_{\mytext{RCAL}} -\hat\mu^{0+}_{\mytext{RCAL}}
+ z_c \sqrt{ \hat V(\hat\mu^{1-}_{\mytext{RCAL}} -\hat\mu^{0+}_{\mytext{RCAL}} ) / n}, \; \infty \Big) ,
\end{align*}
where the estimated variance is defined as
\begin{align*}
& \hat V(\hat\mu^{1-}_{\mytext{RCAL}} -\hat\mu^{0+}_{\mytext{RCAL}} ) = \tilde E \Big[ \Big\{  \varphi_{-} (O; \pi(\cdot;\hat\gamma_{\mytext{RCAL}}),\eta_{1-} (\cdot;\hat\alpha_{\mytext{RWL},1-}), q_{1,1-\tau}(\cdot;\hat\beta_{\mytext{RWL},1-}) ) \\
& \quad
- \varphi_{+} (O^{(0)}; \pi(\cdot;\hat\gamma^{(0)}_{\mytext{RCAL}}),\eta_{0+} (\cdot;\hat\alpha_{\mytext{RWL},0+}), q_{0,\tau}(\cdot;\hat\beta_{\mytext{RWL},0+}))
- (\hat\mu^{1-}_{\mytext{RCAL}} - \hat\mu^{0+}_{\mytext{RCAL}}) \Big\}^2 \Big].
\end{align*}
Two-sided doubly robust confidence intervals of asymptotic level $1-c$ can also be derived for $\mu^1-\mu^0$ similarly as (\ref{eq:mu1-ci}) for $\mu^1$.

Our methods and theory can also be directly extended to ATT $=\nu^1-\nu^0$ as defined in Section \ref{sec:setup}.
The parameter $\nu^1 = E(Y^1 | T=1)$ can be easily estimated by $\tilde E( TY ) / \tilde E(T)$.
The parameter $\nu^0$ can be related to $\mu^0$ in a simple manner:
\begin{align*}
\nu^0 & = E(Y^0 | T=1) = E( T Y^0) / E (T)  = \{ \mu^0 - E((1-T)Y) \} / E(T).
\end{align*}
Because $E ( (1-T)Y)$ and $E(T)$ are identifiable from observed data, regardless of unmeasured confounding, population bounds for $\nu^0$ can be directly
transferred from those for $\mu^0$, and the sharpness of the bounds remains the same under sensitivity model (\ref{eq:sen-model}).
For example, the sharp upper bound for $\nu^0$ is $\nu^{0+} =  \{ \mu^{0+} - E((1-T)Y) \} / E(T)$.
Moreover, sample bounds for $\nu^0$ can also be transferred from those for $\mu^0$, and similar statistical properties are achieved.
The upper bound for $\nu^0$ from RCAL can be simplified similarly as (\ref{eq:mu1+imput}) to be
\begin{align*}
\hat\nu^{0+}_{\mytext{RCAL}} & = \left\{ \hat\mu^{0+}_{\mytext{RCAL}} - \tilde E((1-T)Y) \right\} / \tilde E(T)
 = \tilde E \left\{ T \hat \eta_{0+} (X; \hat \alpha_{\mytext{RWL},0+}) \right\}/ \tilde E(T) .
\end{align*}
Under suitable sparsity conditions,
if model (\ref{eq:logit+pi}) or (\ref{eq:linear+eta0}) with $\beta= \bar\beta_{\mytext{WL},0+}$ is correctly specified,
then a doubly robust confidence interval of asymptotic level $1-c$ for the sharp bound $\nu^{0+}$ is
\begin{align*}
\Big( -\infty, \; \hat\nu^{0+}_{\mytext{RCAL}}
+ z_c \sqrt{ \hat V ( \hat\nu^{0+}_{\mytext{RCAL}} ) / n} \, \Big] ,
\end{align*}
where the estimated variance is defined as
\begin{align*}
& \hat V ( \hat\nu^{0+}_{\mytext{RCAL}} ) = \tilde E \Big[ \Big\{
\varphi_{+}  - (1-T)Y - T \hat\nu^{0+}_{\mytext{RCAL}} \Big\}^2 \Big] / \tilde E^2(T) ,
\end{align*}
with $\varphi_{+} = \varphi_{+} (O^{(0)}; \pi(\cdot;\hat\gamma^{(0)}_{\mytext{RCAL}}),\eta_{0+} (\cdot;\hat\alpha_{\mytext{RWL},0+}), q_{0,\tau}(\cdot;\hat\beta_{\mytext{RWL},0+}))$.
Two-sided doubly robust confidence intervals of asymptotic level $1-c$ can also be derived for $\nu^1$ and $\nu^1-\nu^0$.

\section{Additional simulation results} \label{sec:add-simulation}

\vspace*{-.2in}
\begin{figure} [H]
\begin{tabular}{c}
\includegraphics[width=6in, height=2in]{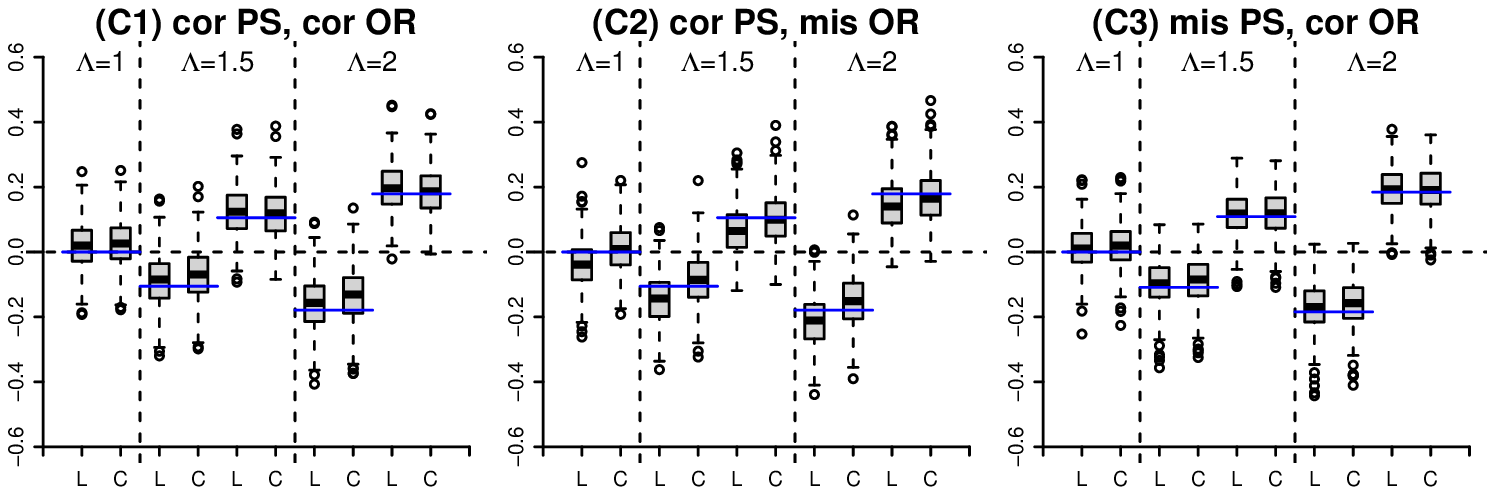} \vspace{-.1in}
\end{tabular}
\caption{\scriptsize   Boxplots of point bounds on $\mu^1$ from regularized likelihood-related estimation (L)
and regularized calibrated estimation without relaxation (C) with $(n,p)=(800,200)$, similarly as in Figure \ref{fig:simu-p10}.
} \label{fig:simu-p200-vanilla} 
\end{figure}

\vspace*{-.2in}
\begin{table} [H]
\caption{\small Coverage proportions of confidence intervals (CIs) from ML and CAL for $(n,p)=(800,10)$} \label{tab:simu-p10}  \vspace{-.05in}
\scriptsize
\begin{center}
\begin{tabular*}{1\textwidth}{@{\extracolsep\fill} c lll lll lll} \hline
      & \multicolumn{3}{c}{(C1) cor PS, cor OR} & \multicolumn{3}{c}{(C2) cor PS, mis OR} & \multicolumn{3}{c}{(C3) mis PS, cor OR} \\
      & $\Lambda=1$ & $\Lambda=1.5$ & $\Lambda=2$ & $\Lambda=1$ & $\Lambda=1.5$ & $\Lambda=2$ & $\Lambda=1$ & $\Lambda=1.5$ & $\Lambda=2$ \\ \hline
      & \multicolumn{9}{c}{Lower 95\% CIs for the sharp lower bound of $\mu^1$} \\
RML  & 0.942 & 0.934 & 0.924 & 0.964 & 0.976 & 0.984 & 0.940 & 0.934 & 0.932 \\
RCAL & 0.942 & 0.922 & 0.910 & 0.944 & 0.936 & 0.926 & 0.936 & 0.928 & 0.920 \\

      & \multicolumn{9}{c}{Upper 95\% CIs for the sharp upper bound of $\mu^1$} \\
RML  & 0.950 & 0.946 & 0.942 & 0.936 & 0.962 & 0.968 & 0.950 & 0.942 & 0.944 \\
RCAL & 0.948 & 0.940 & 0.928 & 0.932 & 0.930 & 0.920 & 0.952 & 0.940 & 0.932 \\

      & \multicolumn{9}{c}{90\% CIs for the sharp lower to upper bound of $\mu^1$} \\
RML  & 0.892 & 0.880 & 0.866 & 0.900 & 0.938 & 0.952 & 0.890 & 0.876 & 0.876 \\
RCAL & 0.890 & 0.862 & 0.838 & 0.876 & 0.866 & 0.846 & 0.888 & 0.868 & 0.852 \\ \hline
\end{tabular*}\\[.05in]
\end{center} 
\end{table}

\vspace*{-.5in}
\begin{table} [H]
\caption{\small Coverage proportions of confidence intervals (CIs) from RML and RCAL without relaxation for $(n,p)=(800,200)$} \label{tab:simu-p200-vanilla}  \vspace{-.05in}
\scriptsize
\begin{center}
\begin{tabular*}{1\textwidth}{@{\extracolsep\fill} c lll lll lll} \hline
      & \multicolumn{3}{c}{(C1) cor PS, cor OR} & \multicolumn{3}{c}{(C2) cor PS, mis OR} & \multicolumn{3}{c}{(C3) mis PS, cor OR} \\
      & $\Lambda=1$ & $\Lambda=1.5$ & $\Lambda=2$ & $\Lambda=1$ & $\Lambda=1.5$ & $\Lambda=2$ & $\Lambda=1$ & $\Lambda=1.5$ & $\Lambda=2$ \\ \hline
      & \multicolumn{9}{c}{Lower 95\% CIs for the sharp lower bound of $\mu^1$} \\
RML   & 0.886 & 0.896 & 0.896 & 0.992 & 0.986 & 0.984 & 0.930 & 0.934 & 0.930 \\
RCAL  & 0.870 & 0.854 & 0.812 & 0.916 & 0.892 & 0.872 & 0.906 & 0.908 & 0.906 \\

      & \multicolumn{9}{c}{Upper 95\% CIs for the sharp upper bound of $\mu^1$} \\
RML   & 0.976 & 0.966 & 0.960 & 0.866 & 0.834 & 0.834 & 0.972 & 0.954 & 0.952 \\
RCAL  & 0.974 & 0.952 & 0.942 & 0.960 & 0.928 & 0.900 & 0.982 & 0.962 & 0.956 \\

      & \multicolumn{9}{c}{90\% CIs for the sharp lower to upper bound of $\mu^1$} \\
RML   & 0.862 & 0.862 & 0.856 & 0.858 & 0.820 & 0.818 & 0.902 & 0.888 & 0.882 \\
RCAL  & 0.844 & 0.806 & 0.754 & 0.876 & 0.820 & 0.772 & 0.888 & 0.870 & 0.862 \\ \hline
\end{tabular*}\\[.05in]
\end{center} 
\end{table}

\newpage
\section{Additional results from empirical application} \label{sec:add-application}

\begin{figure} [H]
\begin{tabular}{c}
\includegraphics[width=6in, height=2in]{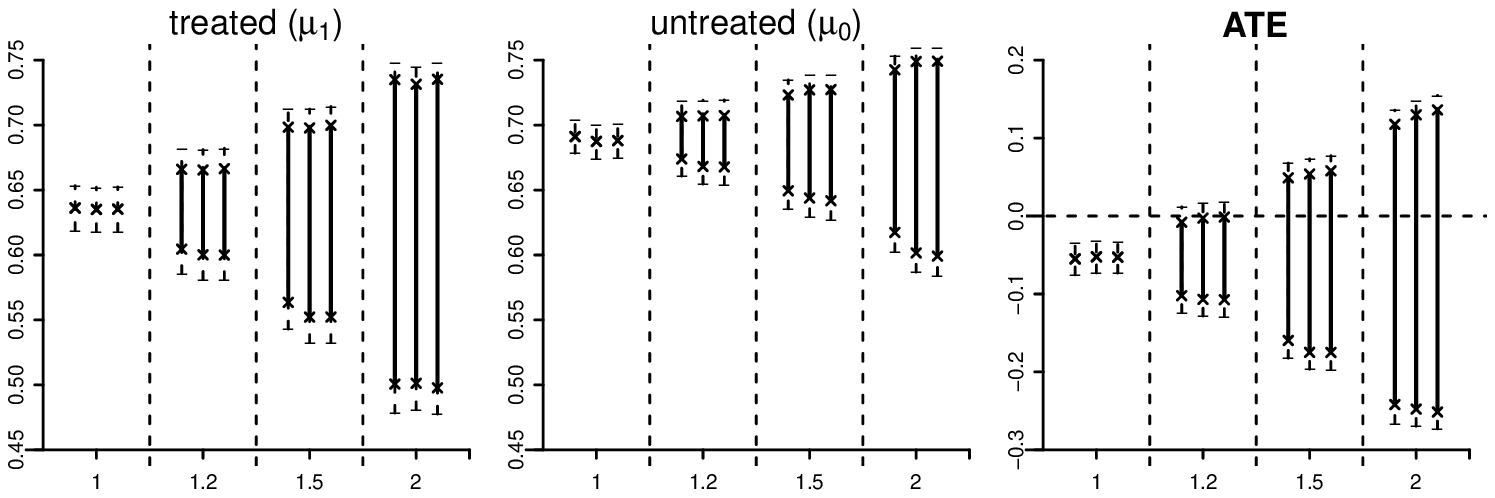} \vspace{-.1in}
\end{tabular}
\caption{\scriptsize Point bounds ($\times$) and 90\% confidence intervals on 30-day survival probabilities and ATE, similarly as Figure \ref{fig:simu-p200}
but from RCAL without relaxation.
} \label{fig:rhc-intx-vanilla} 
\end{figure}

\begin{figure} [H]
\begin{tabular}{c}
\includegraphics[width=6in, height=2in]{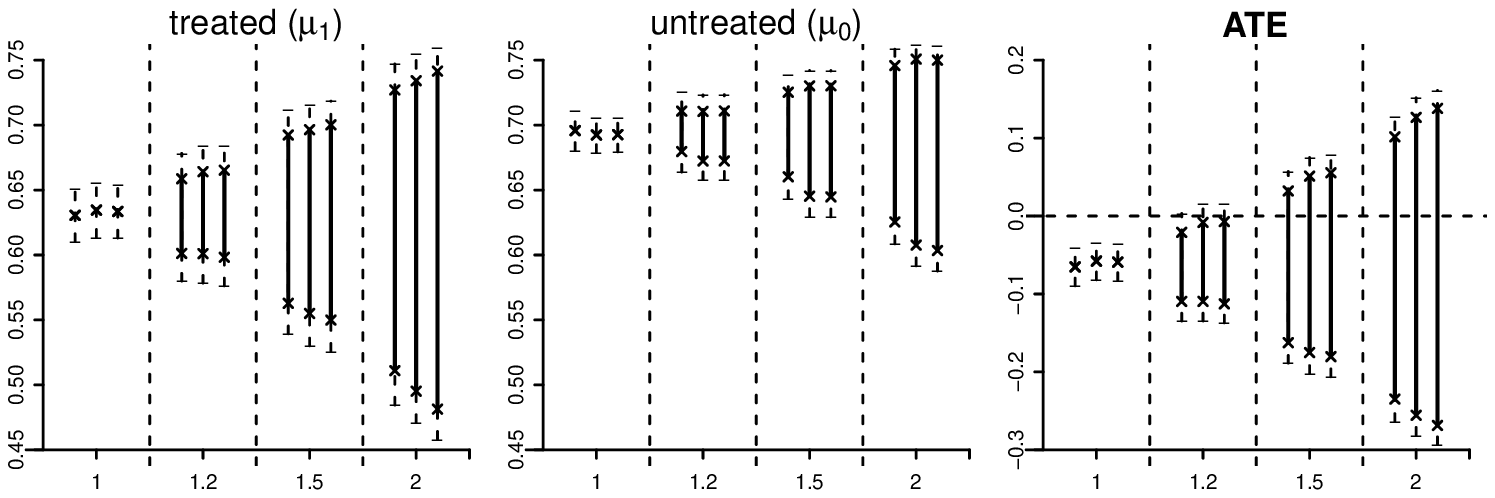} \vspace{-.1in}
\end{tabular}
\caption{\scriptsize Point bounds ($\times$) and 90\% confidence intervals on 30-day survival probabilities and ATE, similarly as Figure \ref{fig:simu-p200}
but using working models with main effects only (non-regularized estimation).
} \label{fig:rhc-main} 
\end{figure}

\begin{table} [H]
\caption{\small Point bounds and SEs for 30-day survival probabilities and ATE,
using working models with main effects and interactions (RCAL with relaxation).} \label{tab:rhc-intx}  \vspace{-.05in}
\scriptsize
\begin{center}
\begin{tabular*}{1\textwidth}{@{\extracolsep\fill} l llll llll} \hline
      & $\Lambda=1$ & & $\Lambda=1.2$  & & $\Lambda=1.5$ & & $\Lambda=2$ \\
      & Bounds & SEs & Bounds & SEs & Bounds & SEs & Bounds & SEs \\ \hline
      & \multicolumn{8}{c}{Lower bounds and SEs for $\mu^1$} \\
RML        & 0.636    & 0.010 & 0.605   & 0.011 & 0.564   & 0.012 & 0.501   & 0.014 \\
RCAL lin   & 0.635    & 0.010 & 0.597   & 0.011 & 0.542   & 0.012 & 0.483   & 0.012\\
RCAL logit & 0.635    & 0.010 & 0.597   & 0.011 & 0.543   & 0.012 & 0.479   & 0.012  \\

      & \multicolumn{8}{c}{Upper bounds and SEs for $\mu^1$} \\
RML        & 0.636    & 0.010  & 0.666    & 0.010 & 0.698   & 0.009 & 0.735   & 0.008 \\
RCAL lin   & 0.635    & 0.010  & 0.668    & 0.010 & 0.702   & 0.009 & 0.737   & 0.008 \\
RCAL logit & 0.635    & 0.010  & 0.669    & 0.010 & 0.704   & 0.009 & 0.741   & 0.008 \\

      & \multicolumn{8}{c}{Lower bounds and SEs for $\mu^0$} \\
RML        & 0.691   & 0.008  & 0.707   & 0.007 & 0.723   & 0.007  & 0.743   & 0.007 \\
RCAL lin   & 0.687   & 0.008  & 0.708   & 0.007 & 0.728   & 0.007  & 0.749   & 0.006  \\
RCAL logit & 0.688   & 0.008  & 0.708   & 0.007 & 0.729   & 0.007  & 0.749   & 0.006 \\

      & \multicolumn{8}{c}{Upper bounds and SEs for $\mu^0$} \\
RML        & 0.691   & 0.008  & 0.674   & 0.008 & 0.649   & 0.008 & 0.617   & 0.009 \\
RCAL lin   & 0.687   & 0.008  & 0.666   & 0.008 & 0.638   & 0.009 & 0.590   & 0.009 \\
RCAL logit & 0.688   & 0.008  & 0.665   & 0.008 & 0.636   & 0.009 & 0.588   & 0.009 \\

      & \multicolumn{8}{c}{Lower bounds and SEs for $\mu^1-\mu^0$} \\
RML        & -0.055   & 0.012   & -0.102   & 0.013 & -0.160   & 0.014 & -0.242   & 0.015 \\
RCAL lin   & -0.052   & 0.012   & -0.111   & 0.013 & -0.186   & 0.013 & -0.266   & 0.013 \\
RCAL logit & -0.053   & 0.012   & -0.111   & 0.013 & -0.186   & 0.013 & -0.270   & 0.013 \\

      & \multicolumn{8}{c}{Upper bounds and SEs for $\mu^1-\mu^0$} \\
RML        & -0.055   & 0.012 & -0.008   & 0.012 & 0.049   & 0.012 & 0.118   & 0.011 \\
RCAL lin   & -0.052   & 0.012 & 0.002   & 0.012  & 0.064   & 0.012 & 0.147   & 0.011 \\
RCAL logit & -0.053   & 0.012 & 0.003   & 0.012  & 0.068   & 0.012 & 0.153   & 0.011 \\  \hline
\end{tabular*}\\[.05in]
\parbox{1\textwidth}{\scriptsize Note: RCAL lin denotes RCAL with relaxation using
linear outcome mean regression (Section \ref{sec:rcal}), and
RCAL logit denotes RCAL with relaxation using logistic outcome mean regression (Section \ref{sec:binary}).}
\end{center} 
\end{table}

\begin{table} [H]
\caption{\small Point bounds and SEs for 30-day survival probabilities and ATE,
using working models with main effects and interactions (RCAL without relaxation).} \label{tab:rhc-intx-vanilla}  \vspace{-.05in}
\scriptsize
\begin{center}
\begin{tabular*}{1\textwidth}{@{\extracolsep\fill} l llll llll} \hline
      & $\Lambda=1$ & & $\Lambda=1.2$  & & $\Lambda=1.5$ & & $\Lambda=2$ \\
      & Bounds & SEs & Bounds & SEs & Bounds & SEs & Bounds & SEs \\ \hline
      & \multicolumn{8}{c}{Lower bounds and SEs for $\mu^1$} \\
RML        & 0.636    & 0.001 & 0.605   & 0.0011 & 0.564   & 0.0012 & 0.501   & 0.0014 \\
RCAL lin   & 0.635    & 0.001 & 0.600   & 0.0011 & 0.552   & 0.0012 & 0.501   & 0.0012 \\
RCAL logit & 0.635    & 0.001 & 0.600   & 0.0011 & 0.552   & 0.0012 & 0.498   & 0.0012 \\

      & \multicolumn{8}{c}{Upper bounds and SEs for $\mu^1$} \\
RML        & 0.636    & 0.001  & 0.666    & 0.001 & 0.698   & 0.009 & 0.735   & 0.008 \\
RCAL lin   & 0.635    & 0.001  & 0.665    & 0.001 & 0.698   & 0.009 & 0.731   & 0.008 \\
RCAL logit & 0.635    & 0.001  & 0.666    & 0.001 & 0.700   & 0.009 & 0.735   & 0.008 \\

      & \multicolumn{8}{c}{Lower bounds and SEs for $\mu^0$} \\
RML        & 0.691   & 0.008  & 0.707   & 0.007 & 0.723   & 0.007 & 0.743   & 0.007 \\
RCAL lin   & 0.687   & 0.008  & 0.707   & 0.007 & 0.727   & 0.007 & 0.749   & 0.006 \\
RCAL logit & 0.688   & 0.008  & 0.707   & 0.007 & 0.727   & 0.007 & 0.749   & 0.006 \\

      & \multicolumn{8}{c}{Upper bounds and SEs for $\mu^0$} \\
RML        & 0.691   & 0.008 & 0.674   & 0.008 & 0.649   & 0.008 & 0.617   & 0.009 \\
RCAL lin   & 0.687   & 0.008 & 0.668   & 0.008 & 0.644   & 0.009 & 0.602   & 0.009 \\
RCAL logit & 0.688   & 0.008 & 0.668   & 0.008 & 0.642   & 0.009 & 0.599   & 0.009 \\

      & \multicolumn{8}{c}{Lower bounds and SEs for $\mu^1-\mu^0$} \\
RML        & -0.055   & 0.0012  & -0.102   & 0.0013 & -0.160   & 0.0014 & -0.242   & 0.0015 \\
RCAL lin   & -0.052   & 0.0012  & -0.107   & 0.0013 & -0.175   & 0.0013 & -0.248   & 0.0013 \\
RCAL logit & -0.053   & 0.0012  & -0.107   & 0.0013 & -0.175   & 0.0013 & -0.251   & 0.0013 \\

      & \multicolumn{8}{c}{Upper bounds and SEs for $\mu^1-\mu^0$} \\
RML        & -0.055   & 0.0012  & -0.008   & 0.0012 & 0.049   & 0.0012 & 0.118   & 0.0011 \\
RCAL lin   & -0.052   & 0.0012  & -0.003   & 0.0012 & 0.054   & 0.0012 & 0.130   & 0.0011 \\
RCAL logit & -0.053   & 0.0012  & -0.001   & 0.0012 & 0.058   & 0.0012 & 0.136   & 0.0011 \\  \hline
\end{tabular*}\\[.05in]
\parbox{1\textwidth}{\scriptsize Note: RCAL lin denotes RCAL without relaxation using
linear outcome mean regression (Section \ref{sec:rcal}), and
RCAL logit denotes RCAL without relaxation using logistic outcome mean regression (Section \ref{sec:binary}).}
\end{center} 
\end{table}

\begin{table} [H]
\caption{\small Point bounds and SEs for 30-day survival probabilities and ATE,
using working models with main effects only (non-regularized estimation).} \label{tab:rhc-main}  \vspace{-.05in}
\scriptsize
\begin{center}
\begin{tabular*}{1\textwidth}{@{\extracolsep\fill} l llll llll} \hline
      & $\Lambda=1$ & & $\Lambda=1.2$  & & $\Lambda=1.5$ & & $\Lambda=2$ \\
      & Bounds & SEs & Bounds & SEs & Bounds & SEs & Bounds & SEs \\ \hline
      & \multicolumn{8}{c}{Lower bounds and SEs for $\mu^1$} \\
ML        & 0.631   & 0.012  & 0.601   & 0.013 & 0.563   & 0.014 & 0.511   & 0.016  \\
CAL lin   & 0.635   & 0.013  & 0.601   & 0.014 & 0.555   & 0.015 & 0.495   & 0.015 \\
CAL logit & 0.634   & 0.012  & 0.598   & 0.013 & 0.550   & 0.015 & 0.481   & 0.014 \\

      & \multicolumn{8}{c}{Upper bounds and SEs for $\mu^1$} \\
ML        & 0.631   & 0.012  & 0.659   & 0.012 & 0.692   & 0.012 & 0.727   & 0.012 \\
CAL lin   & 0.635   & 0.013  & 0.664   & 0.012 & 0.696   & 0.012 & 0.734   & 0.013 \\
CAL logit & 0.634   & 0.012  & 0.665   & 0.012 & 0.700   & 0.011 & 0.742   & 0.011 \\

      & \multicolumn{8}{c}{Lower bounds and SEs for $\mu^0$} \\
ML        & 0.696   & 0.009  & 0.711   & 0.009  & 0.725   & 0.008 & 0.746   & 0.008  \\
CAL lin   & 0.692   & 0.008  & 0.711   & 0.008  & 0.730   & 0.007 & 0.751   & 0.007 \\
CAL logit & 0.693   & 0.008  & 0.711   & 0.008  & 0.730   & 0.007 & 0.750   & 0.007 \\

      & \multicolumn{8}{c}{Upper bounds and SEs for $\mu^0$} \\
ML        & 0.696   & 0.009  & 0.680   & 0.010 & 0.660   & 0.010 & 0.625    & 0.010 \\
CAL lin   & 0.692   & 0.008  & 0.672   & 0.009 & 0.645   & 0.010 & 0.608    & 0.010 \\
CAL logit & 0.693   & 0.008  & 0.672   & 0.009 & 0.645   & 0.009 & 0.603    & 0.010 \\

      & \multicolumn{8}{c}{Lower bounds and SEs for $\mu^1-\mu^0$} \\
ML        & -0.065   & 0.015 & -0.110   & 0.015 & -0.163   & 0.016 & -0.235   & 0.017 \\
CAL lin   & -0.058   & 0.015 & -0.110   & 0.015 & -0.175   & 0.016 & -0.256   & 0.016 \\
CAL logit & -0.059   & 0.014 & -0.113   & 0.015 & -0.180   & 0.016 & -0.269   & 0.015 \\

      & \multicolumn{8}{c}{Upper bounds and SEs for $\mu^1-\mu^0$} \\
ML        & -0.065   & 0.015 & -0.021   & 0.015  & 0.032   & 0.015  & 0.102   & 0.015 \\
CAL lin   & -0.058   & 0.015 & -0.008   & 0.014  & 0.051   & 0.015  & 0.127   & 0.015 \\
CAL logit & -0.059   & 0.014 & -0.007   & 0.014  & 0.056   & 0.014  & 0.138   & 0.014 \\  \hline
\end{tabular*}\\[.05in]
\parbox{1\textwidth}{\scriptsize Note: CAL lin denotes CAL using
linear outcome mean regression (Section \ref{sec:rcal}), and
CAL logit denotes CAL using logistic outcome mean regression (Section \ref{sec:binary}).}
\end{center} 
\end{table}

\begin{figure} [t!]
\begin{tabular}{c}
\includegraphics[width=6in, height=4in]{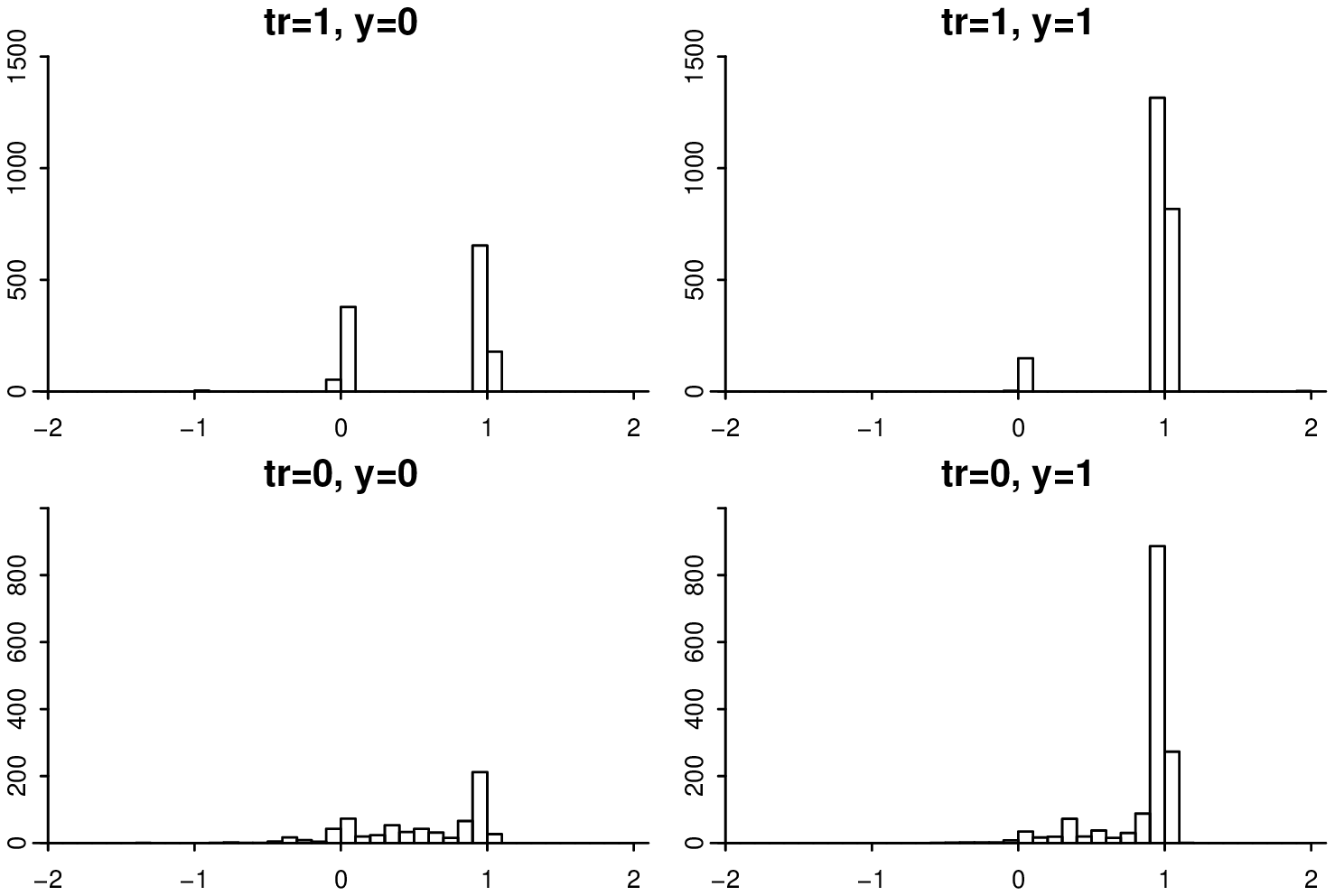} \vspace{-.1in}
\end{tabular}
\caption{\scriptsize Weighted histograms of fitted linear predictors from weighted outcome $60\%$-quantile regression corresponding to $\Lambda=1.5$ (with main effects only) within the treated (top) and untreated (bottom) groups, stratified by the outcomes, 0 or 1 (left or right in each row).
} \label{fig:rhc-main-quantile} 
\end{figure}

\begin{figure} [H]
\begin{tabular}{c}
\includegraphics[width=6in, height=4in]{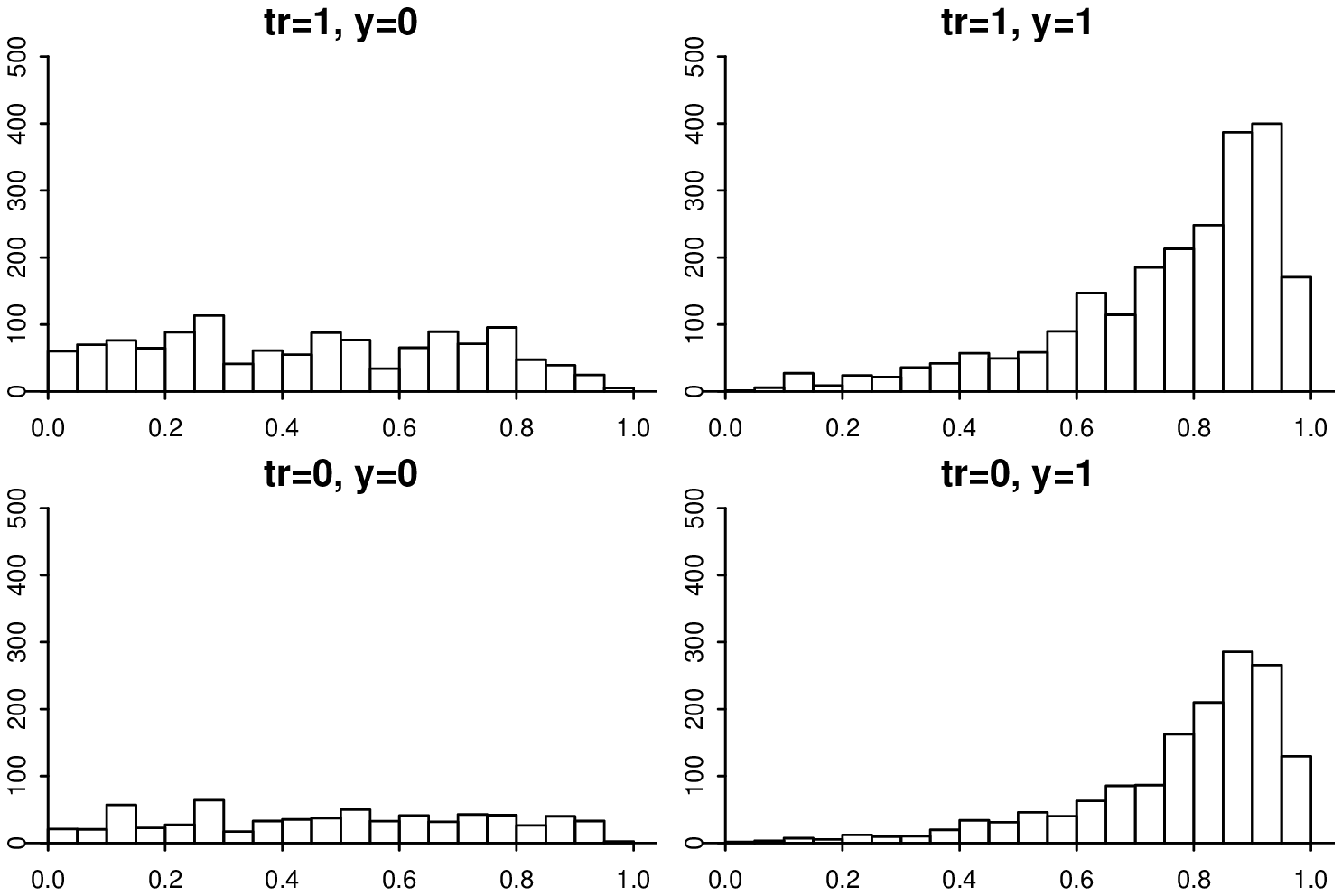} \vspace{-.1in}
\end{tabular}
\caption{\scriptsize Weighted histograms of fitted probabilities from weighted outcome logistic regression (with main effects only) within the treated (top) and untreated (bottom) groups, stratified by the outcomes, 0 or 1 (left or right in each row).
} \label{fig:rhc-main-logistic} 
\end{figure}

We discuss why the point-bound enclosed intervals and confidence intervals for the 30-day survival probability if treated are
wider than those for the survival probability if untreated.
This involves an inside look at the contributing factors of the point bounds by the IPW estimation.
The augmentation term in the augmented IPW estimation is asymptotically zero and hence can be ignored if the propensity score model
is correctly specified.

The upper bound (\ref{eq:dual-hat-mu1+h}) based on IPW estimation can be rewritten as
\begin{align*}
 &\hat \mu_h^{1+} (\hat\gamma) = \frac{1}{n} \sum_{i=1}^n \left\{ \frac{T_i}{\pi(X_i;\hat\gamma)} Y + (\Lambda-\Lambda^{-1})
    T_i \frac{1-\pi(X_i;\hat\gamma)}{\pi(X_i;\hat\gamma)} \rho_\tau (Y, h^\T \hat\beta_{\mytext{WL},1+}(\hat\gamma) ) \right\} \\
 & =\frac{1}{n} \sum_{i=1}^n  \frac{T_i}{\pi(X_i;\hat\gamma)} Y_i \\
 & \quad + (\Lambda-\Lambda^{-1})
 \left\{ \sum_{1\le i\le n, T_i=1} w(X_i;\hat\gamma) \right\} \left\{ \frac{1}{n_1} \sum_{1\le i\le n, T_i=1}
 \tilde w(X_i;\hat\gamma) \rho_\tau (Y_i, h^\T(X_i) \hat\beta_{\mytext{WL},1+}(\hat\gamma))  \right\},
\end{align*}
where $w(X;\hat\gamma) = (1-\pi(X;\hat\gamma )) /  \pi(X;\hat\gamma )$,
and $\tilde w(X; \hat\gamma) = n_1 w(X; \hat\gamma) / \sum_{1\le i\le n, T_i=1} w(X_i;\hat\gamma) $
such that $\sum_{1\le i\le n, T_i=1} \tilde w(X_i;\hat\gamma)=n_1$, with $n_1=\sum_{i=1}^n T_i$, the treated group size.
For $\hat\gamma = \hat\gamma_{\mytext{RCAL}}$ with $n^{-1} \sum_{i=1}^n T_i / \pi(X_i;\hat\gamma)=1$, we have
$\sum_{1\le i\le n, T_i=1} w(X_i;\hat\gamma) = n-n_1$ and hence
\begin{align}
 \hat \mu_h^{1+} (\hat\gamma)
 &=\frac{1}{n} \sum_{i=1}^n  \frac{T_i}{\pi(X_i;\hat\gamma)} Y_i  \nonumber \\
 & \quad + (\Lambda-\Lambda^{-1}) (n-n_1) \left\{ \frac{1}{n_1} \sum_{1\le i\le n, T_i=1}
 \tilde w(X_i;\hat\gamma) \rho_\tau (Y_i, h^\T(X_i) \hat\beta_{\mytext{WL},1+}(\hat\gamma))  \right\}.\label{eq:upper-rewrite}
\end{align}
The first term in (\ref{eq:upper-rewrite}) is the standard IPW estimator.
For fixed $\Lambda$, the second term in (\ref{eq:upper-rewrite}), i.e., the increase to the standard IPW estimator, is determined by the product of
the untreated group size $n-n_1$ and the loss value from weighted quantile regression in the curly bracket.
A similar formula holds for the corresponding upper bound for $\mu^0$.

For the RHC data, the point-bound enclosed bounds on $\mu^1$ are of greater width than those on $\mu^0$.
This can be mainly attributed to the fact the treated group size $n_1$ is 2184 and the untreated group size, $n-n_1$, is 3551,
instead of different magnitudes of the loss values from weighted quantile regression.
For simplicity, we examine the prediction performances from weighted outcome quantile regressions
within the treated and untreated groups
for non-regularized calibrated estimation (CAL) when using main-effect working models.

The weighted quantile loss values evaluated at the non-regularized estimators, $\hat\beta_{\mytext{WL},1+}$ and $\hat\beta_{\mytext{WL},1-}$
or $\hat\beta_{\mytext{WL},0+}$ and $\hat\beta_{\mytext{WL},0-}$, are
$0.120$ and $0.154$ in the treated group, after the weight, $(1-\pi(X;\hat\gamma_{\mytext{CAL}})) /  \pi(X;\hat\gamma_{\mytext{CAL}})$,
is normalized to sum to the group size 2184,
and are $0.119$ and $0.149$ in the untreated group, after the weight, $\pi(X;\hat\gamma^{(0)}_{\mytext{CAL}}) / (1- \pi(X;\hat\gamma^{(0)}_{\mytext{CAL}}))$,
is normalized to sum to the group size 3551. These optimized loss values are comparable between the treated and untreated groups,
suggesting similar accuracies in outcome quantile regression. See Figure \ref{fig:rhc-main-quantile} for a comparison of fitted values
within the treatment groups.

To reaffirm the preceding comparison,
outcome predictive modeling can also be examined through weighted outcome logistic regression used in unconfoundedness estimation (Tan 2020b).
The weighted likelihood loss values evaluated at the non-regularized estimators are comparable, $0.485$ and $0.470$, in
the treated and untreated groups respectively.
See Figure \ref{fig:rhc-main-logistic} for a comparison of fitted probabilities
within the treatment groups.

\section{Technical details}

\subsection{Proof of Lemma \ref{lem:sen-model2}}

(i) The result follows because model (\ref{eq:sen-model}) directly leads to (\ref{eq:sen-model2}) with $U^1 = Y^1$.

(ii) Similarly as in the two expressions for $\lambda^*_1(Y^1,X)$ in (\ref{eq:lambda1}), we have by Bayes' rule,
\begin{align*}
 \frac{ \pi^*(X) P(T=0 |U^1,X)}{(1-\pi^*(X)) P(T=1|U^1,X)} = \frac{\dif P(U^1| X,T=0)}{\dif P(U^1 |X,T=1)}.
\end{align*}
Moreover, because $T \perp Y^1 | (X,U^1)$, we have
\begin{align*}
\frac{\dif P(Y^1| X,U^1, T=0)}{\dif P(Y^1 |X, U^1, T=1)} =1.
\end{align*}
Combining the preceding displays shows that under model (\ref{eq:sen-model2}),
\begin{align*}
\Lambda^{-1} \le \frac{ \pi^*(X) P(T=0 |U^1,X)}{(1-\pi^*(X)) P(T=1|U^1,X)} = \frac{\dif P(Y^1, U^1| X,T=0)}{\dif P(Y^1, U^1 |X,T=1)} \le  \Lambda.
\end{align*}
From the first inequality above, we have
$ \Lambda^{-1}  \frac{\dif P}{\dif \nu} (Y^1, U^1 |X,T=1) \le \frac{\dif P}{\dif \nu} (Y^1, U^1| X,T=0)$,
where $\nu$ is a dominating measure. Marginalizing over $U^1$ then gives
$ \Lambda^{-1}  \frac{\dif P}{\dif \nu} (Y^1  |X,T=1) \le \frac{\dif P}{\dif \nu} (Y^1 | X,T=0)$
and hence $ \Lambda^{-1} \le \frac{\dif P(Y^1| X,T=0)}{\dif P(Y^1 |X,T=1)} $. Similarly,
it can be shown that $ \frac{\dif P(Y^1| X,T=0)}{\dif P(Y^1 |X,T=1)} \le \Lambda $. Therefore, model (\ref{eq:sen-model}) holds.

\subsection{Proof of Proposition \ref{pro:DG}}
We show (\ref{eq:mu1+Q}) directly based on Dorn \& Guo (2022). Similarly, (\ref{eq:mu1-Q}) can be shown.
Conceptually, (\ref{eq:mu1-Q}) and (\ref{eq:mu1+Q}) can be expected from (\ref{eq:mu1-Qh}) and (\ref{eq:mu1+Qh}).
From Proposition 2 in Dorn \& Guo (2022), a solution, $\bar \lambda_1(Y^1,X)$, to (\ref{eq:mu1+}) subject to (\ref{eq:lambda1-givenX-a})--(\ref{eq:lambda1-givenX-b})
can be obtained such that
\begin{align*}
\bar\lambda_1(Y^1, X) = \left\{ \begin{array}{ll}
\Lambda & \text{if }  Y^1 > q^*_{1,\tau} (X) ,\\
\Lambda^{-1} & \text{if }  Y^1 < q^*_{1,\tau}(X).
\end{array} \right.
\end{align*}
Then $\mu^{1+}$ can be written as
\begin{align}
\mu^{1+}
& = E \bigg( TY + T \frac{1-\pi^*(X)}{\pi^*(X)}  \nonumber \\
&  \times \left[ \Lambda 1\{ Y > q^*_{1,\tau}(X)\} + \Lambda^{-1} 1\{ Y < q^*_{1,\tau}(X)\}
 + \bar\lambda_1(Y^1,X) 1\{ Y = q^*_{1,\tau}(X)\} \right] Y \bigg). \label{eq:pro-DG-prf1}
\end{align}
Note that $\bar \lambda_1(Y^1,X)$ also satisfies the constraint (\ref{eq:lambda1-givenX-b}) and hence
\begin{align*}
E  \left\{ T \frac{1-\pi^*(X)}{\pi^*(X)} q^*_{1,\tau}(X) \right\} =
E \left\{ T \frac{1-\pi^*(X)}{\pi^*(X)} \bar \lambda_1(Y, X) q^*_{1,\tau}(X)  \right\} ,
\end{align*}
which can be written as
\begin{align}
& E  \left\{ T \frac{1-\pi^*(X)}{\pi^*(X)} q^*_{1,\tau}(X) \right\}
= E \bigg( TY + T \frac{1-\pi^*(X)}{\pi^*(X)} \nonumber  \\
& \times \left[ \Lambda 1\{ Y > q^*_{1,\tau}(X)\} + \Lambda^{-1} 1\{ Y < q^*_{1,\tau}(X)\}
 + \bar\lambda_1(Y^1,X) 1\{ Y = q^*_{1,\tau}(X)\} \right] q^*_{1,\tau}(X) \bigg). \label{eq:pro-DG-prf2}
\end{align}
Combining (\ref{eq:pro-DG-prf1}) and (\ref{eq:pro-DG-prf2}) yields
\begin{align}
\mu^{1+}
& = E \bigg( TY + T \frac{1-\pi^*(X)}{\pi^*(X)} \left[ \Lambda \{Y -q^*_{1,\tau}(X)\}_+ - \Lambda^{-1} \{ q^*_{1,\tau}(X) -Y\}_+ \right] \bigg) \nonumber \\
& \quad + E  \left\{ T \frac{1-\pi^*(X)}{\pi^*(X)} q^*_{1,\tau}(X) \right\}. \label{eq:pro-DG-prf3}
\end{align}
By decomposing $TY$ as $\frac{TY}{\pi^*(X)} - T \frac{1-\pi^*(X)}{\pi^*(X)} Y$
and using $q^*_{1,\tau}(X)-Y = - \{Y -q^*_{1,\tau}(X)\}_+ + \{ q^*_{1,\tau}(X) -Y\}_+$, we obtain
\begin{align}
\mu^{1+}
& = E \bigg( \frac{TY}{\pi^*(X)}
+ T \frac{1-\pi^*(X)}{\pi^*(X)} \left[ \Lambda \{Y -q^*_{1,\tau}(X)\}_+ - \Lambda^{-1} \{ q^*_{1,\tau}(X) -Y\}_+ \right] \bigg) \nonumber \\
& \quad + E  \left[ T \frac{1-\pi^*(X)}{\pi^*(X)} \{q^*_{1,\tau}(X)-Y\} \right] \nonumber \\
& = E \bigg( \frac{TY}{\pi^*(X)} + T \frac{1-\pi^*(X)}{\pi^*(X)}
\left[ (\Lambda-1) \{Y -q^*_{1,\tau}(X)\}_+ + (1- \Lambda^{-1}) \{ q^*_{1,\tau}(X) -Y\}_+ \right] \bigg) , \label{eq:pro-DG-prf4}
\end{align}
which gives the expression (\ref{eq:mu1+Q}) for $\mu^{1+}$.

\subsection{Proof of Proposition \ref{pro:mu1-h}}

We show the result about $\mu^{1+}_h$. Similarly, the result about $\mu^{1-}_h$ can be shown.

Our proof uses the following lemma (a generalized Neyman--Pearson lemma), which can be obtained from Property 2.2 in Francis \& Wright (1969).
For completeness, we give a direct proof of Lemma \ref{lem:gNP}. For any function $\lambda_1=\lambda_1(Y^1,X)$ and $\beta \in \bbR^{1+m}$, define
\begin{align*}
& L (\lambda_1) = E\left\{ TY+ T \frac{1-\pi^*(X)}{\pi^*(X)} \lambda_1 (Y,X) Y  \right\}, \\
& D(\beta) = E \left\{\frac{T}{\pi^*(X)} Y + (\Lambda-\Lambda^{-1})  T \frac{1-\pi^*(X)}{\pi^*(X)} \rho_\tau (Y, q (X;\beta) ) \right\} ,
\end{align*}
where $q (X;\beta) = h^\T(X) \beta$, with dependency on $h$ suppressed in the notation.

\begin{lem} \label{lem:gNP}
Suppose that there exist $\bar\lambda_1 = \bar\lambda_1(Y^1,X)$ and $\bar\beta \in \bbR^{1+m}$ such that
the constraints (\ref{eq:lambda1-givenX-a}) and (\ref{eq:lambda1-marginX}) are satisfied by $\lambda_1= \bar\lambda_1$ and
\begin{align}
\bar\lambda_1(Y^1, X) = \left\{ \begin{array}{ll}
\Lambda & \text{if }  Y^1 > q (X;\bar\beta),\\
\Lambda^{-1} & \text{if }  Y^1 < q (X;\bar\beta).
\end{array} \right.  \label{eq:lem-gNP-split}
\end{align}
 Then for any function $\lambda_1(Y^1,X)$ satisfying  (\ref{eq:lambda1-givenX-a}) and (\ref{eq:lambda1-marginX})
and $\beta \in \bbR^{1+m}$,
\begin{align*}
L(\lambda_1 ) \le L (\bar\lambda_1)  = D(\bar \beta) \le D(\beta).
\end{align*}
In other words, $\bar\lambda_1$ is a solution to the optimization problem in (\ref{eq:mu1+h}) and
$\bar\beta$ is a solution to the optimization problem in (\ref{eq:beta1+}).
\end{lem}

\noindent\textbf{Proof of Lemma \ref{lem:gNP}.}\;
First, we show $L(\lambda_1 ) \le L (\bar\lambda_1)$ for any function $\lambda_1(Y^1,X)$ satisfying  (\ref{eq:lambda1-givenX-a}) and (\ref{eq:lambda1-marginX}).
This follows directly from the inequality
\begin{align}
E \left\{ T \frac{1-\pi^*(X)}{\pi^*(X)} (\bar\lambda_1(Y,X) -\lambda_1(Y,X)) ( Y- q (X;\bar \beta)) \right\} \ge 0, \label{eq:lem-gNP-prf1}
\end{align}
and the equality
\begin{align}
E \left\{ T \frac{1-\pi^*(X)}{\pi^*(X)} (\bar\lambda_1(Y,X) -\lambda_1(Y,X)) q (X;\bar \beta) \right\} = 0 . \label{eq:lem-gNP-prf2}
\end{align}
Inequality (\ref{eq:lem-gNP-prf1}) holds because $\bar\lambda_1$ and $\lambda_1$ satisfy (\ref{eq:lambda1-givenX-a}),
and hence
$\bar\lambda_1(Y,X) = \Lambda \ge \lambda_1(Y,X)$ if  $Y > q (X;\bar\beta)$
and $\bar\lambda_1(Y,X) = \Lambda^{-1} \le \lambda_1(Y,X)$ if  $Y < q (X;\bar\beta)$.
Equality (\ref{eq:lem-gNP-prf2}) holds because both $\bar\lambda_1$ and $\lambda_1$ satisfy (\ref{eq:lambda1-marginX}):
\begin{align*}
& \quad E \left\{ T \frac{1-\pi^*(X)}{\pi^*(X)} \bar\lambda_1(Y,X) q (X;\bar \beta) \right\}
= E \left\{ T \frac{1-\pi^*(X)}{\pi^*(X)} \lambda_1(Y,X) q (X;\bar \beta) \right\}\\
& = E \left\{ T \frac{1-\pi^*(X)}{\pi^*(X)}  q (X;\bar \beta) \right\}.
\end{align*}

Second, we show that $L (\bar\lambda_1)  = D(\bar \beta)$. By similar calculation leading to (\ref{eq:pro-DG-prf3})--(\ref{eq:pro-DG-prf4})
in the proof of Proposition \ref{pro:DG}, we have
\begin{align}
& \quad E\left\{ T \frac{1-\pi^*(X)}{\pi^*(X)} \bar \lambda_1 (Y,X) Y  \right\} \nonumber \\
& = E \bigg( T \frac{1-\pi^*(X)}{\pi^*(X)} \left[ \Lambda \{Y - q (X;\bar\beta) \}_+ - \Lambda^{-1} \{ q (X;\bar\beta)  -Y\}_+ \right] \bigg) \nonumber \\
& \quad + E  \left\{ T \frac{1-\pi^*(X)}{\pi^*(X)} q (X;\bar\beta)  \right\}  \label{eq:lem-gNP-prf3} \\
& = E \bigg( T \frac{1-\pi^*(X)}{\pi^*(X)} Y + T \frac{1-\pi^*(X)}{\pi^*(X)}
\left[ (\Lambda-1) \{Y -q (X;\bar\beta)\}_+ + (1- \Lambda^{-1}) \{ q (X;\bar\beta) -Y\}_+ \right] \bigg) \label{eq:lem-gNP-prf4} ,
\end{align}
which yields $L (\bar\lambda_1)  = D(\bar \beta)$ after adding $E(TY)$.

Third, we show that $L (\bar\lambda_1) \le  D(\beta)$. By similar calculation leading to (\ref{eq:pro-DG-prf3}) as well as (\ref{eq:lem-gNP-prf3})
but splitting the sample space according to the sign of $Y- q (X;\beta)$, we have
\begin{align*}
& \quad E\left\{ T \frac{1-\pi^*(X)}{\pi^*(X)} \bar \lambda_1 (Y,X) Y  \right\} \nonumber \\
& = E \bigg( T \frac{1-\pi^*(X)}{\pi^*(X)} \left[ \bar\lambda_1 (Y,X) \{Y - q (X;\beta) \}_+ -\bar\lambda_1 (Y,X) \{ q (X;\beta)  -Y\}_+ \right] \bigg) \\
& \quad + E  \left\{ T \frac{1-\pi^*(X)}{\pi^*(X)} q (X; \beta)  \right\}  .
\end{align*}
Then using $\Lambda^{-1}\le \bar\lambda_1 (Y,X) \le \Lambda$ in the preceding inequality and rewriting similarly
as from (\ref{eq:pro-DG-prf3}) to (\ref{eq:pro-DG-prf4}) or  from (\ref{eq:lem-gNP-prf3}) to (\ref{eq:lem-gNP-prf4}) gives
\begin{align*}
& \quad E\left\{ T \frac{1-\pi^*(X)}{\pi^*(X)} \bar \lambda_1 (Y,X) Y  \right\} \nonumber \\
& \le E \bigg( T \frac{1-\pi^*(X)}{\pi^*(X)} \left[ \Lambda \{Y - q (X;\beta) \}_+ - \Lambda^{-1} \{ q (X;\beta)  -Y\}_+ \right] \bigg) \\
& \quad + E  \left\{ T \frac{1-\pi^*(X)}{\pi^*(X)} q (X; \beta)  \right\}  \\
& = E \bigg( T \frac{1-\pi^*(X)}{\pi^*(X)} Y + T \frac{1-\pi^*(X)}{\pi^*(X)}
\left[ (\Lambda-1) \{Y -q (X;\beta)\}_+ + (1- \Lambda^{-1}) \{ q (X;\beta) -Y\}_+ \right] \bigg) ,
\end{align*}
which yields $L (\bar\lambda_1)  \le D(\beta)$ after adding $E(TY)$.
{\hfill $\Box$ \vspace{.1in}}

Next we return to the proof of Proposition \ref{pro:mu1-h}, which is not automatically implied by Lemma \ref{lem:gNP}.
If $D(\beta)$ is differentiable in $\beta$, then the result can be obtained from Property 2.8 in Francis \& Wright (1969).
We give a proof using subgradients, to accommodate the case where $D(\beta)$ is non-differentiable at some points.
For simplicity, define a simplified version of $D(\beta)$ as
\begin{align*}
& \tilde D(\beta) = E \left\{ T \frac{1-\pi^*(X)}{\pi^*(X)} \rho_\tau (Y, q (X;\beta) ) \right\} .
\end{align*}
The function $\tilde D(\beta)$ is convex in $\beta$.
It suffices to show that if $\bar\beta$ is a minimizer of $\tilde D(\beta)$, then there exists $\bar\lambda_1 = \bar\lambda_1(Y^1,X)$
such that $(\bar\lambda_1,\bar\beta)$ satisfy the conditions of Lemma \ref{lem:gNP}.

By the convexity of $\tilde D(\beta)$, if $\bar\beta$ is a minimizer of $\tilde D(\beta)$, then
$ 0 \in \partial \tilde D(\bar\beta)$, where $\partial \tilde D(\cdot)$ denotes a subgradient of $\tilde D(\cdot)$.
By Proposition 2.2 in Bertsekas (1973), there exists a (measurable) function $g(Y^1,X)$ such that
\begin{align}
0 = E \left\{  T \frac{1-\pi^*(X)}{\pi^*(X)} g(Y ,X) h(X) \right\},  \label{eq:pro-mu1-h-prf1}
\end{align}
where, based on subgradients of $\rho_\tau(y,q)$ in $q$,
\begin{align*}
g (Y^1, X) \left\{ \begin{array}{ll}
= \tau & \text{if }  Y^1 > q (X;\bar\beta),\\
\in [\tau-1,\tau],  & \text{if }  Y^1 = q (X;\bar\beta),\\
= \tau-1 & \text{if }  Y^1 < q (X;\bar\beta).
\end{array} \right.
\end{align*}
Note that $\tau = \Lambda /(\Lambda+1)= (\Lambda-1) / (\Lambda-\Lambda^{-1})$ and $1-\tau = (1-\Lambda^{-1}) / (\Lambda-\Lambda^{-1})$,
and define $\bar\lambda_1 (Y^1, X) =1+ (\Lambda-\Lambda^{-1}) g(Y^1,X)$. Then
\begin{align*}
\bar\lambda_1 (Y^1, X) \left\{ \begin{array}{ll}
= \Lambda & \text{if }  Y^1 > q (X;\bar\beta),\\
\in [\Lambda^{-1}, \Lambda],  & \text{if }  Y^1 = q (X;\bar\beta),\\
= \Lambda^{-1} & \text{if }  Y^1 < q (X;\bar\beta).
\end{array} \right.
\end{align*}
Moreover, (\ref{eq:pro-mu1-h-prf1}) can be rewritten in terms of $\bar\lambda_1$ as
\begin{align*}
E \left\{  T \frac{1-\pi^*(X)}{\pi^*(X)} \bar\lambda_1(Y ,X) h(X) \right\} =E \left\{  T \frac{1-\pi^*(X)}{\pi^*(X)}  h(X) \right\}.
\end{align*}
Therefore, (\ref{eq:lambda1-givenX-a}) and (\ref{eq:lambda1-marginX}) are satisfied by $\bar\lambda_1$,
and (\ref{eq:lem-gNP-split}) is satisfied by $(\bar\lambda_1, \beta)$, i.e., the conditions of Lemma 2 are satisfied.
This completes the proof of Proposition \ref{pro:mu1-h}.

\subsection{Proofs of Propositions \ref{pro:DG-DR} and \ref{pro:mu1-h-DR}}

We show the result about $\mu^{1+} ( q_{1,\tau} )$ in Proposition \ref{pro:DG-DR} in detail. Similarly, the result
about $\mu^{1-} ( q_{1,1-\tau} )$ can be shown. Moreover, the same proof can be applied to Proposition \ref{pro:mu1-h-DR},
by treating $h^\T \bar\beta_{\mytext{W},1+} (\pi)$ and $h^\T \bar\beta_{\mytext{W},1-} (\pi)$
as $q_{1,\tau}$ and $q_{1,1-\tau}$ respectively.

First, $\mu^{1+} ( q_{1,\tau} ) = E \{ \varphi_{+} (O; \pi^*,\eta_{1+}, q_{1,\tau}) \}$, because
$\mu^{1+} ( q_{1,\tau} ) = E \{ \varphi_{+,\mytext{IPW}} (O; \pi^*, q_{1,\tau}) \}$ by definition and
\begin{align*}
& \quad E \left[ \left\{ \frac{T}{\pi^*(X)} -1\right\} E_{\eta_1} \{ \tilde Y_{+} (q_{1,\tau})| T=1,X \}  \right] \\
& = E  \left[ E \left\{ \frac{T}{\pi^*(X)} -1 \Big|X  \right\} E_{\eta_1} \{ \tilde Y_{+} (q_{1,\tau})| T=1,X \}  \right] =0.
\end{align*}
by the law of iterated expectations.

Second, we show that $\mu^{1+} ( q_{1,\tau} ) = E \{ \varphi_{+} (O; \pi,\eta^*_1, q_{1,\tau}) \}$. Note that
$\varphi_{+} (O; \pi,\eta^*_1, q_{1,\tau}) $ can be rewritten as
\begin{align*}
\varphi_{+} (O; \pi, \eta^*_1, q_{1,\tau})
& = - (\Lambda-\Lambda^{-1}) T \rho_\tau (Y, q_{1,\tau} (X)) + E  \{ \tilde Y_{+} (q_{1,\tau})| T=1,X \} \\
& \quad + \frac{T}{\pi(X)} \left[ \tilde Y_{+} (q_{1,\tau}) - E \{ \tilde Y_{+} (q_{1,\tau})| T=1,X \} \right].
\end{align*}
The expectation of the third term (which is the only term depending on $\pi$) on the right-hand side above is 0 by the law of iterated expectations:
\begin{align*}
& \quad E \left( \frac{T}{\pi(X)} \left[ \tilde Y_{+} (q_{1,\tau}) - E \{ \tilde Y_{+} (q_{1,\tau})| T=1,X \} \right] \right) \\
& = E \left( \frac{\pi^*(X)}{\pi(X)} E \left[ \tilde Y_{+} (q_{1,\tau}) - E \{ \tilde Y_{+} (q_{1,\tau})| T=1,X \} \Big| T=1, X\right] \right) =0.
\end{align*}
Similarly, replacing $T/\pi(X)$ by $T / \pi^*(X)$ in the above display shows that
\begin{align*}
& \quad E \left( \frac{T}{\pi^*(X)} \left[ \tilde Y_{+} (q_{1,\tau}) - E \{ \tilde Y_{+} (q_{1,\tau})| T=1,X \} \right] \right) =0.
\end{align*}
Hence $E \{ \varphi_{+} (O; \pi,\eta^*_1, q_{1,\tau})\} $ is identical to
 $E \{ \varphi_{+} (O; \pi^*,\eta^*_1, q_{1,\tau})\}= 0$.

\subsection{Inside Theorem~\ref{thm:gam-conv}}

The following result is taken from Tan (2020b), Lemma 1.

\begin{lem} \label{lem:prob-gam}
(i) Denote by $\Omega_{01}$ the event that
\begin{align*}
\sup_{j=0,1,\ldots,p} \left| \tilde E \left[ \left\{-T \me^{-f^\T(X) \bar\gamma_{\mytext{CAL}} } + (1-T) \right\} f_j(X) \right] \right| \le C_{01} \lambda_0 .
\end{align*}
where $C_{01} = \sqrt{2} (\me^{-B_0}+1) C_0$.
Under Assumption~\ref{ass:gam}(i)--(ii), $P(\Omega_{01}) \ge 1- 2\epsilon$.\\
(ii) Denote by $\Omega_{02}$ the event that
\begin{align*}
\sup_{j,k=0,1,\ldots,p} | (\tilde \Sigma_f)_{jk} - (\Sigma_f)_{jk} | \le C_{02} \lambda_0,
\end{align*}
where $C_{02} = 4 \me^{-B_0} C_0^2$.
Under Assumption~\ref{ass:gam}(i)--(ii), $P( \Omega_{02} ) \ge 1- 2 \epsilon^2$.
\end{lem}

The error bounds stated in Theorem \ref{thm:gam-conv}
are valid in the event $\Omega_0 = \Omega_{01}\cap \Omega_{11}$, with probability at least $1-4\epsilon$,
by the proof of Corollary 2 in Tan (2020a).

\subsection{Proof of Theorem \ref{thm:beta-conv}} \label{sec:beta-proof}

Suppose that Assumptions \ref{ass:gam}--\ref{ass:beta} are satisfied. Then
Theorem \ref{thm:beta-conv} follows from Lemma \ref{lem:beta-global} and Corollary \ref{cor:beta-global}, which depend on
probability Lemmas \ref{lem:emp-concen}--\ref{lem:Sigma-h} and analytical Lemmas \ref{lem:depend-gam}--\ref{lem:beta-local}.

For simplicity, write
$\hat\gamma=\hat\gamma_{\mytext{RCAL}}$,
$\bar\gamma=\bar\gamma_{\mytext{CAL}}$,
$\hat\beta=\hat\beta_{\mytext{RWL},1+}$, and
$\bar\beta=\bar\beta_{\mytext{WL},1+}$ in this and next two sections.
In addition, our proof is applicable to a general loss function,
\begin{align*}
\ell (\beta; \gamma) &= \tilde  E \left\{ T w(X;\gamma) \rho (Y, h^\T(X) \beta) \right\}, 
\end{align*}
where $\rho(y,u)$ is assumed to be convex and $L$-Lipschitz in $u$. But $\rho(y,u)$ may not be twice differentiable in $u$.
For $\rho=\rho_\tau$ from weighted $\tau$-quantile regression, $L=1$.

Denote the expected loss function for $\beta$ as
$\bar \ell (\beta; \gamma) =  E \{ \ell(\beta;\gamma)\} = E \{ T w(X; \gamma) \rho (Y,h^\T\beta)\}$.
Moreover, denote
\begin{align*}
Q (\beta,\bar\beta; \bar\gamma) & = (\beta-\bar\beta)^\T \Sigma_h (\beta-\bar\beta)= E \left\{ T w(X; \bar\gamma) | h^\T(X) (\beta-\bar\beta) |^2 \right\}, \\
D (\beta,\bar\beta; \bar\gamma) &= \ell(\beta;\bar\gamma) -\ell(\bar\beta; \bar\gamma) - \{ \bar \ell(\beta;\bar\gamma) - \bar\ell(\bar\beta; \bar\gamma) \} \\
& = (\tilde E-E) \left[ T w(X; \bar\gamma) \{\rho (Y,h^\T\beta)-\rho (Y,h^\T \bar\beta) \} \right] ,
\end{align*}
where $(\tilde E-E)(Z)$ denotes $n^{-1}\sum_{i=1}^n \{Z_i - E(Z)\}$ for a variable $Z$ that is a function of $(Y,T,X)$.

First, we provide two probability lemmas for controlling probabilities.

\begin{lem} \label{lem:emp-concen}
For any $r>0$, denote by $\Omega_{11}=\Omega_{11}(r)$ the event that
\begin{align*}
& \quad \sup_{\|\beta-\bar\beta\|_1 \le r} | D (\beta,\bar\beta; \bar\gamma) |
= \sup_{\|\beta-\bar\beta\|_1 \le r} \left| (\tilde E - E) \left[  T w(X;\bar\gamma) \left\{ \rho (Y,h^\T\beta) - \rho (Y, h^\T \bar\beta) \right\} \right] \right| \\
& \le r C_{11} \lambda_0,
\end{align*}
where $C_{11}=8 L \me^{-B_0} C_1$. Under Assumptions~\ref{ass:gam}(ii) and \ref{ass:beta}(i), $P(\Omega_{11} ) \ge 1-2\epsilon$.
\end{lem}

\begin{prf} \label{prf:lem-emp-concen}
For $\|\beta - \bar\beta\|_1 \le r$, direct calculation shows that
\begin{align}
& \quad T w(X;\bar\gamma)| \rho (Y,h^\T\beta) - \rho (Y, h^\T \bar\beta) | \nonumber \\
& \le L T w(X;\bar\gamma) | h^\T(X) (\beta - \bar\beta) |
 \le L T w(X;\bar\gamma) \|h(X)\|_\infty \|\beta - \bar\beta\|_1
 \le L \me^{-B_0} C_1 r . \label{eq:lem-emp-concen-prf1}
\end{align}
By the symmetrization and contraction theorems (e.g., Buhlmann \& van de Geer 2011, Theorems 14.3 and 14.4), we have
\begin{align*}
& \quad E \sup_{\|\beta-\bar\beta\|_1 \le r} \left| (\tilde E - E) \left[  T w(X;\bar\gamma) \left\{ \rho (Y,h^\T\beta) - \rho (Y, h^\T \bar\beta) \right\} \right] \right| \\
& \le  2 E \sup_{\|\beta-\bar\beta\|_1 \le r} \left| \frac{1}{n} \sum_{i=1}^n \sigma_i \left[ T_i w(X_i;\bar\gamma) \left\{ \rho (Y_i,h^\T(X_i)\beta) -
\rho (Y_i, h^\T(X_i) \bar\beta) \right\} \right] \right| \\
& \le  4 L E \sup_{\|\beta-\bar\beta\|_1 \le r} \left| \frac{1}{n} \sum_{i=1}^n \sigma_i T_i w(X_i;\bar\gamma) h^\T(X_i) (\beta - \bar\beta) \right| \\
& \le  4 L r \times \sup_{j=0,1,\ldots,m} \left| \frac{1}{n} \sum_{i=1}^n \sigma_i T_i w(X_i;\bar\gamma) h_j(X_i) \right|,
\end{align*}
where $(\sigma_1,\ldots,\sigma_n)$ are independent Rademacher variables with $P(\sigma_i=\pm 1) =1/2$,
independently of $\{(Y_i,T_i,X_i): i=1,\ldots,n\}$. By applying Hoeffding's moment inequality (Buhlmann \& van de Geer 2011, Lemma 14.14) to the mean-zero variables
$\sigma_i T_i w(X_i;\bar\gamma) h_j(X_i)$, we find from the preceding inequality
\begin{align*}
& \quad E \sup_{\|\beta-\bar\beta\|_1 \le r} \left| (\tilde E - E) \left[  T w(X;\bar\gamma) \left\{ \rho (Y,h^\T\beta) - \rho (Y, h^\T \bar\beta) \right\} \right] \right| \\
& \le  4 L r \times \me^{-B_0} C_1 \sqrt{\frac{2\log(2+2p)}{n}}.
\end{align*}
By applying Massart's inequality (Buhlmann \& van de Geer 2001, Theorem 14.2; mean-zero condition not needed) with the bound (\ref{eq:lem-emp-concen-prf1}), we have with probability at least $1- 2 \epsilon$,
\begin{align*}
& \quad \sup_{\|\beta-\bar\beta\|_1 \le r} \left| (\tilde E - E) \left[  T w(X;\bar\gamma) \left\{ \rho (Y,h^\T\beta) - \rho (Y, h^\T \bar\beta) \right\} \right] \right| \\
& \le  L \me^{-B_0} C_1 r \left\{ 4\sqrt{\frac{2\log(2+2p)}{n}} + \sqrt{ \frac{8\log(1/(2\epsilon))}{n} } \right\}\\
& \le  L \me^{-B_0} C_1 r \left\{ 4\sqrt{\frac{2\log(2+2p)}{n}} + 2 \sqrt{ \frac{8\log(1/(2\epsilon))}{n} } \right\}\\
& \le  8 L \me^{-B_0} C_1 r \sqrt{\frac{\log((1+p)/\epsilon) }{n}}  ,
\end{align*}
where the last inequality uses $\sqrt{a} + \sqrt{b} \le \sqrt{ 2(a + b)}$.
\end{prf}

Denote  $\Sigma_h = E [ T w(X; \bar\gamma) h(X) h^\T (X)] $,
and $\tilde \Sigma_h = \tilde E [ T w(X; \bar\gamma) h(X) h^\T (X)] $,
the sample version of $\Sigma_h$. The following result is similar to Tan (2020), Lemma 1(ii).

\begin{lem} \label{lem:Sigma-h}
Denote by $\Omega_{12}$ the event that
\begin{align*}
\sup_{j,k=0,1,\ldots,m} | (\tilde \Sigma_h)_{jk} - (\Sigma_h)_{jk} | \le C_{12} \lambda_0 ,
\end{align*}
where $C_{12} = 4 \me^{-B_0} C_1^2$.
Under Assumptions~\ref{ass:gam}(ii) and \ref{ass:beta}(i), $P( \Omega_{12} ) \ge 1- 2 \epsilon^2$.
\end{lem}

Next, we provide several analytical lemmas within various events.

\begin{lem} \label{lem:depend-gam}
In the event $\Omega_0 \cap \Omega_{12}$, we have
\begin{align*}
& \quad \left|  \ell (\beta; \hat\gamma) - \ell (\bar\beta; \hat\gamma) -\{ \ell (\beta; \bar\gamma) - \ell (\bar\beta; \bar\gamma)\} \right| \\
& =\left| \tilde E \left\{ T ( w(X;\hat\gamma)-w(X;\bar\gamma)) (\rho(Y, h^\T\beta)- \rho(Y, h^\T \bar\beta) ) \right\} \right| \\
& \le (M_{01} |S_{\bar\gamma}|\lambda_0^2 )^{1/2} Q^{1/2} (\beta,\bar\beta; \bar\gamma) + M_{02} \lambda_0  \|\beta-\bar\beta\|_1,
\end{align*}
where $M_{01} = L^2 \me^{2 C_0 M_0\varrho_0} M_0$, $M_{02} =L \me^{C_0 M_0\varrho_0} (M_0 C_{12} \varrho_0)^{1/2}$,
and $C_{12}$ is defined in Lemma \ref{lem:Sigma-h}.
\end{lem}

\begin{prf}
By the mean value theorem for some $u\in [0,1]$,
\begin{align*}
& \quad | w(X;\hat\gamma)-w(X;\bar\gamma)| = \left| \me^{-f^\T(X) \hat\gamma} - \me^{-f^\T (X) \bar\gamma} \right| \\
& = \left| \me^{-u f^\T (X) \hat\gamma - (1-u)f^\T (X) \bar\gamma } f^\T(X) (\hat\gamma -\bar\gamma) \right| \\
&\le \me^{| f^\T(X) (\hat\gamma -\bar\gamma) |} \me^{-f^\T (X) \bar\gamma} | f^\T(X) (\hat\gamma -\bar\gamma) | . 
\end{align*}
In the event $\Omega_0$, we have
$\| \hat\gamma - \bar\gamma \|_1 \le  M_0 |S_{\bar\gamma}| \lambda_0 \le  M_0 \varrho_0$ and
$ | f^\T(X) (\hat\gamma -\bar\gamma) | \le C_0 M_0 \varrho_0$.
Then by the Lipschitz property of $\rho(\cdot)$ and the Cauchy--Schwartz inequality, we have
\begin{align}
& \quad \left| \tilde E \left\{ T ( w(X;\hat\gamma)-w(X;\bar\gamma)) (\rho(Y, h^\T\beta)- \rho(Y, h^\T \bar\beta) ) \right\} \right| \nonumber \\
& \le L \me^{C_0 M_0\varrho_0} \tilde E \left\{ T \me^{-f^\T (X) \bar\gamma} | f^\T(X) (\hat\gamma -\bar\gamma) | \cdot | h^\T(X) (\beta-\bar\beta) | \right\} \nonumber \\
& \le L \me^{C_0 M_0\varrho_0} \tilde E^{1/2} \left\{ T \me^{-f^\T (X) \bar\gamma} | f^\T(X) (\hat\gamma -\bar\gamma) |^2 \right\} \tilde E^{1/2}
\left\{ T \me^{-f^\T (X) \bar\gamma}  | h^\T(X) (\beta-\bar\beta) |^2 \right\} . \label{eq:lem-depend-gam-prf1}
\end{align}
Moreover, in the event $\Omega_0$, we have
\begin{align*}
\tilde E \left\{ T \me^{-f^\T (X) \bar\gamma} | f^\T(X) (\hat\gamma -\bar\gamma) |^2 \right\} \le M_0 |S_{\bar\gamma}| \lambda_0^2 .
\end{align*}
In the event $\Omega_{12}$ from Lemma \ref{lem:Sigma-h}, we have
\begin{align*}
& \quad  \tilde E \left\{ T \me^{-f^\T (X) \bar\gamma}  | h^\T(X) (\beta-\bar\beta) |^2 \right\} \\
& \le  E \left\{ T \me^{-f^\T (X) \bar\gamma}  | h^\T(X) (\beta-\bar\beta) |^2 \right\}  + C_{12} \lambda_0 \|\beta-\bar\beta\|_1^2.
\end{align*}
Combining the preceding three displays yields
\begin{align*}
& \quad \left| \tilde E \left\{ T ( w(X;\hat\gamma)-w(X;\bar\gamma)) (\rho(Y, h^\T\beta)- \rho(Y, h^\T \bar\beta) ) \right\} \right| \\
& \le L \me^{C_0 M_0\varrho_0} (M_0 |S_{\bar\gamma}|\lambda_0^2 )^{1/2} \left\{ Q^{1/2} (\beta,\bar\beta; \bar\gamma) + (C_{12}\lambda_0)^{1/2} \|\beta-\bar\beta\|_1 \right\},
\end{align*}
which leads to the desired results with $|S_{\bar\gamma}| \lambda_0 \le \varrho_0$.
\end{prf}

\begin{lem} \label{lem:basic-ineq-b}
In the event $\Omega_0\cap\Omega_{12}$, for any $\beta$ satisfying
\begin{align}
\ell (\beta; \hat\gamma) + A_1 \lambda_0 \|\beta_{1:p}\|_1 \le \ell (\bar\beta; \hat\gamma) + A_1 \lambda_0 \| \bar\beta_{1:p}\|_1, \label{eq:basic-ineq-b}
\end{align}
we have
\begin{align}
& \quad \bar \ell(\beta;\bar\gamma) - \bar\ell(\bar\beta; \bar\gamma) + (A_1-M_{02}) \lambda_0 \|\beta-\bar\beta\|_1 \nonumber \\
& \le 2 A_1 \lambda_0 \sum_{j\in \Sigma_h} |\beta_j -\bar\beta_j| +
|D (\beta,\bar\beta; \bar\gamma)|  + (M_{01} |S_{\bar\gamma}|\lambda_0^2 )^{1/2} Q^{1/2} (\beta,\bar\beta; \bar\gamma), \label{eq:basic-ineq-b2}
\end{align}
where $(M_{01},M_{02})$ are defined in Lemma \ref{lem:depend-gam}.
\end{lem}

\begin{prf}
Equation (\ref{eq:basic-ineq-b}) can be rewritten as
\begin{align*}
& \quad \ell (\beta; \bar\gamma) - \ell (\bar\beta; \bar\gamma) + A_1 \lambda_0 \|\beta_{1:p}\|_1 \\
& \le A_1 \lambda_0 \| \bar\beta_{1:p}\|_1  -
\left[ \ell (\beta; \hat\gamma) - \ell (\bar\beta; \hat\gamma) -\{ \ell (\beta; \bar\gamma) - \ell (\bar\beta; \bar\gamma)\} \right].
\end{align*}
In the event $\Omega_0\cap\Omega_{12}$, applying Lemma \ref{lem:depend-gam} yields
\begin{align*}
& \quad \ell (\beta; \bar\gamma) - \ell (\bar\beta; \bar\gamma) + A_1 \lambda_0 \|\beta_{1:p}\|_1 \\
& \le A_1 \lambda_0 \| \bar\beta_{1:p}\|_1 +
 (M_{01} |S_{\bar\gamma}|\lambda_0^2 )^{1/2} Q^{1/2} (\beta,\bar\beta; \bar\gamma) + M_{02} \lambda_0  \|\beta-\bar\beta\|_1 ,
\end{align*}
which can be rearranged as
\begin{align*}
& \quad \bar\ell (\beta; \bar\gamma) - \bar\ell (\bar\beta; \bar\gamma) + A_1 \lambda_0 \|\beta_{1:p}\|_1 \\
& \le A_1 \lambda_0 \| \bar\beta_{1:p}\|_1 +
 (M_{01} |S_{\bar\gamma}|\lambda_0^2 )^{1/2} Q^{1/2} (\beta,\bar\beta; \bar\gamma) + M_{02} \lambda_0  \|\beta-\bar\beta\|_1 - D (\beta,\bar\beta; \bar\gamma),
\end{align*}
where $D (\beta,\bar\beta; \bar\gamma) =
\ell (\beta; \bar\gamma) - \ell (\bar\beta; \bar\gamma) - \{\bar\ell (\beta; \bar\gamma) - \bar\ell (\bar\beta; \bar\gamma)\} $.
Applying to the preceding inequality the identity $| \beta_j | = |\bar\beta_j | + |\beta_j - \bar\beta_j |$ for $j\not\in \Sigma_h$
and the triangle inequality
\begin{align*}
| \beta_j | & \ge |\bar\beta_j| - |\beta_j - \bar\beta_j |  , \quad j  \in \Sigma_h \backslash \{0\},
\end{align*}
and simple manipulation gives
\begin{align*}
& \quad \bar\ell (\beta; \bar\gamma) - \bar\ell (\bar\beta; \bar\gamma) + A_1 \lambda_0 \sum_{j\not\in \Sigma_h} |\beta_j -\bar\beta_j|  \\
& \le A_1 \lambda_0 \sum_{j\in \Sigma_h\backslash \{0\} } |\beta_j -\bar\beta_j| +
 (M_{01} |S_{\bar\gamma}|\lambda_0^2 )^{1/2} Q^{1/2} (\beta,\bar\beta; \bar\gamma) + M_{02} \lambda_0  \|\beta-\bar\beta\|_1 - D (\beta,\bar\beta; \bar\gamma) \\
& \le A_1 \lambda_0 \sum_{j\in \Sigma_h } |\beta_j -\bar\beta_j| +
 (M_{01} |S_{\bar\gamma}|\lambda_0^2 )^{1/2} Q^{1/2} (\beta,\bar\beta; \bar\gamma) + M_{02} \lambda_0  \|\beta-\bar\beta\|_1 - D (\beta,\bar\beta; \bar\gamma).
\end{align*}
The result follows by adding $A_1 \lambda_0 \sum_{j\in \Sigma_h} |\beta_j -\bar\beta_j|$ to both sides above.
\end{prf}

\begin{lem} \label{lem:beta-local}
Suppose that $A_1 > (M_{02}+C_{11}) (\xi_1+1)/(\xi_1-1)$ with $(C_{11}, M_{02})$ from Lemmas \ref{lem:emp-concen} and \ref{lem:depend-gam}.
Fix $0 < r \le \zeta_1$.
In the event $\Omega_0\cap\Omega_{11}(r) \cap\Omega_{12}$, for
any $\beta$ satisfying (\ref{eq:basic-ineq-b}) and $\| \hat\beta-\bar\beta\|_1 =r$, we have
\begin{align}
\bar \ell(\beta;\bar\gamma) - \bar\ell(\bar\beta; \bar\gamma) >0 \label{eq:beta-strict}
\end{align}
and
\begin{align}
& \quad  \bar \ell(\beta;\bar\gamma) - \bar\ell(\bar\beta; \bar\gamma) + A_{11} \lambda_0 \|\beta-\bar\beta\|_1 \nonumber \\
& \le \kappa_1^{-1} \xi_{11}^{-2} M_{01} |S_{\bar\gamma}|\lambda_0^2 +  \kappa_1^{-1} \nu_1^{-2} \xi_{12}^2  |S_{\bar\beta}|\lambda_0^2 , \label{eq:beta-local}
\end{align}
where  $A_{11} = A_1-M_{02}- C_{11}$, $\xi_{11} = 1- 2 A_1 / \{(\xi_1+1) A_{11} \} \in (0,1)$,
$\xi_{12} = (\xi_1+1) A_{11}$, and $M_{01}$ is defined in Lemma \ref{lem:depend-gam}.
\end{lem}

\begin{prf}
With $\|\beta - \bar\beta\|_1 = r$, we have in the event $\Omega_{11}(r)$ from Lemma \ref{lem:emp-concen},
\begin{align*}
|D (\beta,\bar\beta; \bar\gamma)|  \le r C_{11} \lambda_0.
\end{align*}
Moreover, with $\beta$ satisfying (\ref{eq:basic-ineq-b}), we apply Lemma \ref{lem:basic-ineq-b} and obtain from (\ref{eq:basic-ineq-b2})
that in the event $\Omega_0\cap\Omega_{11}(r) \cap\Omega_{12}$,
\begin{align}
& \quad \bar \ell(\beta;\bar\gamma) - \bar\ell(\bar\beta; \bar\gamma) + (A_1-M_{02}) \lambda_0 \|\beta-\bar\beta\|_1 \nonumber \\
& \le 2 A_1 \lambda_0 \sum_{j\in S_{\bar\beta}} |\beta_j -\bar\beta_j| + r C_{11} \lambda_0
+ (M_{01} |S_{\bar\gamma}|\lambda_0^2 )^{1/2} Q^{1/2} (\beta,\bar\beta; \bar\gamma) \label{eq:basic-ineq-b3} \\
& = 2 A_1 \lambda_0 \sum_{j\in S_{\bar\beta}} |\beta_j -\bar\beta_j| + C_{11} \lambda_0 \|\beta-\bar\beta\|_1
+ (M_{01} |S_{\bar\gamma}|\lambda_0^2 )^{1/2} Q^{1/2} (\beta,\bar\beta; \bar\gamma), \nonumber
\end{align}
where the last step uses $\|\beta -\bar\beta\|_1=r$. Rearranging the preceding inequality gives
\begin{align*}
\Delta & := \bar \ell(\beta;\bar\gamma) - \bar\ell(\bar\beta; \bar\gamma) + A_{11} \lambda_0 \|\beta-\bar\beta\|_1 \\
& \le 2 A_1 \lambda_0 \sum_{j\in S_{\bar\beta}} |\beta_j -\bar\beta_j|
+ (M_{01} |S_{\bar\gamma}|\lambda_0^2 )^{1/2} Q^{1/2} (\beta,\bar\beta; \bar\gamma),
\end{align*}
where $A_{11} = A_1-M_{02}- C_{11}$. This leads to two possible cases: either
\begin{align}
\xi_{11} \Delta \le (M_{01} |S_{\bar\gamma}|\lambda_0^2 )^{1/2} Q^{1/2} (\beta,\bar\beta; \bar\gamma), \label{eq:beta-case1}
\end{align}
or $(1-\xi_{11}) \Delta \le 2 A_1 \lambda_0 \sum_{j\in S_{\bar\beta}} |\beta_j -\bar\beta_j|$, that is,
\begin{align}
 \Delta \le (\xi_1+1) A_{11} \lambda_0\sum_{j\in S_{\bar\beta}} |\beta_j -\bar\beta_j|=
 \xi_{12} \lambda_0\sum_{j\in S_{\bar\beta}} |\beta_j -\bar\beta_j|, \label{eq:beta-case2}
\end{align}
where $\xi_{11} = 1- 2 A_1 / \{(\xi_1+1) A_{11} \} \in (0,1)$ because
$ A_1 > (M_{02}+ C_{11}) (\xi_1+1)/(\xi_1-1)$,
and $\xi_{12} = (\xi_1+1) A_{11}$. We deal with the two cases separately as follows.

In the case of (\ref{eq:beta-case1}), we have
\begin{align*}
\bar \ell(\beta;\bar\gamma) - \bar\ell(\bar\beta; \bar\gamma) \le \Delta
\le \xi_{11}^{-1} (M_{01} |S_{\bar\gamma}|\lambda_0^2 )^{1/2} Q^{1/2} (\beta,\bar\beta; \bar\gamma),
\end{align*}
which, with $\bar \ell(\beta;\bar\gamma) - \bar\ell(\bar\beta; \bar\gamma) \ge \kappa_1 Q (\beta,\bar\beta;\bar\gamma)$
for $ \|\beta -\bar\beta\|_1 =  r \le \zeta_1$ under the margin condition, Assumption \ref{ass:beta}(ii), implies that
$Q (\beta,\bar\beta; \bar\gamma) \le (\kappa_1 \xi_{11})^{-2} (M_{01} |S_{\bar\gamma}|\lambda_0^2 )$.
Returning to (\ref{eq:beta-case1}) shows
\begin{align}
\Delta \le \kappa_1^{-1} \xi_{11}^{-2} (M_{01} |S_{\bar\gamma}|\lambda_0^2 ). \label{eq:beta-case1b}
\end{align}

In the case of (\ref{eq:beta-case2}), we have
\begin{align}
\sum_{j\not\in S_{\bar\beta}} |\beta_j -\bar\beta_j|\le \xi_1 \sum_{j\in S_{\bar\beta}} |\beta_j -\bar\beta_j|, \label{eq:beta-cone}
\end{align}
which, under the compatibility condition, Assumption \ref{ass:beta}(iii), implies
\begin{align}
\sum_{j\in S_{\bar\beta}} |\beta_j -\bar\beta_j| \le \nu_1^{-1} |S_{\bar\beta}|^{1/2} Q^{1/2} (\beta,\bar\beta; \bar\gamma). \label{eq:beta-cone2}
\end{align}
Substituting this into (\ref{eq:beta-case2}) yields
\begin{align}
\bar \ell(\beta;\bar\gamma) - \bar\ell(\bar\beta; \bar\gamma) \le \Delta
\le (\xi_{12} \lambda_0 ) \nu_1^{-1} |S_{\bar\beta}|^{1/2} Q^{1/2} (\beta,\bar\beta; \bar\gamma), \label{eq:beta-case2b}
\end{align}
which, with $\bar \ell(\beta;\bar\gamma) - \bar\ell(\bar\beta; \bar\gamma) \ge \kappa_1 Q (\beta,\bar\beta;\bar\gamma)$
for $ \|\beta -\bar\beta\|_1 = r \le \zeta_1 $ under Assumption \ref{ass:beta}(ii), implies that
$Q (\beta,\bar\beta; \bar\gamma) \le \kappa_1^{-2} \nu_1^{-2} \xi_{12}^2  |S_{\bar\beta}|\lambda_0^2 $.
Returning to (\ref{eq:beta-case2b}) then shows
\begin{align}
\Delta \le \kappa_1^{-1} \nu_1^{-2} \xi_{12}^2  |S_{\bar\beta}|\lambda_0^2 . \label{eq:beta-case2c}
\end{align}
Combining (\ref{eq:beta-case1b}) and (\ref{eq:beta-case2c}) leads to inequality (\ref{eq:beta-local}).

To show (\ref{eq:beta-strict}), suppose that $\bar \ell(\beta;\bar\gamma) - \bar\ell(\bar\beta; \bar\gamma)=0 $. Then $Q(\beta,\bar\beta; \bar\gamma) =0$ by the margin condition.
We return to the two cases: (\ref{eq:beta-case1}) or (\ref{eq:beta-case2}).
In the case of (\ref{eq:beta-case1}), this implies that $\Delta=0$ and hence $\| \beta-\hat\beta \|_1=0$, a contradiction with $\| \beta-\hat\beta \|_1= r >0$.
In the case of (\ref{eq:beta-case2}), inequality (\ref{eq:beta-cone2}) implies that $\sum_{j\in S_{\bar\beta}} |\beta_j -\bar\beta_j|=0$,
from which (\ref{eq:beta-cone}) implies that $\sum_{j\not\in S_{\bar\beta}} |\beta_j -\bar\beta_j|=0$, and hence
we also have $\| \beta-\hat\beta \|_1=0$, a contradiction. Therefore, $\bar \ell(\beta;\bar\gamma) - \bar\ell(\bar\beta; \bar\gamma) >0 $
for any $\beta$ satisfying (\ref{eq:basic-ineq-b}) and $\|\beta-\bar\beta\|_1=r$.
\end{prf}

\begin{lem} \label{lem:beta-global}
Suppose that $A_1 > (M_{02}+C_{11}) (\xi_1+1)/(\xi_1-1)$  with $(C_{11}, M_{02})$ from Lemmas \ref{lem:emp-concen} and \ref{lem:depend-gam}. Take
\begin{align*}
r &= A_{11}^{-1} \left( \kappa_1^{-1} \xi_{11}^{-2} M_{01} |S_{\bar\gamma}|\lambda_0 +  \kappa_1^{-1} \nu_1^{-2} \xi_{12}^2  |S_{\bar\beta}|\lambda_0 \right),
\end{align*}
where $(A_{11}, M_{01}, \xi_{11}, \xi_{12})$ are defined as in Lemma \ref{lem:beta-local}.
Then in the event $\Omega_0\cap\Omega_{11}(r) \cap\Omega_{12}$,
\begin{align*}
& \quad  \bar \ell(\hat\beta;\bar\gamma) - \bar\ell(\bar\beta; \bar\gamma) + A_{11} \lambda_0 \|\hat\beta-\bar\beta\|_1 \nonumber \\
& \le \frac{3\xi_1+1}{2}
\left( \kappa_1^{-1} \xi_{11}^{-2} M_{01} |S_{\bar\gamma}|\lambda_0^2 +  \kappa_1^{-1} \nu_1^{-2} \xi_{12}^2  |S_{\bar\beta}|\lambda_0^2 \right).
\end{align*}
\end{lem}

\begin{prf}
By Assumptions \ref{ass:gam}(iv) and \ref{ass:beta}(iv),
\begin{align*}
r \le A_{11}^{-1} \left( \kappa_1^{-1} \xi_{11}^{-2} M_{01} \varrho_0 +  \kappa_1^{-1} \nu_1^{-2} \xi_{12}^2 \varrho_1 \right) \le \zeta_1.
\end{align*}
Consider two possible cases, $\|\hat\beta -\bar\beta\|_1 > r$ or $\le r$, both in the event $\Omega_0\cap\Omega_{11}(r) \cap\Omega_{12}$.

In the case of $\|\hat\beta -\bar\beta\|_1 > r$, let
$\tilde\beta = \bar\beta + \{r / \| \hat\beta-\bar\beta\|_1\} (\hat\beta-\bar\beta)$, satisfying $\|\tilde\beta-\bar\beta\|_1=r$.
Moreover, $\hat\beta$ by definition satisfies (\ref{eq:basic-ineq-b}), which, by the convexity of
$\ell (\beta; \hat\gamma) + A_1 \lambda_0 \|\beta_{1:p}\|_1$, implies that
$\tilde\beta$ also satisfies (\ref{eq:basic-ineq-b}).
Then we apply Lemma \ref{lem:beta-local} with $\beta=\tilde\beta$
and obtain from (\ref{eq:beta-strict}) and (\ref{eq:beta-local}) that in the event $\Omega_0\cap\Omega_{11}(r) \cap\Omega_{12}$,
\begin{align*}
\|\tilde\beta-\bar\beta\|_1 < A_{11}^{-1} \left( \kappa_1^{-1} \xi_{11}^{-2} M_{01} |S_{\bar\gamma}|\lambda_0 +
\kappa_1^{-1} \nu_1^{-2} \xi_{12}^2  |S_{\bar\beta}|\lambda_0 \right) = r,
\end{align*}
which contradicts $\|\tilde\beta-\bar\beta\|_1 = r$. In other words, the case of $\|\hat\beta -\bar\beta\|_1 > r$ does not hold in the event
$\Omega_0\cap\Omega_{11}(r) \cap\Omega_{12}$,

In the case of $\|\hat\beta -\bar\beta\|_1 \le r$, we apply Lemmas \ref{lem:emp-concen} and \ref{lem:basic-ineq-b} with $\beta = \hat\beta$
and obtain that in the event $\Omega_0\cap\Omega_{11}(r) \cap\Omega_{12}$,
\begin{align*}
\Delta & := \bar \ell(\hat\beta;\bar\gamma) - \bar\ell(\bar\beta; \bar\gamma) + (A_1-M_{02}) \lambda_0 \|\hat\beta-\bar\beta\|_1 \nonumber \\
& \le 2 A_1 \lambda_0 \sum_{j\in S_{\bar\beta}} |\hat\beta_j -\bar\beta_j| + r C_{11} \lambda_0
+ (M_{01} |S_{\bar\gamma}|\lambda_0^2 )^{1/2} Q^{1/2} (\hat\beta,\bar\beta; \bar\gamma),
\end{align*}
which is the same as line (\ref{eq:basic-ineq-b3}) with $\beta=\hat\beta$.
Then we consider two subcases:
\begin{align}
(1-\xi_{11}) \Delta &\le 2 A_1 \lambda_0 \sum_{j\in S_{\bar\beta}} |\hat\beta_j -\bar\beta_j| + r C_{11} \lambda_0, \label{eq:beta-subcase1} \\
\text{or} \qquad \xi_{11} \Delta & \le (M_{01} |S_{\bar\gamma}|\lambda_0^2 )^{1/2} Q^{1/2} (\hat\beta,\bar\beta; \bar\gamma). \label{eq:beta-subcase2}
\end{align}
The split is mainly chosen to simplify the resulting bound on $\Delta$.
In the case of (\ref{eq:beta-subcase1}), we use $\|\hat\beta -\bar\beta\|_1 \le r$ and obtain
$\Delta \le (1-\xi_{11})^{-1} (2 A_1 + C_{11}) r \lambda_0$, which can be simplified as
\begin{align*}
\Delta & \le (\xi_1+1) (1+ C_{11}/(2 A_1))
\left( \kappa_1^{-1} \xi_{11}^{-2} M_{01} |S_{\bar\gamma}|\lambda_0^2 +  \kappa_1^{-1} \nu_1^{-2} \xi_{12}^2  |S_{\bar\beta}|\lambda_0^2 \right)\\
& \le \frac{3\xi_1+1}{2} \left( \kappa_1^{-1} \xi_{11}^{-2} M_{01} |S_{\bar\gamma}|\lambda_0^2 +  \kappa_1^{-1} \nu_1^{-2} \xi_{12}^2  |S_{\bar\beta}|\lambda_0^2 \right).
\end{align*}
The second step above uses $C_{11} < A_1(\xi_1-1)/(\xi_1+1)$ by the condition on $A_1$.
In the case of (\ref{eq:beta-subcase2}), we obtain similarly as from (\ref{eq:beta-case1}) to (\ref{eq:beta-case1b}),
\begin{align*}
\Delta \le  \kappa_1^{-1} \xi_{11}^{-2} (M_{01} |S_{\bar\gamma}|\lambda_0^2 ).
\end{align*}
Combining the proceeding two displays yields the desired result.
\end{prf}

\begin{cor} \label{cor:beta-global}
In the setting of Lemma \ref{lem:beta-global},  we also have in the event $\Omega_0\cap\Omega_{11}(r) \cap\Omega_{12}$,
\begin{align}
(\hat\beta-\bar\beta)^\T \Sigma_h (\hat\beta-\bar\beta) &\le  \kappa_1^{-1} \{ \bar \ell(\hat\beta;\bar\gamma) - \bar\ell(\bar\beta; \bar\gamma) \},
\label{eq:beta-Q}\\
 (\hat\beta-\bar\beta)^\T \tilde \Sigma_h (\hat\beta-\bar\beta) &\le  \kappa_1^{-1} \{ \bar \ell(\hat\beta;\bar\gamma) - \bar\ell(\bar\beta; \bar\gamma) \}
 + C_{12} \zeta_1 r \lambda_0,
 \label{eq:beta-Q2} \\
|\ell(\hat\beta;\bar\gamma) - \ell(\bar\beta; \bar\gamma) | &\le \bar \ell(\hat\beta;\bar\gamma) - \bar\ell(\bar\beta; \bar\gamma) + C_{11} r \lambda_0,
\label{eq:beta-ell}
\end{align}
where $(C_{11}, C_{12})$ are defined in Lemmas \ref{lem:emp-concen} and \ref{lem:Sigma-h} respectively.
\end{cor}

\begin{prf}
From the proof of Lemma \ref{lem:beta-global}, we have $\|\hat\beta -\bar\beta\|_1\le r \le \zeta_1$.
Then (\ref{eq:beta-Q}) holds directly by the margin condition, Assumption \ref{ass:beta}(ii).
Moreover, (\ref{eq:beta-Q2}) follows because by the definition of $\Omega_{12}$ in Lemma \ref{lem:Sigma-h},
\begin{align*}
\left| (\hat\beta-\bar\beta)^\T (\tilde\Sigma_h -\Sigma_h) (\hat\beta-\bar\beta) \right|
& \le \left\{ \sup_{j,k=0,\ldots,m} | (\tilde\Sigma_h)_{jk} - (\Sigma_h)_{jk} | \right\} \|\hat\beta-\bar\beta\|_1^2 \\
& \le C_{12} \lambda_0 r^2 \le C_{12} \zeta_1 r \lambda_0.
\end{align*}
Finally, (\ref{eq:beta-ell}) follows because by the definition of $\Omega_{11}(r)$ in Lemma \ref{lem:emp-concen},
\begin{align*}
& \quad \left| \ell(\hat\beta;\bar\gamma) - \ell(\bar\beta; \bar\gamma) - \{ \bar \ell(\hat\beta;\bar\gamma) - \bar\ell(\bar\beta; \bar\gamma) \} \right| \\
& = \left| (\tilde E-E) \left[ T w(X; \bar\gamma) \{\rho (Y,h^\T\beta)-\rho (Y,h^\T \bar\beta) \} \right] \right| \le rC_{11} \lambda_0.
\end{align*}
\end{prf}

\subsection{Proof of Theorem \ref{thm:alpha-conv}}

Suppose that Assumptions \ref{ass:gam}--\ref{ass:alpha} are satisfied. Then
Theorem \ref{thm:alpha-conv} follows from Lemma \ref{lem:alpha-conv}, which depends on
probability Lemmas \ref{lem:prob-alpha}--\ref{lem:prob-alpha2} and analytical Lemmas \ref{lem:depend-gam2}--\ref{lem:alpha-compat}.

We use the following notation in this and next section. Write $\hat\alpha=\hat\alpha_{\mytext{RWL},1+}$ and $\bar\alpha =\bar\alpha_{\mytext{WL},1+}$.
Denote $\Omega_{11} = \Omega_{11}(r)$ with $r$ defined in Lemma \ref{lem:beta-global}
and $\Omega_1 = \Omega_{11} \cap \Omega_{12}$.
By abuse of notation, write $ \rho_\Lambda (y,u) = ( \Lambda-\Lambda^{-1} ) \rho_\tau (y,u)$, which is $(\Lambda-\Lambda^{-1})$-Lipschitz. Denote
\begin{align*}
\tilde Q (\alpha,\bar\alpha; \gamma) & =\tilde E \left\{ T w(X; \gamma) | f^\T(X) (\alpha-\bar\alpha) |^2 \right\}, \\
\tilde Q (\alpha,\bar\alpha; \bar\gamma) & = (\alpha-\bar\alpha)^\T \tilde \Sigma_f (\alpha-\bar\alpha)
= \tilde E \left\{ T w(X; \bar\gamma) | f^\T(X) (\alpha-\bar\alpha) |^2 \right\},
\end{align*}
where $\tilde \Sigma_f = \tilde E \{ T w(X; \bar\gamma) f(X) f^\T (X)\}$, the sample version of $\Sigma_f$.

First, we state two probability lemmas, taken from Tan (2020b), Lemmas 2--4.

\begin{lem} \label{lem:prob-alpha}
Denote by $\Omega_{21}$ the event that
\begin{align*}
& \quad \sup_{j=0,1,\ldots,p} \left| \tilde E \left[ T w(X; \bar\gamma )
\{ Y + \rho_\Lambda(Y, h^\T\bar\beta) - f^\T(X)\bar\alpha\} f_j(X) \right] \right| \le C_{21} \lambda_0 ,
\end{align*}
where $C_{21} = \me^{-B_0} C_0 \sqrt{ 8 (D_0^2 + D_1^2) } $.
Under Assumptions~\ref{ass:gam}(i)--(ii) and \ref{ass:alpha}(i), $P( \Omega_{21} ) \ge 1- 2 \epsilon$.
\end{lem}

Denote  $\Sigma_{f2} = E [ T w(X; \bar\gamma ) \{Y + \rho_\Lambda(Y, h^\T\bar\beta) - f^\T(X)\bar\alpha \}^2 f(X) f^\T (X)] $,
and $\tilde \Sigma_{f2} = \tilde E [ T w(X; \bar\gamma)$ $ \{Y + \rho_\Lambda(Y, h^\T\bar\beta) - f^\T(X)\bar\alpha\}^2 f(X) f^\T (X)] $,
the sample version of $\tilde \Sigma_{f2}$.
Similarly, denote  $\Sigma_{f1} = E [ T w(X; \bar\gamma ) | Y + \rho_\Lambda(Y, h^\T\bar\beta) - f^\T(X)\bar\alpha | f(X) f^\T (X)] $,
and $\tilde \Sigma_{f1} = \tilde E [ T w(X; \bar\gamma) | Y + \rho_\Lambda(Y, h^\T\bar\beta) - f^\T(X)\bar\alpha | f(X) f^\T (X)] $,
the sample version of $\Sigma_{f1}$.

\begin{lem} \label{lem:prob-alpha2}
(i) Denote by  $\Omega_{22}$ the event that
\begin{align*}
\sup_{j,k=0,1,\ldots,p} | (\tilde \Sigma_{f2})_{jk} - (\Sigma_{f2})_{jk} | \le  C_{02} ( D_0^2 \lambda_0^2 +  D_0 D_1 \lambda_0 ),
\end{align*}
where $C_{02} = 4 \me^{-B_0} C_0^2 $ as in Lemma \ref{lem:prob-gam}(ii).
Under Assumptions~\ref{ass:gam}(i)--(ii) and \ref{ass:alpha}(i), $P(\Omega_{22}) \ge 1- 2\epsilon^2$.\\
(ii) Denote by $\Omega_{23}$ the event that
\begin{align*}
\sup_{j,k=0,1,\ldots,p} | (\tilde \Sigma_{f1})_{jk} - (\Sigma_{f1})_{jk} | \le C_{02} \sqrt{D_0^2 +D_1^2} \lambda_0.
\end{align*}
Under Assumptions~\ref{ass:gam}(i)--(ii) and \ref{ass:alpha}(i), $P( \Omega_{23} ) \ge 1- 2 \epsilon^2$.
\end{lem}

Next, we provide several analytical lemmas within various events.

\begin{lem} \label{lem:depend-gam2}
In the event $\Omega_0\cap\Omega_1$, we have
\begin{align*}
& \quad \left| \tilde E \left[ w(X;\hat\gamma) \{\rho_\Lambda (Y,h^\T\hat\beta) - \rho_\Lambda (Y,h^\T\bar\beta)\}f^\T(X) (\hat\alpha-\bar\alpha)  \right]  \right| \\
& \le (M_{11} |S_{\bar\beta}| \lambda_0^2 )^{1/2}\; \tilde Q^{1/2} (\hat\alpha,\bar\alpha; \bar\gamma),
\end{align*}
where $ M_{11} = ( \Lambda-\Lambda^{-1} )^2 \me^{2 C_0 M_0\varrho_0} M_1$.
\end{lem}

\begin{prf}
In the event $\Omega_0$, we use $|f^\T(X) (\hat\gamma-\bar\gamma)| \le C_0 M_0 \varrho_0$ and obtain
\begin{align*}
& \quad \left| \tilde E \left[ w(X;\hat\gamma) \{\rho_\Lambda (Y,h^\T\hat\beta) - \rho_\Lambda (Y,h^\T\bar\beta)\}f^\T(X) (\hat\alpha-\bar\alpha)  \right]  \right| \\
& \le \tilde E \left[ \me^{|f^\T(\hat\gamma-\bar\gamma)|}
w(X;\bar\gamma) \left| \{\rho_\Lambda (Y,h^\T\hat\beta) - \rho_\Lambda (Y,h^\T\bar\beta)\}f^\T(X) (\hat\alpha-\bar\alpha)  \right|  \right] \\
& \le \me^{C_0 M_0 \varrho_0} \tilde E \left[
w(X;\bar\gamma)  \left|  \{\rho_\Lambda (Y,h^\T\hat\beta) - \rho_\Lambda (Y,h^\T\bar\beta)\}f^\T(X) (\hat\alpha-\bar\alpha)  \right| \right] .
\end{align*}
Then by the Lipschitz property of $\rho_\Lambda (\cdot)$ and the Cauchy--Schwartz inequality, we have
\begin{align*}
& \quad \left| \tilde E \left[ w(X;\hat\gamma) \{\rho_\Lambda (Y,h^\T\hat\beta) - \rho_\Lambda (Y,h^\T\bar\beta)\}f^\T(X) (\hat\alpha-\bar\alpha)  \right]  \right| \\
& \le ( \Lambda-\Lambda^{-1} ) \me^{C_0 M_0 \varrho_0} \tilde E \left\{
w(X;\bar\gamma)  \left|h^\T (X) (\hat\beta -\bar\beta)f^\T(X) (\hat\alpha-\bar\alpha)  \right| \right\} \\
& \le ( \Lambda-\Lambda^{-1} ) \me^{C_0 M_0 \varrho_0} \tilde E^{1/2} \left\{ w(X;\bar\gamma) |h^\T (X) (\hat\beta -\bar\beta)|^2 \right\}
\tilde E^{1/2} \left\{ T w(X;\bar\gamma) |f^\T(X) (\hat\alpha-\bar\alpha) |^2 \right\}.
\end{align*}
Moreover, in the event $\Omega_0\cap\Omega_1$, we have
\begin{align*}
\tilde E \left\{ T w(X;\bar\gamma) | h^\T(X) (\hat\gamma -\bar\gamma) |^2 \right\} \le M_1 |S_{\bar\beta}| \lambda_0^2 .
\end{align*}
Combining the preceding two displays yields the desired result.
\end{prf}

\begin{lem} \label{lem:basic-ineq-a}
In the event $\Omega_0\cap\Omega_1 \cap\Omega_{21}\cap\Omega_{22}$, if $\lambda_0\le 1$, then
\begin{align}
& \quad \me^{-\varrho_{01}} \tilde Q ( \hat\alpha, \bar\alpha; \gamma) + (A_2- C_{21}) \lambda_0 \| \hat\alpha_{1:p} \|_1 \nonumber \\
& \le \tilde E \left[ w(X; \bar\gamma) \{Y + \rho_\Lambda(Y,h^\T \bar\beta) - f^\T(X) \bar\alpha\} f^\T(X) \right] (\hat\alpha-\bar\alpha)
+ 2 A_2 \lambda_0 \sum_{j\in S_{\bar\alpha}} |\hat\alpha_j-\bar\alpha_j|  \nonumber \\
& \quad + \left\{ (M_{03} |S_{\bar\gamma}| \lambda_0^2 )^{1/2} + (M_{11} |S_{\bar\beta}| \lambda_0^2 )^{1/2} \right\}
\tilde Q^{1/2} (\hat\alpha,\bar\alpha; \bar\gamma), \label{eq:basic-ineq-a}
\end{align}
where $M_{03} =\me^{2\varrho_{01}} (D_0^2 + D_1^2) (M_0 + \varrho_{02}) + \me^{2\varrho_{01}} (D_0^2 + D_0 D_1) \varrho_{02}$
with $(D_0,D_1)$ from Assumption \ref{ass:alpha}(i),
$\varrho_{01} = C_0 M_0\varrho_0$, $\varrho_{02} = C_{02} M_0^2  \varrho_0 $, $C_{21}$ is defined in Lemma \ref{lem:prob-alpha},
and $M_{11}$ is defined in Lemma \ref{lem:depend-gam2}.
\end{lem}

\begin{prf}
Similarly as in Tan (2020b), Lemma 6, we have from the definition of $\hat\alpha$,
\begin{align}
& \quad \tilde Q ( \hat\alpha, \bar\alpha; \hat\gamma) + A_2 \lambda_0 \| \hat\alpha_{1:p} \|_1 \nonumber \\
& \le \tilde E \left[ w(X; \hat\gamma) \{Y + \rho_\Lambda(Y,h^\T \hat\beta) - f^\T(X) \bar\alpha\} f^\T(X) \right] (\hat\alpha-\bar\alpha)
+ A_2 \lambda_0 \| \bar\alpha_{1:p} \|_1 .  \label{eq:basic-ineq-a-prf1}
\end{align}
Similarly as in Tan (2020b), Lemma 7, the first term on the left-hand side of (\ref{eq:basic-ineq-a-prf1}) can be lower bounded in the event $\Omega_0$ as
\begin{align*}
\tilde Q ( \hat\alpha, \bar\alpha; \hat\gamma) \ge \me^{-C_0 M_0 \varrho_0} \tilde Q ( \hat\alpha, \bar\alpha; \bar\gamma).
\end{align*}
The first term on the right-hand side of (\ref{eq:basic-ineq-a-prf1}) can be decomposed as
\begin{align}
& \quad \tilde E \left[ w(X; \hat\gamma) \{Y + \rho_\Lambda(Y,h^\T \hat\beta) - f^\T(X) \bar\alpha\} f^\T(X) \right] (\hat\alpha-\bar\alpha) \nonumber  \\
& = \tilde E \left[ w(X; \bar\gamma) \{Y + \rho_\Lambda(Y,h^\T \bar\beta) - f^\T(X) \bar\alpha\} f^\T(X) \right] (\hat\alpha-\bar\alpha) \nonumber  \\
& \quad + \tilde E \left[ \{w(X; \hat\gamma)-w(X;\bar\gamma) \} \{Y + \rho_\Lambda(Y,h^\T \bar\beta) - f^\T(X) \bar\alpha\} f^\T(X) \right] (\hat\alpha-\bar\alpha) \nonumber  \\
& \quad + \tilde E \left[ w(X; \hat\gamma) \{ \rho_\Lambda(Y,h^\T \hat\beta) -\rho_\Lambda(Y,h^\T \bar\beta)  \} f^\T(X) \right] (\hat\alpha-\bar\alpha). \label{eq:basic-ineq-a-prf2}
\end{align}
Similarly as in Tan (2020b), Lemma 8, the second term on the right-hand side of (\ref{eq:basic-ineq-a-prf2}) can be bounded in the event $\Omega_0\cap\Omega_{22}$ as
\begin{align*}
& \quad \left| \tilde E \left[ \{w(X; \hat\gamma)-w(X;\bar\gamma) \} \{Y + \rho_\Lambda(Y,h^\T \bar\beta) - f^\T(X) \bar\alpha\} f^\T(X) \right] (\hat\alpha-\bar\alpha)\right| \\
& \le \me^{\varrho_{01}}
\tilde E^{1/2} \left[ T w(X;\bar\gamma) \{Y + \rho_\Lambda(Y,h^\T \bar\beta) - f^\T(X) \bar\alpha\}^2 |f^\T(X) (\hat\gamma-\bar\gamma) |^2  \right] \\
& \quad \times \tilde E^{1/2} \left\{ T w(X;\bar\gamma) |f^\T(X) (\hat\alpha-\bar\alpha) |^2 \right\} \\
& \le \me^{\varrho_{01}} \left\{ (D_0^2 + D_0 D_1)C_{02} \lambda_0 ( M_0 |S_{\bar\gamma}|\lambda_0 )^2 + (D_0^2 + D_1^2) C_{02} \lambda_0 ( M_0 |S_{\bar\gamma}|\lambda_0)^2
   + M_0 |S_{\bar\gamma}|\lambda_0^2 \right\}  \\
& \quad \times \tilde E^{1/2} \left\{ T w(X;\bar\gamma) |f^\T(X) (\hat\alpha-\bar\alpha) |^2 \right\} \\
& \le (M_{03} |S_{\bar\gamma}| \lambda_0^2 )^{1/2} \tilde E^{1/2} \left\{ T w(X;\bar\gamma) |f^\T(X) (\hat\alpha-\bar\alpha) |^2 \right\} ,
\end{align*}
where Lemma \ref{lem:prob-alpha2}(i) is simplified with $\lambda_0\le 1$,
$M_{03} =\me^{2\varrho_{01}} (D_0^2 + D_1^2) (M_0 + \varrho_{02}) + \me^{2\varrho_{01}} (D_0^2 + D_0 D_1) \varrho_{02}$,
$\varrho_{01} = C_0 M_0\varrho_0$, and $\varrho_{02} = C_{02} M_0^2  \varrho_0 $.
Note that $\lambda_0$ here differs from $\lambda_0$ in Tan (2020b), which is defined as $\max(C_{01}, C_{02})\lambda_0$.
Then applying Lemma \ref{lem:depend-gam2} to the third term on the right-hand side of (\ref{eq:basic-ineq-a-prf2})
shows that in the event $\Omega_0 \cap\Omega_1$,
\begin{align}
& \quad \me^{-\varrho_{01}} \tilde Q ( \hat\alpha, \bar\alpha; \gamma) + A_2 \lambda_0 \| \hat\alpha_{1:p} \|_1 \nonumber \\
& \le \tilde E \left[ w(X; \bar\gamma) \{Y + \rho_\Lambda(Y,h^\T \bar\beta) - f^\T(X) \bar\alpha\} f^\T(X) \right] (\hat\alpha-\bar\alpha)
+ A_2 \lambda_0 \| \bar\alpha_{1:p} \|_1  \nonumber \\
& \quad + (M_{03} |S_{\bar\gamma}| \lambda_0^2 )^{1/2} \; \tilde Q^{1/2} (\hat\alpha,\bar\alpha; \bar\gamma)
 + (M_{11} |S_{\bar\beta}| \lambda_0^2 )^{1/2} \; \tilde Q^{1/2} (\hat\alpha,\bar\alpha; \bar\gamma). \label{eq:basic-ineq-a-prf3}
\end{align}
The first term on the right-hand side of (\ref{eq:basic-ineq-a-prf3}) is upper bounded by $C_{21} \lambda_0 \|\hat\alpha-\bar\alpha\|_1$
in the event $\Omega_{21}$ from Lemma \ref{lem:prob-alpha}.
Then similar reasoning as in Tan (2020b), Lemma 9, leads to the desired result.
Note that $\lambda_1$ in Tan (2020b) is defined to be $\ge C_{21} \lambda_0$.
\end{prf}

\begin{lem} \label{lem:alpha-compat}
In the event $\Omega_{02}$ (from Lemma \ref{lem:prob-gam}), the population compatibility condition in Assumption~\ref{ass:alpha}(ii)
implies a sample compatibility condition for $\tilde\Sigma_f$:
for any vector $b=(b_0,b_1,\ldots,b_p)^\T \in \bbR^{1+p} $ such that $\sum_{j\not \in S_{\bar\alpha}} |b_j| \le \xi_2 \sum_{j\in S_{\bar\alpha}} |b_j|$, we have
\begin{align*}
(1-\varrho_2) \nu_2^2  \Big(\sum_{j\in S_{\bar\alpha}} |b_j|\Big)^2 \le |S_{\bar\alpha}| \left( b^\T \tilde \Sigma_f b  \right) ,
\end{align*}
provided that $(1+\xi_2)^2 \nu_2^{-2} C_{02} |S_{\bar\alpha}| \lambda_0 \le \varrho_2 \, (<1)$.
\end{lem}

\begin{prf}
This result is taken from Tan (2020b), Lemma 11.
\end{prf}

\begin{lem} \label{lem:alpha-conv}
Suppose that $A_2 > C_{21} (\xi_2+1)/(\xi_2-1)$, with $C_{21}$ from Lemma \ref{lem:prob-alpha}.
In the event $\Omega_0\cap\Omega_1 \cap\Omega_{21}\cap\Omega_{22}$, if $\lambda_0\le 1$, then
\begin{align*}
& \quad \tilde Q ( \hat\alpha, \bar\alpha; \hat\gamma) + (A_2-C_{21})\me^{\varrho_{01}}  \lambda_0 \| \hat\alpha - \bar\alpha\|_1   \\
& \le 2 \me^{2\varrho_{01}} \xi_{21}^{-2} \left( M_{03} |S_{\bar\gamma}| \lambda_0^2   + M_{11} |S_{\bar\beta}| \lambda_0^2 \right) +
 \me^{2\varrho_{01}} \nu_{21}^{-2} \xi_{22}^2 |S_{\bar\alpha}|  \lambda_0^2 ,
\end{align*}
where $\xi_{21} = 1-2A_2 /\{ (\xi_2+1)(A_2-C_{21})\} \in (0,1)$, $\xi_{22}= (\xi_2+1)(A_2-C_{21})$,
$\nu_{21} = (1-\varrho_2)^{1/2} \nu_2$, $M_{11}$ is defined in Lemma \ref{lem:depend-gam2},
and $(\varrho_{01},M_{03})$ are defined in Lemma \ref{lem:basic-ineq-a}.
\end{lem}

\begin{prf}
Denote $b=  \hat\alpha - \bar\alpha$ and $\Delta = \me^{-\varrho_{01}} \tilde Q ( \hat\alpha, \bar\alpha; \hat\gamma) + (A_2-C_{21}) \lambda_0 \| b \|_1 $.
In the event $\Omega_0 \cap \Omega_{11}\cap\Omega_{12} \cap \Omega_{22}$, inequality (\ref{eq:basic-ineq-a})
from Lemma~\ref{lem:basic-ineq-a} leads to two possible cases: either
\begin{align}
\xi_{21} \Delta  \le  \left\{ (M_{03} |S_{\bar\gamma}| \lambda_0^2 )^{1/2} + (M_{11} |S_{\bar\beta}| \lambda_0^2 )^{1/2} \right\}
 \tilde Q^{1/2} (\hat\alpha,\bar\alpha; \bar\gamma), \label{eq:alpha-conv-prf1}
\end{align}
or $(1-\xi_{21}) \Delta \le 2 A_2 \lambda_0 \sum_{j\in S_{\bar\alpha}} |b_j|$, that is,
\begin{align}
\Delta
\le (\xi_2+1) (A_2-C_{21}) \lambda_0 \sum_{j\in S_{\bar\alpha}} |b_j| = \xi_{22} \lambda_0 \sum_{j\in S_{\bar\alpha}} |b_j| , \label{eq:alpha-conv-prf2}
\end{align}
where $\xi_{21} = 1-2A_2 /\{ (\xi_2+1)(A_2-C_{21})\} \in (0,1)$ because $A_2 > C_{21} (\xi_2+1)/(\xi_2-1)$ and $\xi_{22}= (\xi_2+1)(A_2-C_{21})$.
We deal with the two cases separately as follows.

In the case of (\ref{eq:alpha-conv-prf1}), we have
$\tilde Q (\hat\alpha,\bar\alpha; \bar\gamma) \le \me^{2\varrho_{01}} \xi_{21}^{-2}
\{ (M_{03} |S_{\bar\gamma}| \lambda_0^2 )^{1/2} + (M_{11} |S_{\bar\beta}| \lambda_0^2 )^{1/2} \}^2$ by simple manipulation.
Returning to (\ref{eq:alpha-conv-prf1}) shows
\begin{align}
\Delta & \le  \me^{\varrho_{01}}  \xi_{21}^{-2} \left\{ (M_{03} |S_{\bar\gamma}| \lambda_0^2 )^{1/2} + (M_{11} |S_{\bar\beta}| \lambda_0^2 )^{1/2} \right\}^2 \nonumber \\
& \le 2 \me^{\varrho_{01}} \xi_{21}^{-2} \left( M_{03} |S_{\bar\gamma}| \lambda_0^2   + M_{11} |S_{\bar\beta}| \lambda_0^2 \right).\label{eq:alpha-conv-prf3}
\end{align}

In the case of (\ref{eq:alpha-conv-prf2}), we have $\sum_{j \not\in S_{\bar\alpha}} |b_j| \le \xi_2 \sum_{j\in S_{\bar\alpha}} |b_j|$,
which, by the empirical compatibility condition in Lemma~\ref{lem:alpha-compat}, implies that
$\sum_{j\in S_{\bar\alpha}} |b_j| \le (1-\varrho_2)^{-1/2}\nu_2^{-1} |S_{\bar\alpha}|^{1/2} \tilde Q^{1/2} (\hat\alpha,\bar\alpha; \bar\gamma)$.
Substituting this into (\ref{eq:alpha-conv-prf2}) yields
\begin{align*}
\Delta \le (\xi_{22} \lambda_0) (1-\varrho_2)^{-1/2}\nu_2^{-1} |S_{\bar\alpha}|^{1/2}  \tilde Q^{1/2} (\hat\alpha,\bar\alpha; \bar\gamma).
\end{align*}
Then similar reasoning as from (\ref{eq:alpha-conv-prf1}) to (\ref{eq:alpha-conv-prf3}) shows
\begin{align}
\Delta \le \me^{\varrho_{01}} (1-\varrho_2)^{-1} \nu_2^{-2} \xi_{22}^2 |S_{\bar\alpha}|  \lambda_0^2 .\label{eq:alpha-conv-prf4}
\end{align}
Combining (\ref{eq:alpha-conv-prf3}) and (\ref{eq:alpha-conv-prf4}) yields the desired result.
\end{prf}

\subsection{Proof of Theorem \ref{thm:mu1+expan}}

Denote $\hat\varphi_{+} = \varphi_{+}(O; \pi(\cdot;\hat\gamma), \eta_{1+}(\cdot;\hat\alpha), q_{1,\tau}(\cdot;\hat\beta) )$
and $\bar\varphi_{+}=\varphi_{+}(O; \pi(\cdot;\bar\gamma), \eta_{1+}(\cdot;\bar\alpha), q_{1,\tau}(\cdot;\bar\beta) )$.
The estimator $\hat \mu^{1+}_{\mytext{RCAL}}$ can be decomposed as
\begin{align*}
\hat \mu^{1+}_{\mytext{RCAL}} & = \hat\mu^{1+} (\bar\gamma,\bar\alpha, \bar\beta)
+ \tilde E ( \hat \varphi_{+} - \bar\varphi_{+} ) \\
& =  \hat\mu^{1+} (\bar\gamma,\bar\alpha, \bar\beta) + \Delta_1 + \Delta_2 + \Delta_3,
\end{align*}
where
\begin{align*}
\Delta_1 & = \tilde E \left[\left\{1- \frac{T}{\pi(X;\bar\gamma)} \right\} f^\T(X) (\hat\alpha-\bar\alpha) \right] ,\\
\Delta_2 & = \tilde E \left[ T \left\{Y + \varrho_\Lambda(Y,h^\T \bar\beta) - f^\T(X) \bar\alpha \right\}
\left\{ \frac{1}{\pi(X;\hat\gamma)} - \frac{1}{\pi(X;\bar\gamma)} \right\} \right], \\
\Delta_3 & = \tilde E \left[ T \left\{\frac{1}{\pi(X;\hat\gamma)} -1 \right\} \{\varrho_\Lambda(Y,h^\T \hat\beta) - \varrho_\Lambda(Y,h^\T \bar\beta) \} \right] .
\end{align*}
We show that if $\lambda_0\le 1$, then in the event $\Omega_0\cap\Omega_1\cap\Omega_{21}\cap\Omega_{22}\cap\Omega_{21}$,
\begin{align}
& \left| \hat \mu^{1+}_{\mytext{RCAL}}- \hat\mu^{1+} (\bar\gamma,\bar\alpha, \bar\beta) \right|
\le M_{31} |S_{\bar\gamma}| \lambda_0^2 + M_{32} |S_{\bar\beta}| \lambda_0^2  + M_{33} |S_{\bar\alpha}| \lambda_0^2 , \label{eq:expan-prf}
\end{align}
where $M_{31} = C_{01} M_2 + C_{21} M_0 + \me^{C_0 M_0 \varrho_0} \sqrt{D_0^2+D_1^2} ( M_0/2+ C_{23} M_0^2 \varrho_0) +
  (\Lambda-\Lambda^{-1}) (M_{01}+ 3 M_1) /2$,
$M_{32} = C_{01} M_2 + (\Lambda-\Lambda^{-1}) (3 M_1/2)$,
$M_{33} = C_{01} M_2 $, $(C_{01},C_{21},C_{23})$ are defined in Lemmas \ref{lem:prob-gam}(i), \ref{lem:prob-alpha}, and \ref{lem:prob-alpha2}(ii),
and $M_{01}$ is defined in Lemma \ref{lem:depend-gam}.

Similarly as in the proof of Theorem 3 in Tan (2020b), the term $\Delta_1$ can be bounded in the event $\Omega_0\cap\Omega_1\cap\Omega_{21}\cap\Omega_{22}$
(with $\Omega_{01} \subset \Omega_0$ from Lemma \ref{lem:prob-gam}) as
\begin{align*}
| \Delta_1 | &\le \left\| E \left[\left\{1- \frac{T}{\pi(X;\bar\gamma)} \right\} f \right] \right\|_\infty \|\hat\alpha-\bar\alpha\|_1 \\
& \le C_{01} \lambda_0 \times M_2 (|S_{\bar\gamma}| + |S_{\bar\beta}| + |S_{\bar\alpha}|) \lambda_0 ,
\end{align*}
and the term $\Delta_2$ can be bounded in the event $\Omega_0 \cap \Omega_{21} \cap \Omega_{23}$ as
\begin{align}
 |\Delta_2 | & \le
 \left| \tilde E \left[ T \me^{-f^\T(X)\bar\gamma} \left\{Y + \varrho_\Lambda(Y,h^\T \bar\beta) - f^\T(X) \bar\alpha \right\}f^\T (\hat\gamma-\bar\gamma) \right] \right| \nonumber \\
 & \quad + \me^{C_0 M_0 \varrho_0} \tilde E \left[
 T \me^{-f^\T(X)\bar\gamma} \left| Y + \varrho_\Lambda(Y,h^\T \bar\beta) - f^\T(X) \bar\alpha \right| (f^\T (\hat\gamma-\bar\gamma))^2 \right]/2 \label{eq:pro-mu1+expan-prf1}\\
 & \le C_{21} \lambda_0 \times M_0 |S_{\bar\gamma}| \lambda_0 +
   \me^{C_0 M_0 \varrho_0} \sqrt{D_0^2+D_1^2} \left\{ 2 C_{23} \lambda_0 (M_0 |S_{\bar\gamma}| \lambda_0)^2 + M_0 |S_{\bar\gamma}| \lambda_0^2 \right\} /2 \nonumber\\
 & \le \left\{ C_{21} M_0 + \me^{C_0 M_0 \varrho_0} \sqrt{D_0^2+D_1^2} ( M_0/2+ C_{23} M_0^2 \varrho_0) \right\} |S_{\bar\gamma}| \lambda_0^2 ,\nonumber
\end{align}
where Lemma \ref{lem:prob-alpha2}(ii) is simplified with $\lambda_0\le 1$.
Moreover, the term $\Delta_3$ can be bounded by the triangle inequality as
\begin{align}
 |\Delta_3 | & \le
 \left| \tilde E \left[ T \me^{-f^\T(X)\bar\gamma} \{ \varrho_\Lambda(Y,h^\T \hat\beta) - \varrho_\Lambda(Y,h^\T \bar\beta)\} \right] \right| \nonumber \\
 & \quad + \left| \tilde E \left[
 T \left\{ \me^{-f^\T(X)\hat\gamma} -\me^{-f^\T(X)\bar\gamma} \right\} \{ \varrho_\Lambda(Y,h^\T \hat\beta) - \varrho_\Lambda(Y,h^\T \bar\beta)\} \right] \right| .
 \label{eq:pro-mu1+expan-prf2}
\end{align}
By inequality (\ref{eq:beta-ell}) in Corollary \ref{cor:beta-global},
the first term on the right-hand side of (\ref{eq:pro-mu1+expan-prf2}) can be bounded in the event $\Omega_0\cap\Omega_1$ as
\begin{align*}
\left| \tilde E \left[ T \me^{-f^\T(X)\bar\gamma} \{ \varrho_\Lambda(Y,h^\T \hat\beta) - \varrho_\Lambda(Y,h^\T \bar\beta)\} \right] \right|
\le (\Lambda-\Lambda^{-1}) M_1 (|S_{\bar\gamma}| + |S_{\bar\beta}|) \lambda_0^2.
\end{align*}
By (\ref{eq:lem-depend-gam-prf1}) using the Lipschitz property of $\rho(\cdot)$ together with (\ref{eq:beta-Q2}) in Corollary \ref{cor:beta-global}, the second term on the right-hand side of (\ref{eq:pro-mu1+expan-prf2})
can be bounded in the event $\Omega_0\cap\Omega_1$ as
\begin{align*}
& \quad \left| \tilde E \left[
 T \left\{ \me^{-f^\T(X)\hat\gamma} -\me^{-f^\T(X)\bar\gamma} \right\} \{ \varrho_\Lambda(Y,h^\T \hat\beta) - \varrho_\Lambda(Y,h^\T \bar\beta)\} \right] \right| \\
 & \le (\Lambda-\Lambda^{-1}) ( M_{01} |S_{\bar\gamma}| \lambda_0^2 ) ^{1/2} \{ M_1 (|S_{\bar\gamma}| + |S_{\bar\beta}| ) \lambda_0^2 \}^{1/2} \\
 & \le (\Lambda-\Lambda^{-1}) \left\{ M_{01} |S_{\bar\gamma}| \lambda_0^2 + M_1 (|S_{\bar\gamma}| + |S_{\bar\beta}| ) \lambda_0^2 \right\} /2.
\end{align*}
Combining the preceding bounds on $(\Delta_1,\Delta_2,\Delta_3)$ leads to the desired result (\ref{eq:expan-prf}).

Note that the control of $\Delta_1$, the leading term of $\Delta_2$ on the right-hand side of (\ref{eq:pro-mu1+expan-prf1}), and
the leading term of $\Delta_3$ on the right-hand side of (\ref{eq:pro-mu1+expan-prf2}) depends on
the fact that the target values $(\bar\gamma,\bar\alpha,\bar\beta)$ satisfy the population versions of the calibration equations (\ref{eq:cal-eq}).

\subsection{Proof of Proposition \ref{pro:relaxed}}

The result can be proved in a similar manner as Proposition \ref{pro:mu1-h}. See Francis \& Wright (1969), Section 3 for related results.
Alternatively, we find the dual representation directly.
To simplify notation, assume that $T_i=1$ for $i=1,\ldots,n_1 (\le n)$ and rewrite (\ref{eq:dhat-mu1+h}) as
\begin{align}
\dhat \mu_h^{1+} (\hat\gamma) = \frac{1}{n} \sum_{i=1}^{n_1} Y_i +
\max_{\lambda_1 }
\frac{1}{n} \sum_{i=1}^{n_1} w_i Y_i \lambda_{1i} ,  \label{eq:pro-relaxed-prf1}
\end{align}
with $\lambda_1 = (\lambda_{1i})_{i=1,\ldots,n_1}$ subject to
\begin{align*}
& \Lambda^{-1} \le \lambda_{1i} \le \Lambda, \quad i=1,\ldots,n_1,\\
&  \sum_{i=1}^{n_1} w_i \lambda_{1i} = \sum_{i=1}^{n_1} w_i ,\\
& \left| \sum_{i=1}^{n_1} w_i h_{ji} \lambda_{1i} - \sum_{i=1}^{n_1} w_i h_{ji} \right| \le n \tilde\lambda_\beta, \quad j=1,\ldots, m,
\end{align*}
where $\tilde \lambda_\beta =  (\Lambda-\Lambda^{-1}) \lambda_\beta$, $w_i = (1-\pi(X_i;\hat\gamma))/ \pi(X_i;\hat\gamma)$, and $h_{ji} = h_j(X_i)$,

The primal problem in (\ref{eq:pro-relaxed-prf1}) is linear programming (i.e., objective and constraints are linear in $\lambda_{1i}$'s)
and a feasible solution exists: $\lambda_{1i} \equiv 1 $. Hence the strong duality holds (Boyd \& Vandenberghe 2004).
Consider a primal problem in the form
\begin{align}
\text{max}_x \quad & c^\T x \label{eq:pro-relaxed-prf2} \\
\text{subject to} \quad & A_1 x \preceq b_1, \nonumber \\
& A_2 x = b_2,  \nonumber
\end{align}
and the corresponding dual problem
\begin{align}
\text{min}_{u_1,u_2} \quad & b_1^\T u_1 + b_2^\T u_2 \label{eq:pro-relaxed-prf3} \\
\text{subject to} \quad & A_1^\T u_1 + A_2^\T u_2 = c, \nonumber \\
& u_1 \succeq 0. \nonumber
\end{align}
where $\preceq$ denotes componentwise inequality.
The primal problem in (\ref{eq:pro-relaxed-prf1}), after excluding the constant $ n^{-1} \sum_{i=1}^{n_1} Y_i$ and rescaling, can be expressed
in the form (\ref{eq:pro-relaxed-prf2}) as
\begin{align}
\text{max}_{\lambda_1 } \quad &
\sum_{i=1}^{n_1} w_i Y_i \lambda_{1i} \label{eq:pro-relaxed-prf4} \\
\text{subject to} \quad
&  \lambda_{1i} \le \Lambda,  \quad i=1,\ldots,n_1, \nonumber \\
& - \lambda_{1i} \le -\Lambda^{-1} , \quad i=1,\ldots,n_1, \nonumber \\
& \sum_{i=1}^{n_1} w_i h_{ji} \lambda_{1i} \le \sum_{i=1}^{n_1} w_i h_{ji} + n \tilde\lambda_\beta, \quad j=1,\ldots, m, \nonumber  \\
& - \sum_{i=1}^{n_1} w_i h_{ji} \lambda_{1i} \le  -\sum_{i=1}^{n_1} w_i h_{ji} + n \tilde\lambda_\beta, \quad j=1,\ldots, m,  \nonumber  \\
&  \sum_{i=1}^{n_1} w_i \lambda_{1i} = \sum_{i=1}^{n_1} w_i .  \nonumber
\end{align}
Denote the dual variable $u_1$ associated with the inequality constraints as
$\alpha_+ = (\alpha_{i+})_{i=1,\ldots,n_1} $,
$\alpha_- = (\alpha_{i-})_{i=1,\ldots,n_1} $,
$\beta_{(0)+} =(\beta_{j+})_{j=1,\ldots,p} $,
$\beta_{(0)-} =(\beta_{j-})_{j=1,\ldots,p} $.
Denote the dual variable $u_2$ associated with the equality constraint as $\beta_0$.
The constraints in the dual problem (\ref{eq:pro-relaxed-prf3}) can be shown to be
\begin{align*}
& \alpha_{i+}  - \alpha_{i-} + \beta_{(0)+}^\T w_i  h_{(0)i} - \beta_{(0)-}^\T  w_i  h_{(0)i} + \beta_0 w_i = w_i Y_i , \quad i=1,\ldots,n_1,  \\
& \alpha_+ \succeq 0, \quad \alpha_- \succeq 0, \quad \beta_{(0)+} \succeq 0, \quad \beta_{(0)-} \succeq 0, \nonumber
\end{align*}
where $h_{(0)i} = (h_{1i},\ldots,h_{mi})^\T$.
Let $\beta_{(0)} = \beta_{(0)+} - \beta_{(0)-}$. Then the preceding constraints can be simplified as
\begin{align}
& \alpha_{i+}  - \alpha_{i-} + \beta_{(0)}^\T w_i  h_{(0)i} + \beta_0 w_i = w_i Y_i , \quad i=1,\ldots,n_1, \label{eq:pro-relaxed-prf5}\\
& \alpha_+ \succeq 0, \quad \alpha_- \succeq 0, , \nonumber
\end{align}
The objective in the dual problem (\ref{eq:pro-relaxed-prf3}) can be calculated as
\begin{align}
& \sum_{i=1}^{n_1} \alpha_{i+} \Lambda - \sum_{i=1}^{n_1} \alpha_{i-} \Lambda^{-1}
 + \beta_{(0)+}^\T \left\{ \sum_{i=1}^{n_1} w_i h_{(0)i} + n\tilde\lambda_\beta  \right\}
 + \beta_{(0)-}^\T \left\{- \sum_{i=1}^{n_1} w_i h_{(0)i} + n\tilde\lambda_\beta  \right\}  + \beta_0 \sum_{i=1}^{n_1} w_i \nonumber \\
& =\sum_{i=1}^{n_1} (\alpha_{i+} \Lambda - \alpha_{i-} \Lambda^{-1})
 + \sum_{i=1}^{n_1} w_i (\beta_0+\beta_{(0)}^\T h_{(0)i}) + n\tilde\lambda_\beta \|\beta_{(0)}\|_1  , \label{eq:pro-relaxed-prf6}
\end{align}
where $\| \beta_{(0)} \|_1 = \beta_{(0)+} + \beta_{(0)-}$ because $\beta_{(0)+} \succeq 0$ and $\beta_{(0)-} \succeq 0$.

To find the desired dual representation in $(\beta_0, \beta_{(0)})$, we
minimize the objective (\ref{eq:pro-relaxed-prf6}) over $(\alpha_+, \alpha_-)$ for fixed $(\beta_0, \beta_{(0)} )$.
The constraints (\ref{eq:pro-relaxed-prf5}) only require that $\alpha_{i+}  - \alpha_{i-}$ is fixed for each $i$,
whereas the objective (\ref{eq:pro-relaxed-prf6}) is additive in $\alpha_{i+} \Lambda - \alpha_{i-} \Lambda^{-1}$.
Hence the objective (\ref{eq:pro-relaxed-prf6}) can be minimized separately over the pairs $(\alpha_{i-}, \alpha_{i+})$.
If $\alpha_{i+}  - \alpha_{i-} \ge 0$, then $\alpha_{i+} \Lambda - \alpha_{i-} \Lambda^{-1}$ is minimized by setting $\alpha_{i-}=0$
and hence by (\ref{eq:pro-relaxed-prf5})
\begin{align*}
& \alpha_{i+} + \beta_{(0)}^\T  w_i  h_{(0)i} + \beta_0 w_i = w_i Y_i ,\\
& \alpha_{i+} \Lambda - \alpha_{i-} \Lambda^{-1}  = \alpha_{i+} \Lambda = \Lambda w_i( Y_i - \beta_0- \beta_{(0)}^\T h_{(0)i}) \ge  0 .
\end{align*}
If $\alpha_{i+}  - \alpha_{i-} \le 0$, then $\alpha_{i+} \Lambda - \alpha_{i-} \Lambda^{-1}$ is minimized by setting $\alpha_{i+}=0$
and hence by (\ref{eq:pro-relaxed-prf5}),
\begin{align*}
& - \alpha_{i-}  + \beta_{(0)}^\T  w_i  h_{(0)i} + \beta_0 w_i  = w_i Y_i , \\
& \alpha_{i+} \Lambda - \alpha_{i-} \Lambda^{-1}  = - \alpha_{i-} \Lambda^{-1} = \Lambda^{-1}  w_i( Y_i - \beta_0- \beta_{(0)}^\T h_{(0)i}) \le 0.
\end{align*}
Combining the preceding two cases and substituting the relationships to (\ref{eq:pro-relaxed-prf6}) shows that
the objective (\ref{eq:pro-relaxed-prf6}) is minimized over $(\alpha_+, \alpha_-)$ for fixed $(\beta_0, \beta_{(0)} )$ to be
\begin{align*}
& \sum_{i=1}^{n_1} (\alpha_{i+} \Lambda - \alpha_{i-} \Lambda^{-1})
 + \sum_{i=1}^{n_1} w_i (\beta_0+\beta_{(0)}^\T h_{(0)i}) + n\tilde\lambda_\beta \|\beta_{(0)}\|_1   \\
& = \sum_{i=1}^{n_1} \left\{ \Lambda w_i( Y_i - \beta_0 - \beta_{(0)}^\T h_{(0)i})_+
- \Lambda^{-1}  w_i( -Y_i + \beta_0+ \beta_{(0)}^\T h_{(0)i})_+ \right\} \\
& \quad + \sum_{i=1}^{n_1} w_i (\beta_0+\beta_{(0)}^\T h_{(0)i})   + n\tilde\lambda_\beta \|\beta_{(0)}\|_1 \\
& = \sum_{i=1}^{n_1} \left\{ (\Lambda-1) w_i( Y_i - \beta_0 - \beta_{(0)}^\T h_{(0)i})_+
+ (1- \Lambda^{-1})  w_i( -Y_i + \beta_0+ \beta_{(0)}^\T h_{(0)i})_+ \right\} \\
& \quad  + \sum_{i=1}^{n_1} w_i Y_i + n\tilde\lambda_\beta \|\beta_{(0)}\|_1, \\
& = (\Lambda-\Lambda^{-1}) \sum_{i=1}^{n_1} w_i \rho_\tau ( Y_i - \beta_0 - \beta_{(0)}^\T h_{(0)i}) + \sum_{i=1}^{n_1} w_i Y_i + n\tilde\lambda_\beta \|\beta_{(0)}\|_1,
\end{align*}
where $\tau= (\Lambda-1)/(\Lambda-\Lambda^{-1}) = \Lambda/(\Lambda+1)$, and the second step uses the decomposition
\begin{align*}
\beta_0 +\beta_{(0)}^\T h_{(0)i} =
Y_i + (- Y_i + \beta_0 +\beta_{(0)}^\T h_{(0)i})_+  -  (Y_i - \beta_0 - \beta_{(0)}^\T h_{(0)i})_+ .
\end{align*}
By the strong duality, the optimal value from the primal problem (\ref{eq:pro-relaxed-prf4}) is equal to
\begin{align*}
& \min_{\beta_0, \beta_{(0)}} \left\{ (\Lambda-\Lambda^{-1}) \sum_{i=1}^{n_1} w_i \rho_\tau ( Y_i - \beta_0 - \beta_{(0)}^\T h_{(0)i}) + \sum_{i=1}^{n_1} w_i Y_i + n\tilde\lambda_\beta \|\beta_{(0)}\|_1 \right\} \\
& = \sum_{i=1}^{n_1} w_i Y_i +  (\Lambda-\Lambda^{-1})  \min_{\beta_0, \beta_{(0)}} \left\{\sum_{i=1}^{n_1} w_i \rho_\tau ( Y_i - \beta_0 - \beta_{(0)}^\T h_{(0)i}) + n \lambda_\beta \|\beta_{(0)}\|_1 \right\} .
\end{align*}
Substituting this into (\ref{eq:pro-relaxed-prf1}) gives the desired result.

\subsection{Proof of Proposition \ref{pro:dhat-expan}}

The result follows because
\begin{align*}
& \quad \left| \lambda_\beta\| (\hat\beta_{\mytext{RWL},1+})_{1:p} \|_1 - \lambda_\beta \| (\bar\beta_{\mytext{WL},1+})_{1:p} \|_1 \right| \\
& \le \lambda_\beta \| (\hat\beta_{\mytext{RWL},1+})_{1:p}- (\bar\beta_{\mytext{WL},1+})_{1:p} \|_1
\le A_1 M_1 (|S_{\bar\gamma}|+ |S_{\bar\beta}|) \lambda_0^2
\end{align*}
by the triangle inequality and the error bound (\ref{eq:beta-error2}).

\subsection{Proofs of Lemma \ref{lem:sen-model3} and related results}

First, we show Lemma \ref{lem:sen-model3}. By Lemma 2.1 in Yadlowsky et al.~(2022),
model (\ref{eq:sen-model3}) with $U^1$ replaced by $U$ satisfying $T \perp (Y^0,Y^1) | (X,U)$ is
equivalent to the following model: for any $y$ and $y^\prime$,
\begin{align}
\lambda^*_1 (y, X) \le \Gamma \lambda^*_1 ( y^\prime, X) . \label{eq:sen-model3c}
\end{align}
This result remains valid if $T \perp (Y^0,Y^1) | (X,U)$ is modified to $T \perp Y^1 | (X,U^1)$.
Then model (\ref{eq:sen-model3b}) directly implies model (\ref{eq:sen-model3c}),
by combining $\lambda^*_1 (y, X) \le r(X) \Gamma$ and $r(X) \le \lambda^*_1(y^\prime, X)$ under model (\ref{eq:sen-model3b}).
Conversely, model (\ref{eq:sen-model3c}) implies model (\ref{eq:sen-model3b}) by
taking $r(X) = \inf_y \lambda^*_1 (y,X)$, which satisfies
 $\Gamma^{-1} \le r(X) \le 1$ because $E \{ \lambda^*_1(Y,X) | T=1,X\} = 1$.
In fact, if $r(x_0) >1$ for some $x_0$, then $\lambda^*_1(y,x_0) >1$ for all $y$,
which contradicts  $E \{ \lambda^*_1(Y,X) | T=1,X=x_0\} = 1$.
If $ r(x) < \Gamma^{-1}$ for some $x_0$, then there exists some $y_0$ such that $\lambda^*_1 (y_0,x_0) < \Gamma^{-1}$.
Under model (\ref{eq:sen-model3c}), this implies that
$\lambda^*_1(y,x_0) \le \Gamma \lambda^*_1 (y_0,x_0) <1$ for all $y$, which  contradicts  $E \{ \lambda^*_1(Y,X) | T=1,X=x_0\} = 1$.
Therefore, model (\ref{eq:sen-model3b}) is equivalent to model (\ref{eq:sen-model3c})
and also to model (\ref{eq:sen-model3}).

Second, we use Lemma \ref{lem:sen-model3} to show that
model (\ref{eq:sen-model}) implies model (\ref{eq:sen-model3c}) with $\Gamma=\Lambda^2$,
and model (\ref{eq:sen-model3b}) with $\Gamma=\Lambda$ implies model (\ref{eq:sen-model}).
In fact, model (\ref{eq:sen-model}) implies model (\ref{eq:sen-model3c}) with $\Gamma=\Lambda^2$,
by combining $\lambda^*_1(y,X) \le \Lambda$ and $\Lambda^{-1} \le \lambda^*_1 (y^\prime,X)$.
Model (\ref{eq:sen-model3b}) with $\Gamma=\Lambda$ says
$ r (X) \le \lambda^*_1 (Y^1, X) \le r(X) \Lambda$ and $\Lambda^{-1} \le r(X) \le 1$,
which directly  implies model (\ref{eq:sen-model}): $ \Lambda^{-1} \le \lambda^*_1 (Y^1, X) \le \Lambda$.

\subsection{Proof of Proposition \ref{pro:matching}}

(i) Denote $p^*(X) =  P( Y \le q^*_1(X) | T=1, X)$.
Then $r^*(X) =\{ p^*(X) + \Gamma (1- p^*(X)) \}^{-1}$.
By definition of $\tau^*(X)$, direct calculation shows
\begin{align*}
\tau^*(X) & =  \frac{r^*(X) \Gamma -1} {r^*(X) \Gamma - r^*(X)} \\
& =\left\{ \frac{\Gamma}{ p^*(X) + \Gamma (1- p^*(X))} - 1 \right\}  \frac{p^*(X) + \Gamma (1- p^*(X)) }{\Gamma-1} = p^*(X).
\end{align*}
This gives the desired result.

(iii) First, we show that
\begin{align}
 & \tilde Y_{+} (q^*_1) = q^*_1 (X) +  r^*(X) \left\{ \Gamma (Y- q^*_1(X))_+ - (q^*_1(X) - Y)_+  \right\}, \label{eq:pro-matching-prf}
\end{align}
where $\tilde Y_{+} (q_1)$ is defined with $(\Lambda_1(X), \Lambda_2(X))$ set to $(\Lambda^*_1(X), \Lambda^*_2(X))$.
In fact, by direct calculation using result (i), we have
\begin{align*}
& q^*_1 (X) +  r^*(X) \left\{ \Gamma (Y- q^*_1(X))_+ - (q^*_1(X) - Y)_+  \right\}\\
& = Y + (q^*_1 (X)-Y) + r^*(X) \left\{ \Gamma (Y- q^*_1(X))_+ - (q^*_1(X) - Y)_+  \right\} \\
& = Y + (r^*(X)\Gamma-1) \Gamma (Y- q^*_1(X))_+ + (1 - r^*(X)) (q^*_1(X) - Y)_+  \\
& = Y + (\Lambda^*_2(X) -1) \Gamma (Y- q^*_1(X))_+ + (1- \Lambda^*_1(X)) (q^*_1(X) - Y)_+  = \tilde Y_{+} (q^*_1) .
\end{align*}
Next, writing
\begin{align*}
(1-T) q^*_1(X) = T  \frac{1-\pi^*(X)}{\pi^*(X)}q^*_1(X)  - \left\{\frac{T}{\pi(X)} -1\right\} q^*_1(X) ,
\end{align*}
and then using (\ref{eq:pro-matching-prf}), we have by direct calculation
\begin{align*}
& \psi_{+} (O; \pi^*, q^*_1, r^*) =TY + (1-T) q^*_1 (X) +  T \frac{1-\pi^*(X)}{\pi^*(X)} r^*(X) \left\{ \Gamma (Y- q^*_1(X))_+ - (q^*_1(X) - Y)_+  \right\}\\
& = TY +  T \frac{1-\pi^*(X)}{\pi^*(X)}  \tilde Y_{+} (q^*_1)
- \left\{\frac{T}{\pi(X)} -1\right\} q^*_1(X) ,
\end{align*}
which yields $\varphi_{+, (\Lambda^*_1,\Lambda^*_2)} (O; \pi^*,\eta^*_1,q^*_1 ) $ because
\begin{align}
& E \{ \tilde Y_{+} (q^*_1)  | T=1, X \} \nonumber \\
& = q^*_1 (X) +  r^*(X) E \left\{ \Gamma (Y- q^*_1(X))_+ - (q^*_1(X) - Y)_+  | T=1, X\right\} = q^*_1(X),\label{eq:pro-matching-prf2}
\end{align}
with $q^*_1(X)$ satisfying $ E \{ \Gamma (Y- q^*_1(X))_+ - (q^*_1(X) - Y)_+ | T=1, X \})= 0$.

(ii) This follows from (iii), because
\begin{align*}
 & \mu^{1+}_\Gamma  =  E \{ \psi_{+} (O; \pi^*, q^*_1, r^* )\}, \\
 & \mu^{1+}_{(\Lambda^*_1,\Lambda^*_2)} = E\{ \varphi_{+, (\Lambda^*_1,\Lambda^*_2)} (O; \pi^*,\eta^*_1,q^*_1 ) \},
\end{align*}
with $q^* _1(X) = q^*_{1,\tau^*}(X)$ from result (i).
Alternatively, the result also follows from (\ref{eq:mu1+Q-ext2}), (\ref{eq:mu1+Q-latent}), and (\ref{eq:pro-matching-prf2}).

\vspace{.3in}
\centerline{\bf\Large Supplement References}

\begin{description}\addtolength{\itemsep}{-.15in}

\item Belloni, A. and Chernozhukov, V. (2011) $\ell_1$-penalized quantile regression in high-dimensional sparse models, {\em Annals of Statistics}, 39, 82-130.

\item Bickel, P., Ritov, Y., and Tsybakov, A.B. (2009) Simultaneous analysis of Lasso and Dantzig selector, {\em Annals of Statistics}, 37, 1705-1732.

\item Manski, C.F. (1988) {\em Analog Estimation Methods in Econometrics}, New York: Chapman \& Hall.

\item Pollard, D. (1991) Asymptotics for least absolute deviation regression estimators, {\em Econometric Theory}, 7, 186-199.

\item White, H. (1982) Maximum likelihood estimation of misspecified models, {\em Econometrica}, 50, 1-25.
\end{description}

\end{document}